\documentclass[a4paper,10pt]{article}
\usepackage[utf8x]{inputenc}
\usepackage{graphicx}
\usepackage{amsmath}
\usepackage{amssymb}
\usepackage{bbold}
\usepackage{geometry}
\usepackage{afterpage}
\usepackage{cite}
\def\lsim{\;\raise0.3ex\hbox{$<$\kern-0.75em\raise-1.1ex\hbox{$\sim$}}\;}
\def\gsim{\;\raise0.3ex\hbox{$>$\kern-0.75em\raise-1.1ex\hbox{$\sim$}}\;}

\begin{document}
\begin{titlepage}
 \vspace*{0.1cm}
 \rightline{TTP 13-009}\rightline{DESY 13-056}

\begin{center}
 {\bf Dimension $7$ operators in the $b\to s$ transition.}

\vspace{0.5cm}
{G.~Chalons\footnote{guillaume.chalons@kit.edu}, F.~Domingo\footnote{florian.domingo@desy.de}}\\

\vspace{4mm}
$^1${\it Institut f\"ur Theoretische Teilchenphysik, \\
Karlsruhe Institute of Technology, Universit\"at Karlsruhe \\
Engesserstraße 7, 76128 Karlsruhe, Germany}

\vspace{1mm}
$^2${\it Deutsches Elektronen SYnchrotron,\\ Notkestraße 85, 22607 Hamburg, Germany}
\vspace{10mm}

\end{center}

\begin{abstract}
\noindent We extend the low-energy effective field theory relevant for $b\to s$ transitions up to operators of mass-dimension $7$ and compute the 
associated anomalous-dimension matrix. We then compare our findings to the known results for dimension $6$ operators and derive a 
solution for the renormalization group equations involving operators of dimension $7$. We finally apply our analysis to a particularly
simple case where the Standard Model is extended by an electroweak-magnetic operator and consider limits on this scenario from the 
decays $B_s\to\mu^+\mu^-$ and $B\to K\nu\bar{\nu}$. 
\end{abstract}

\end{titlepage}

\section{Introduction}
Flavour-violating processes are well-known as a central test of the Standard Model (SM). The pattern conceded to flavour transitions is indeed
particularly constrained in this model, the Cabibbo-Kobayashi-Maskawa (CKM) matrix encoding the only source of flavour-breaking effects 
while only the charged currents of the weak interaction convey flavour-violation at tree-level. Consequently, flavour transitions also 
define closely watched observables in the quest for new physics and, due to the absence of positive deviations from the SM predictions, set
serious constraints on the forms that physics beyond the SM (BSM) could take. One of the latest results is the observation by the LHCb 
collaboration of the decay ${B}_s^0\to\mu^+\mu^-$ \cite{Aaij:2012nna}, with a branching ratio very compatible with the SM expectations (refer to
\cite{Buras:2013uqa} for a recent summary):
\begin{equation}
 BR({B}_s^0\to\mu^+\mu^-)^{\mbox{\small exp.}}=(3.23 \pm 0.27)\cdot 10^{-9}\ \ \ ;\ \ \ BR({B}_s^0\to\mu^+\mu^-)^{\mbox{\small SM}}=(3.56 \pm 0.18)\cdot 10^{-9}
\end{equation}

\noindent On the theoretical side, dedicated tools have been devised in order to study flavour-violating processes, in the form of low-energy 
effective field theories (EFT; refer to e.g.\ \cite{Buchalla:1995vs} for a review). These EFT's allow for a separation of the long distance, low-energy 
strong interaction effects and the short-distance, flavour-changing physics. In the context of $B$-physics, at the level of terms of dimension $4$ and
smaller, the relevant EFT consists in a QED$\times$QCD 
model with five quark-flavours ($u$, $d$, $c$, $s$, $b$) and $3$ charged leptons ($e$, $\mu$, $\tau$), as well as neutrinos.
Then, in the low-energy processes involving those fields, at a scale $\mu_b\sim M_B$, the impact of higher-energy (top/Electroweak/Higgs/BSM;
we assume here that there are no further low-energy `invisible' fields) physics can be essentially encoded within operators of dimension $>4$, 
collectively defining the `effective hamiltonian'. The strong-interaction problem then consists in evaluating the $S$-matrix elements driven by
these operators among physical (mesonic/baryonic) states: this question is answered, either through lattice-QCD, QCD sum rules,
heavy-quark expansions and other theoretical descriptions of the non-perturbative strong-interaction effects, or phenomenologically, through an 
identification of the decay-constants by comparison with a few standard channels. The short-distance problem is summarized within the couplings 
multiplying the operators. Those must be matched at high-energy with the predictions of the `more fundamental' theory (Standard Model, 
supersymmetry-inspired models, etc.): scattering amplitudes in both the EFT and the `full-theory' are equated at the matching scale $\mu_0\gsim M_W$, 
hence defining a boundary condition for the parameters of the EFT in terms of those of the underlying model.

\noindent To relate these two scales, $\mu_b$ and $\mu_0$, one relies on the Renormalization Group Equations (RGE) driven by the renormalization
of the EFT: leading logarithmic contributions can thus be consistently resummed. It is mostly in this part of the procedure that we will be 
interested in the following.

\noindent To fix notations, let us write the lagrangian density of the EFT under consideration:
\begin{equation}
 {\cal L}_{eff}=-\frac{1}{4}F_{\mu\nu}F^{\mu\nu}-\frac{1}{4}G_{\mu\nu}^aG^{a\,\mu\nu}\ 
+\ \imath\bar{f}\left[\gamma^{\mu}D_{\mu}-m_f\right]f\ 
-\ {\cal H}_{eff}^{(\mbox{\tiny dim}\geq5)}
\end{equation}
with $f=u,d,c,s,b,e,\mu,\tau,\nu_{e,\mu,\tau}$, $D_{\mu}=\partial_{\mu}-\imath e Q_f A_{\mu}-\imath g_S G^a_{\mu}$ the covariant derivative, 
$m_f$ the mass of the fermion $f$, $Q_f$, its charge; $F_{\mu\nu}=\partial_{\mu}A_{\nu}-\partial_{\nu}A_{\mu}$ and $G_{\mu\nu}=G_{\mu\nu}^aT^a
=(\partial_{\mu}G^a_{\nu}-\partial_{\nu}G^a_{\mu}+g_Sf^{abc}G_{\mu}^bG_{\nu}^c)T^a$ represent the electromagnetic and gluonic field tensors, 
respectively; $T^a$ and $f^{abc}$ denote the $SU(3)_c$ generators and structure constants; $e$ and $g_S$ respectively stand for the elementary 
electric charge and the strong coupling constant; $\alpha\equiv\frac{e^2}{4\pi}$, $\alpha_S\equiv\frac{g_S^2}{4\pi}$; ${\cal H}_{eff}^{(
\mbox{\tiny dim}\geq5)}$ is the effective hamiltonian.

\noindent The question arising at this point is that of the operators that one needs to consider in order to describe $b\to s$ transitions. Sensibly, people 
have considered, up to now, only operators of the lowest possible mass-dimension, that is dimension-$6$ operators\footnote{In fact, the magnetic and 
chromo-magnetic operators have mass-dimension $5$. However, they always appear, e.g.\ in the SM, with an additional mass-suppression, lowering their
scale to an apparent dimension $6$.}. This approach is justified by the suppression factor of $\frac{m_b}{M_Z}\sim5\cdot10^{-2}$ which is expected 
for higher-dimensional operators. Moreover, it was (with reason) regarded as sufficient to confine to the smallest subset of dimension-6 operators 
that would close under renormalization and for which the classical models (SM, supersymmetry-inspired, etc.) would generate a non-trivial 
contribution:
\begin{equation}
 \ {\cal H}_{eff}^{(\mbox{\tiny dim}=6)}=\sum_i C_i(\mu)O_i^{(\mbox{\tiny dim}=6)}(\mu)
\end{equation}
where the list of operators $O_i$ can be read in e.g.\ \cite{Buchalla:1995vs}, \cite{Chetyrkin:1996vx} or \cite{Gambino:2003zm} (with small 
variations; note that a factor $G_F/\sqrt{2}$ is conventionally factored out in the usual notations). The determination at leading order of 
the anomalous-dimension matrix for the four-quark operators of this subset is quite old: 
\cite{Gaillard:1974nj,Altarelli:1974exa,Shifman:1976ge,Gilman:1979bc,Guberina:1979ix}. The renormalization of the magnetic and chromomagnetic
operators can be found in \cite{Shifman:1976de}, while \cite{Grinstein:1987vj} included the mixing of those with the four-quark operators
(a two-loop effect). One may refer to \cite{Gilman:1979ud,Eeg:1988yd,Flynn:1988ve} for the analysis of the semi-leptonic operators. Later works
have focussed on next-to-leading order ($O(\alpha_S)$) \cite{Buras:1992tc,Ciuchini:1993vr,Misiak:1994zw,Chetyrkin:1997gb,Buras:2002tp,Gambino:2003zm}, 
electroweak ($O(\alpha)$; see e.g.\ \cite{Gambino:2001au}) and finally next-to-next-to-leading QCD ($O(\alpha_S^2)$; refer e.g.\ to the summary in 
\cite{Misiak:2006zs}) effects: this formidable amout of work allows for a theoretical prediction, e.g.\ in the SM $\bar{B}\to X_s\gamma$ decay,
competitive with experimental bounds.

\noindent In this paper we choose to adopt a different, more unprejudiced if somewhat more anecdotical, approach to the renormalization of the 
$b\to s$ EFT: we shall consider all possible (on-shell) operators up to mass-dimension $7$ and compute the corresponding anomalous-dimension matrix
at one-loop QCD order. As far as we know, no attempt has ever been made in that 
direction, due to the suppression of the order $\frac{m_b}{M_Z}\sim5\cdot10^{-2}$ which one expects for higher-dimensional operators. In fact, if 
one considers the matching conditions in the particular case of the SM, with its restricted flavour-changing currents, the suppression would be even 
larger. The inclusion of higher-dimension operators hence admitedly appears in this concrete case more as a curiosity than a compelling necessity.
There are however several reasons why analysing dimension $7$ effects may not be completely irrelevant:
\begin{enumerate}
 \item From the point of view of precision physics, one observes that $\alpha_S(M_Z)\sim \frac{m_b}{M_Z}$: with increasing precision in the SM 
evaluation as well as experimental measurements, observables shall eventually become sensitive to dimension $7$ effects.
 \item Certain new-physics contributions are actually `hidden' dimension $7$ effects; a simple example lies in the famous Higgs-penguin contributions 
to dimension $6$ $bsll$-operators \cite{Buras:2002vd}, relevant e.g.\ in supersymmetric models at large $\tan\beta$: the corresponding coefficients 
actually contain a factor $m_b$, formally increasing their order to dimension $7$. Note however that other dimension $7$ effects are not expected to 
receive an equally large $\tan\beta$-enhancement, although they should be relevant already at subleading order \cite{Thorsten}.
 \item New-physics in dimension $6$ operators is already stringently constrained, essentially enforcing the usual `Minimal Flavour Violation'
condition. A possible strategy to account for this absence of new-physics in flavour observables would be to reject it on operators of higher 
dimension. Similar proposals have been made in the neutrino sector to concile neutrino masses, baryon/lepton-number and lepton-flavour
violating processes \cite{Weldon:1980gi}. Note that, here, we do not propose a mechanism that would ensure the suppression of new physics
by rejecting it on operators of dimension $\geq7$ (although this might be achievable, e.g.\ by assigning adequate charges under a 
discrete symmetry), but we simply mention this possibility as a motivation to consider such operators.
 \item If one parametrizes new physics blindly in an expansion of SM dimension $6$ operators \cite{Buchmuller:1985jz,Grzadkowski:2010es}, it turns
out that certain operators would only have a dimension-$7$ signature at low energy. Including dimension $7$ operators for the $b\to s$ transition
thus naturally enters an unprejudiced analysis of physics BSM.
\end{enumerate}

\noindent In the next section, we shall derive a list of all the (on-shell) operators of mass-dimension $5$, $6$ and $7$ which may intervene in the 
$b\to s$ EFT. We shall then detail the calculation of their ultraviolet (UV)-divergences at one-loop QCD order, before we proceed to the 
renormalization and establish the RGE's. The following section will be dedicated to the solution of these RGE's in some specific cases. Finally, we 
shall illustrate our discussion by presenting a concrete, if naive, case where our analysis of dimension $7$ operators apply. A short conclusion 
will eventually summarize our achievements.

\section{Operators of dimension $5-7$ in the $b\to s$ transition}
\subsection{List of operators}\label{list}
The first step of our analysis consists in establishing a list of all the operators of dimension $5$, $6$ and $7$ intervening in the 
$b\to s$ transition (but Lorentz + gauge invariant!). For simplicity, all the fields shall be taken on-shell, i.e.\ satisfy their equation of 
motion: this will be sufficient, at least for the leading-order calculation that we aim at. For a discussion concerning the relevance of on-shell
EFT's and in particular the question of applying only `naive' classical equations of motions (dismissing ghosts and gauge-fixing terms), we refer the reader
to \cite{Simma:1993ky}. We can obviously distinguish among three categories of operators: four-quark, two-quark + two lepton (semi-leptonic) and 
$\bar{b}s$-Gauge operators.

\noindent Let us start with four-fermion operators: the fermion fields already account for a mass-dimension $6$, leaving room for at most one 
covariant (for gauge-invariance) derivative, when one restricts to operators of mass-dimension $\leq7$. Considering chiral fermions (that is, 
we include projectors $P_{L,R}\equiv\frac{1\mp\gamma_5}{2}$ in the fermion products), there are at most three ways to contract the spinor 
algebra: $\mathbb{1}$ provides scalar currents, $\gamma^{\mu}$, vector currents and $\sigma^{\mu\nu}\equiv[\gamma^{\mu},\gamma^{\nu}]$, tensor 
currents\footnote{Note that the definition of $\sigma^{\mu\nu}$ which we adopt here for simplicity differs somewhat, by a factor $\imath/2$, 
from the one the reader may have encountered in the literature.}; any higher combination of $\gamma$-matrices can be reduced down to those three 
cases (through the use of the Levi-Civita tensor $\varepsilon_{\mu\nu\rho\sigma}$ and identities of the Dirac algebra): note that $\gamma_5$ only 
gives a sign, when applied on chiral fermions. Moreover, we may always keep the $b$ and $s$ of the flavour transition within the same current: 
other combinations are made redundant by the Fierz identities (see e.g.\ Appendix \ref{Fierz}).

\noindent Those considerations allow us to construct the relevant three classes of dimension $6$ operators:
\begin{enumerate}
 \item scalar $(\bar{b}P_{L,R}s)(\bar{f}P_{L,R}f)$;
 \item vector $(\bar{b}\gamma^{\mu}P_{L,R}s)(\bar{f}\gamma_{\mu}P_{L,R}f)$;
 \item tensor $(\bar{b}\sigma^{\mu\nu}P_{L,R}s)(\bar{f}\sigma_{\mu\nu}P_{L,R}f)$. Note that only two chiral combinations are possible for the tensor 
currents: $L\times L$ and $R\times R$; the other combinations, $L\times R$ and $R\times L$, are identically zero.
\end{enumerate}

\noindent One may then consider authentic dimension $7$ operators by incorporating one covariant derivative in the fermion products. Two possibilities 
appear: 
\begin{enumerate}
 \item contracting the Lorentz index of $D_{\mu}$ with a vector current
 \item contracting it with a tensor current, the second tensorial index being contracted with the second, vector current.
\end{enumerate}
Note however that, using the equations of motion on fermions ($(\imath\displaystyle{\not}\hspace{0.2mm}D-m_f)f=0$) and realizing partial integrations 
(i.e.\ adding a total derivative to the lagrangian density),
the resulting set of operators is largely redundant (together with the dimension $6$ operators). Indeed, the second possibility which we mentioned 
(`$D^{\mu}\gamma^{\nu}\otimes\sigma_{\mu\nu}$') can always be reduced down to operators of the first class (`$D_{\mu}\otimes\gamma^{\mu}$') plus 
dimension $6$ operators (multiplying fermion masses). Additionally, certain combinations of the operators of the first class reduce to dimension $6$
terms. We thus retain only two kinds of linearly independant operators:
\begin{enumerate}
 \item $[\bar{b}\imath(\overrightarrow{D}-\overleftarrow{D})^{\mu}P_{L,R}s](\bar{f}\gamma_{\mu}P_{L,R}f)$;
 \item $(\bar{b}\gamma^{\mu}P_{L,R}s)[\bar{f}\imath(\overrightarrow{D}-\overleftarrow{D})_{\mu}P_{L,R}f]$.
\end{enumerate}
For semi-leptonic operators, i.e.\ when $f$ is a lepton ($f=l$), the analysis is essentially over.

\noindent Let us therefore focus on four-quark operators ($f=q$). One should then also consider the contraction of the colour indices. Two 
possibilities arise:
\begin{enumerate} 
 \item product of two colour-singlet currents: colour-indices contracted between the $\bar{b}$ and $s$ on one side, $\bar{q}$ and $q$ on the other;
 \item product of two octet currents: colour-indices contracted between the $\bar{b}$ and $q$ on one side, $\bar{q}$ and $s$ on the other.
\end{enumerate}
In the special cases where $q=b,s$, however, Fierz identities make this distinction superfluous, so that we may consider only singlet products then
(refer to Appendix \ref{Fierz}). This is our final word for four-fermion operators.

\noindent We now turn to $\bar{b}s$-Gauge operators. The methodology follows that of \cite{Grinstein:1987vj} for the dimension $6$ operators: one may 
simply write all the possibilities to include covariant derivatives (at most four for operators of dimension $\leq7$) within the fermionic current.
Using partial integration and equations of motion, it turns out that all these covariant derivatives can be combined in field-strength tensors.
One thus simply needs to consider the possibilities to combine the indices of these field-strength tensors with those of the fermionic current: 
\begin{itemize}
 \item When only one field-strength is present, it can only contract with a (Lorentz) tensor current, and, in the case of the QCD field strength 
$G_{\mu\nu}^a$, with a $SU(3)_c$ octet current.
 \item When two field-strengths are present, we may either contract their Lorentz indices together -- thus reducing the fermionic current to a scalar -- 
via the metric or a $\varepsilon_{\mu\nu\rho\sigma}$ tensor --, or contract two of their Lorentz indices with a tensor current (the other two through 
the metric, or equivalently $\varepsilon_{\mu\nu\rho\sigma}$). In the case where only one QCD field-strength is involved (among the two), one needs 
again a $SU(3)_c$ octet on the fermionic side. When two QCD field-strengths are involved ($G_{\mu\nu}^aG_{\rho\sigma}^b$), the color indices may be contracted together
($\delta^{ab}$: colour-singlet), with a symmetric tensor ($\{T^a,T^b\}$) or with an antisymmetric tensor ($[T^a,T^b]$) (superposition of colour-octets 
+ singlet), depending of the compatibility of these structures with the Lorentz form of the fermionic current.
\end{itemize}

\noindent We now have all the ingredients to present the list of relevant operators. Note that their normalization is a priori free: our choice can be justified
a posteriori to ensure that all the RGE's intervene at the same, leading-order in $\alpha_S$. Note however that this choice may be slightly misleading,
for instance in the SM, for reasons that we will discuss in the next section, when we solve the RGE's.
\begin{itemize}
 \item $(\bar{b}s)(\bar{l}l)$ operators\footnote{We repeat that, for the tensor operators, only the $L\times L$ and $R\times R$ combinations
are relevant, which may not be obvious from our notation.}:
\begin{equation}\label{bsllop}\begin{cases}
 S^l_{L,R\ L,R}=\frac{\alpha}{\alpha_S} m_b(\bar{b}P_{L,R}s)(\bar{l}P_{L,R}l)\\
 V^l_{L,R\ L,R}=\frac{\alpha}{\alpha_S} m_b(\bar{b}\gamma^{\mu}P_{L,R}s)(\bar{l}\gamma_{\mu}P_{L,R}l)\\
 T^l_{L,R\ L,R}=\frac{\alpha}{\alpha_S} m_b(\bar{b}\sigma^{\mu\nu}P_{L,R}s)(\bar{l}\sigma_{\mu\nu}P_{L,R}l)\ \ \ \ ;\ \ \sigma^{\mu\nu}\equiv[\gamma^{\mu},\gamma^{\nu}]\\
 H^l_{L,R\ L,R}=\frac{\alpha}{\alpha_S}\left[\bar{b}\left(\imath\overrightarrow{\partial}-\imath\overleftarrow{\partial}+2Q_deA+2g_ST^aG^a
\right)^{\mu}P_{L,R}s\right](\bar{l}\gamma_{\mu}P_{L,R}l)\\
 \tilde{H}^l_{L,R\ L,R}=\frac{\alpha}{\alpha_S}(\bar{b}\gamma^{\mu}P_{L,R}s)
\left[\bar{l}\left(\imath\overrightarrow{\partial}-\imath\overleftarrow{\partial}+2Q_leA\right)_{\mu}P_{L,R}l\right]
\end{cases}\end{equation}
 \item $(\bar{b}s)(\bar{q}q)$ operators\footnote{We discard ${\cal S,V,T,H,\tilde{H}}^q$ for $q=b,s$ since they reduce to $S,V,T,H,\tilde{H}^q$ due
to Fierz transformations which are provided in Appendix \ref{Fierz}. The colour indices $\alpha$, $\beta$ of the fermions are displayed when not 
trivially contracted. Note finally that the ambiguous notation `$\overrightarrow{D}_{\mu}q_{\alpha}$' stands actually for $(D_{\mu}q)_{\alpha}$\ldots}:
\begin{equation}\label{bsqqop}\begin{cases}
 &S^q_{L,R\ L,R}=m_b(\bar{b}P_{L,R}s)(\bar{q}P_{L,R}q)\\
 &V^q_{L,R\ L,R}=m_b(\bar{b}\gamma^{\mu}P_{L,R}s)(\bar{q}\gamma_{\mu}P_{L,R}q)\\
 &T^q_{L,R\ L,R}=m_b(\bar{b}\sigma^{\mu\nu}P_{L,R}s)(\bar{q}\sigma_{\mu\nu}P_{L,R}q)\\
 &H^q_{L,R\ L,R}=\left[\bar{b}\imath\left(\overrightarrow{D}-\overleftarrow{D}
\right)^{\mu}P_{L,R}s\right](\bar{q}\gamma_{\mu}P_{L,R}q)\\
 &\tilde{H}^q_{L,R\ L,R}=(\bar{b}\gamma^{\mu}P_{L,R}s)
\left[\bar{q}\imath\left(\overrightarrow{D}-\overleftarrow{D}\right)_{\mu}P_{L,R}q\right]\\
 &{\cal S}^q_{L,R\ L,R}=m_b(\bar{b}_{\alpha}P_{L,R}s_{\beta})(\bar{q}_{\beta}P_{L,R}q_{\alpha})\\
 &{\cal V}^q_{L,R\ L,R}=m_b(\bar{b}_{\alpha}\gamma^{\mu}P_{L,R}s_{\beta})(\bar{q}_{\beta}\gamma_{\mu}P_{L,R}q_{\alpha})\\
 &{\cal T}^q_{L,R\ L,R}=m_b(\bar{b}_{\alpha}\sigma^{\mu\nu}P_{L,R}s_{\beta})(\bar{q}_{\beta}\sigma_{\mu\nu}P_{L,R}q_{\alpha})\\
 &{\cal H}^q_{L,R\ L,R}=\left[\bar{b}_{\alpha}\imath\left(\overrightarrow{D}-\overleftarrow{D}
\right)^{\mu}P_{L,R}s_{\beta}\right](\bar{q}_{\beta}\gamma_{\mu}P_{L,R}q_{\alpha})\\
 &{\cal \tilde{H}}^q_{L,R\ L,R}=(\bar{b}_{\alpha}\gamma^{\mu}P_{L,R}s_{\beta})
\left[\bar{q}_{\beta}\imath\left(\overrightarrow{D}-\overleftarrow{D}\right)_{\mu}P_{L,R}q_{\alpha}\right]
\end{cases}\end{equation}
 \item $(\bar{b}s)$-Gauge operators:
\begin{equation}\label{bsGop}\begin{cases}
 E_{L,R}=\frac{\imath em_b^2}{4\pi\alpha_S}\bar{b}\sigma^{\mu\nu}P_{L,R}s\,F_{\mu\nu}\\
 Q_{L,R}=\frac{\imath m_b^2}{g_S}\,\bar{b}\sigma^{\mu\nu}T^aP_{L,R}s\,G^a_{\mu\nu}\\
{\cal E}^S_{L,R}=\frac{\alpha}{\alpha_S}\bar{b}P_{L,R}s\,F_{\mu\nu}F^{\mu\nu}\\
 \tilde{\cal E}^S_{L,R}=\imath \frac{\alpha}{\alpha_S}\bar{b}P_{L,R}s\,F_{\mu\nu}\tilde{F}^{\mu\nu}\\
 {\cal E}^T_{L,R}=\frac{\alpha}{\alpha_S}\bar{b}\sigma^{\nu\rho}P_{L,R}s\,F_{\mu\nu}F^{\mu}_{\ \rho}\\
{\cal H}^S_{L,R}=\frac{e}{g_S}\bar{b}T^aP_{L,R}s\,F_{\mu\nu}G^{a\,\mu\nu}\\
\tilde{\cal H}^S_{L,R}=\frac{\imath e}{g_S}\bar{b}T^aP_{L,R}s\,F_{\mu\nu}\tilde{G}^{a\,\mu\nu}\\
{\cal H}^T_{L,R}=\frac{e}{g_S}\bar{b}\sigma^{\nu\rho}T^aP_{L,R}s\,F^{\mu}_{\ \nu}G^a_{\mu\rho}\\
 {\cal Q}^D_{L,R}=\bar{b}P_{L,R}s\,G_{\mu\nu}^aG^{a\,\mu\nu}\\
 \tilde{\cal Q}^D_{L,R}=\imath \bar{b}P_{L,R}s\,G_{\mu\nu}^a\tilde{G}^{a\,\mu\nu}\\
{\cal Q}^S_{L,R}=\bar{b}\frac{\{T^a,T^b\}}{2}P_{L,R}s\,G^a_{\mu\nu}G^{b\,\mu\nu}\\
\tilde{\cal Q}^S_{L,R}=\imath \bar{b}\frac{\{T^a,T^b\}}{2}P_{L,R}s\,G^a_{\mu\nu}\tilde{G}^{b\,\mu\nu}\\
{\cal Q}^T_{L,R}=\bar{b}\sigma^{\nu\rho}\frac{[T^a,T^b]}{2}P_{L,R}s\,G^a_{\mu\nu}G^{b\,\mu}_{\,\ \ \rho}
\end{cases}\end{equation}
\end{itemize}
This list determines the effective hamiltonian of our EFT (the generic notation $O^i$ spans the whole list):
\begin{equation}
 {\cal H}_{eff}^{(\mbox{\tiny dim}\leq7)}=\sum_i C_i(\mu)O^i(\mu)+h.c.
\end{equation}
Summing over all the flavours and chiralities, we count $251$ terms.

\noindent For the purpose of illustration, let us sketch the typical matching conditions that one could expect for these operators in a 
Minimal-Flavour-Violating model (the CKM matrix elements are written as $V_{qq'}$). Only the operators $(\bar{b}s)(\bar{q}q)$ with $q=c,u$ would 
receive a contribution at tree-level: $C_{q=c,u}\sim\frac{G_FV_{qs}V_{qb}^*}{m_b}$ (the case $q=u$ may be neglected because of the CKM suppression). 
All the other contributions would arise at the loop-level, $C_{\mbox{\tiny dim 6}}\sim\frac{\alpha_S}{4\pi}\frac{G_FV_{ts}V_{tb}^*}{m_b}$ and 
$C_{\mbox{\tiny dim 7}}\sim\frac{\alpha_S}{4\pi}\frac{G_FV_{ts}V_{tb}^*}{m_b}\frac{m_b}{M_W}$ for the operators present at dimension 
$6$ and $7$ respectively.

\subsection{UV-divergent amplitudes involving the dimension $5-7$ operators}\label{amplitudes}
Before renormalizing the EFT, we compute the divergences associated with QCD loops. We perform this calculation in the most naive conceivable way: 
in the Feynman-t'Hooft gauge and dimensional regularization $D=4-2\varepsilon$; we keep only the divergent terms $\left[2-\frac{D}{2}\right]^{-1}$. 
Note that we will assume that the couplings $C_i$ come together with an electric charge factor $e$ whenever a photon or a lepton appear in the 
external lines: in this fashion $(\bar{b}s)(\bar{l}l)$ shall be regarded as electroweakly suppressed with respect to $(\bar{b}s)(\bar{q}q)$, rejecting 
$(\bar{b}s)(\bar{l}l)$ UV-contributions to $(\bar{b}s)(\bar{q}q)$ operators to a subdominant order. The computation can be organized in `blocks':
\begin{itemize}
\item $(\bar{b}s)(\bar{l}l)$ contributions to the $(\bar{b}s)(\bar{l}l)$ amplitude: see Fig.\ref{bsll_bsll}
\begin{figure}[h!]
\begin{center}
 \includegraphics[width=11.4cm]{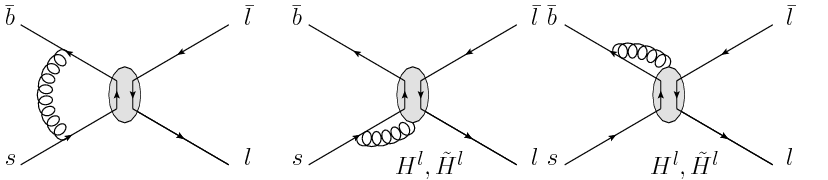}
\end{center}
\caption{Contributions from the $(\bar{b}s)(\bar{l}l)$ operators to the $\bar{b}s\to\bar{l}l$ amplitude. Here as in the following plots, the grey
blob represents a vertex associated with the operators of mass-dimension $5-7$. The subscript `$H^l,\tilde{H}^l$' under some of the diagrams
indicates that such diagrams are relevant only for this kind of operators. Feynman diagrams are plotted with JaxoDraw \cite{Binosi:2008ig}.
\label{bsll_bsll}}\hrule
\end{figure}
 \item $(\bar{b}s)(\bar{q}q)$ contributions to the $(\bar{b}s)$-Gauge amplitude.
\begin{figure}[h]
\begin{center}
 \includegraphics[width=16cm]{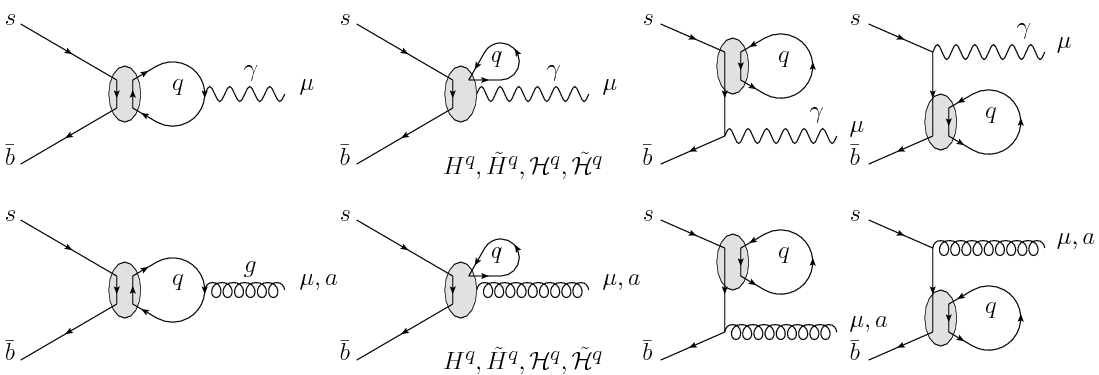}
\end{center}
\caption{Contributions from the $(\bar{b}s)(\bar{q}q)$ operators to the $\bar{b}s\to\gamma/g$ amplitude. Again, as in the following figures,
the label `$H^q,\tilde{H}^q,{\cal H}^q,\tilde{\cal H}^q$' indicates that the corresponding diagrams intervene for such operators only.\label{bsqq_bsG}}\hrule
\end{figure}

\noindent Such contributions involve a quark loop. While the external fermions are explicitly taken on-shell,
through the application of their equations of motion, explicit off-shell terms appear for the external photons and gluons: such terms must
organize as an equation of motion for the gauge-boson line, which provides us with a cross-check of our calculation. They do not matter for
the renormalization so that we will not keep them explicitly in the divergent amplitudes. The amplitudes $(\bar{b}s)-\gamma^*$ or $(\bar{b}s)
-g^*$ are relevant for the divergent amplitudes $(\bar{b}s)(\bar{l}l)$, $(\bar{b}s)(\bar{q}'q')$ and $(\bar{b}s)(gg)$ though, which we will 
present later.
\begin{figure}[t]
\begin{center}
 \includegraphics[width=16cm]{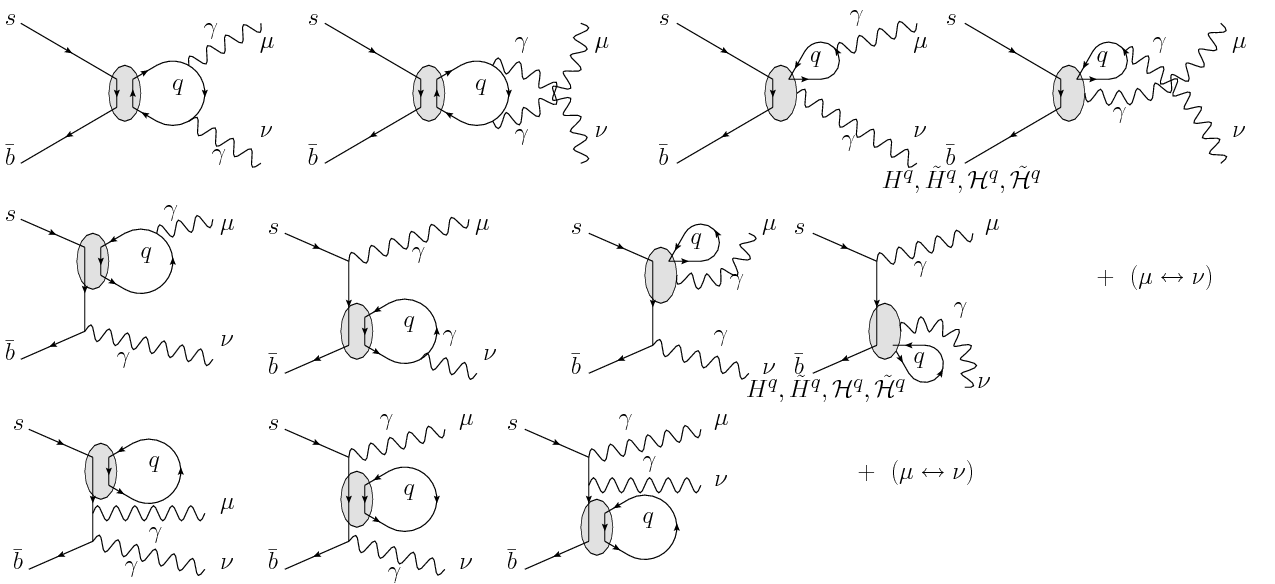}
\end{center}
\caption{Contributions from the $(\bar{b}s)(\bar{q}q)$ operators to the $\bar{b}s\to2\gamma$ amplitude. $+(\mu\leftrightarrow\nu)$ signals
that one must add the `mirror' diagrams exchanging the two photon lines.\label{bsqq_bsGG}}\hrule
\end{figure}\begin{figure}[d]
\begin{center}
 \includegraphics[width=16cm]{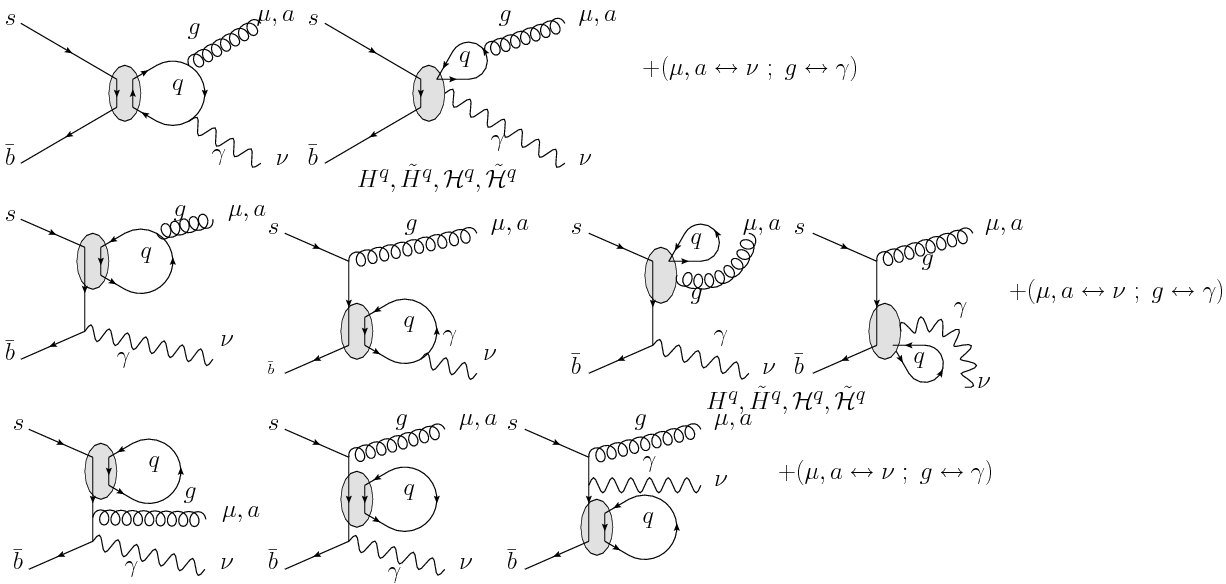}
\end{center}
\caption{Contributions from the $(\bar{b}s)(\bar{q}q)$ operators to the $\bar{b}s\to\gamma g$ amplitude.\label{bsqq_bsGg}}\hrule
\end{figure}
\clearpage
\begin{figure}[t]
\begin{center}
 \includegraphics[width=16cm]{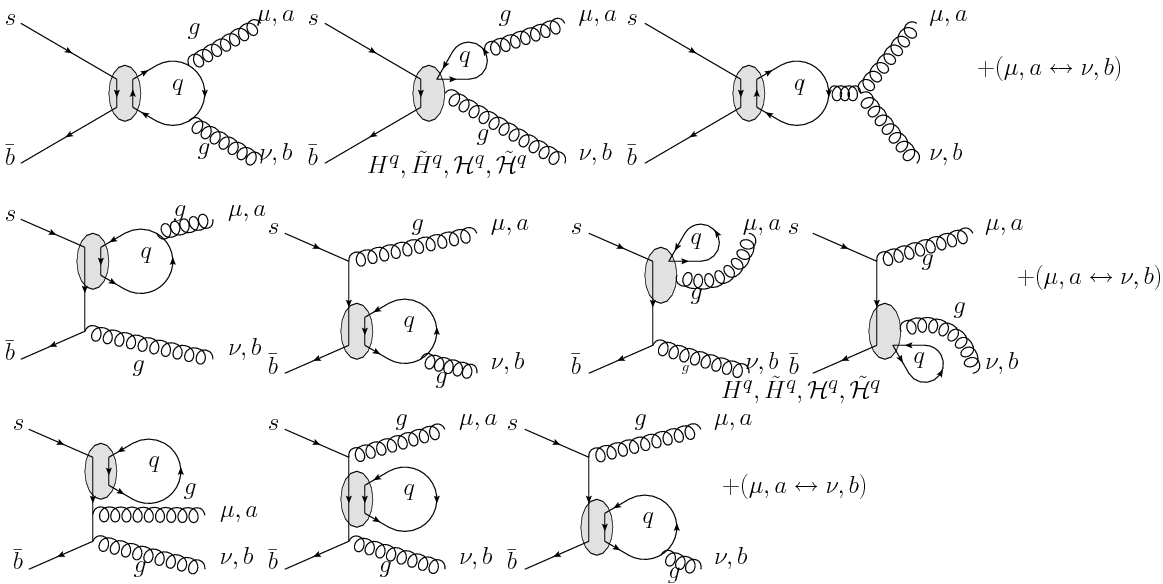}
\end{center}
\caption{Contributions from the $(\bar{b}s)(\bar{q}q)$ operators to the $\bar{b}s\to2g$ amplitude.\label{bsqq_bsgg}}\hrule
\end{figure}

\noindent The contributions to the amplitudes $\bar{b}s\to\gamma/g$ are shown in Fig.\ref{bsqq_bsG}. The diagrams intervening in $\bar{b}s\to2\gamma/\gamma g/2g$
are depicted in Fig.\ref{bsqq_bsGG},\ref{bsqq_bsGg},\ref{bsqq_bsgg}. Note that in the case of the dimension $6$ four-quark operators, no contribution to the 
$\bar{b}s-VV'$ (where $V,V'\in\{\gamma,g\}$) operators is expected: the dimension $6$ basis is stable and does not require dimension $7$ counterterms. The 
corresponding amplitudes in Fig.\ref{bsqq_bsGG},\ref{bsqq_bsGg},\ref{bsqq_bsgg} therefore provide a simple check of the matching conditions determined in 
Fig.\ref{bsqq_bsG}. For the authentic dimension $7$ operators, however, the calculation of the $\bar{b}s-VV'$ amplitudes is fully relevant.

\noindent The operators $(\bar{b}s)(\bar{q}q)$ with $q=b,s$ deserve a particular attention. Beyond the `s-channel' diagrams, similar to those of the 
other $(\bar{b}s)(\bar{q}q)$ operators, they indeed allow for a `t-channel' contribution: these can be viewed as `$\frac{\pi}{2}$-rotated' 
contributions associated with the ${\cal S}^q$, ${\cal V}^q$, ${\cal T}^q$, ${\cal H}^q$, $\tilde{\cal H}^q$ operators, which, in the case of $q=b,s$, 
coincide with ${S}^q$, ${V}^q$, ${T}^q$, ${H}^q$, $\tilde{H}^q$, via Fierz identities. Instead of computing the new `t-channel' contributions\footnote{
We, in fact, checked this relation explicitly.}, we thus 
use our generic result for the ${\cal S}^q$, ${\cal V}^q$, ${\cal T}^q$, ${\cal H}^q$, $\tilde{\cal H}^q$ operators and project it on ${S}^q$, 
${V}^q$, ${T}^q$, ${H}^q$, $\tilde{H}^q$ with the Fierz matrices $\tilde{V}^{b,s}_{\mbox{Fierz}}$ exchanging the two bases and defined in Appendix
\ref{Fierz}. A factor $-1$ appears in this process corresponding to the anticommutation of two fermion fields / the transformation of an open fermion
loop into a closed one.
\item $(\bar{b}s)(\bar{q}q)$ contributions to the $(\bar{b}s)(\bar{f}f)$ amplitude.

The contributions to the $\bar{b}s\to\bar{l}l$ amplitude proceed only from the exchange of an off-shell photon, as depicted on the first diagram
of Fig.\ref{bsqq_bsqq}. For the $\bar{b}s\to\bar{q}'q'$ amplitude, contributions originate similarly from the exchange of an off-shell gluon
(note that the photon exchange would be of a higher-order). Additional `diagonal' contributions are obtained for $q=q'$ through the dressing of
the $(\bar{b}s)(\bar{q}q)$ vertex by a gluon line. All the corresponding diagrams are depicted in Fig.\ref{bsqq_bsqq}.
\begin{figure}[t]
\begin{center}
 \includegraphics[width=16cm]{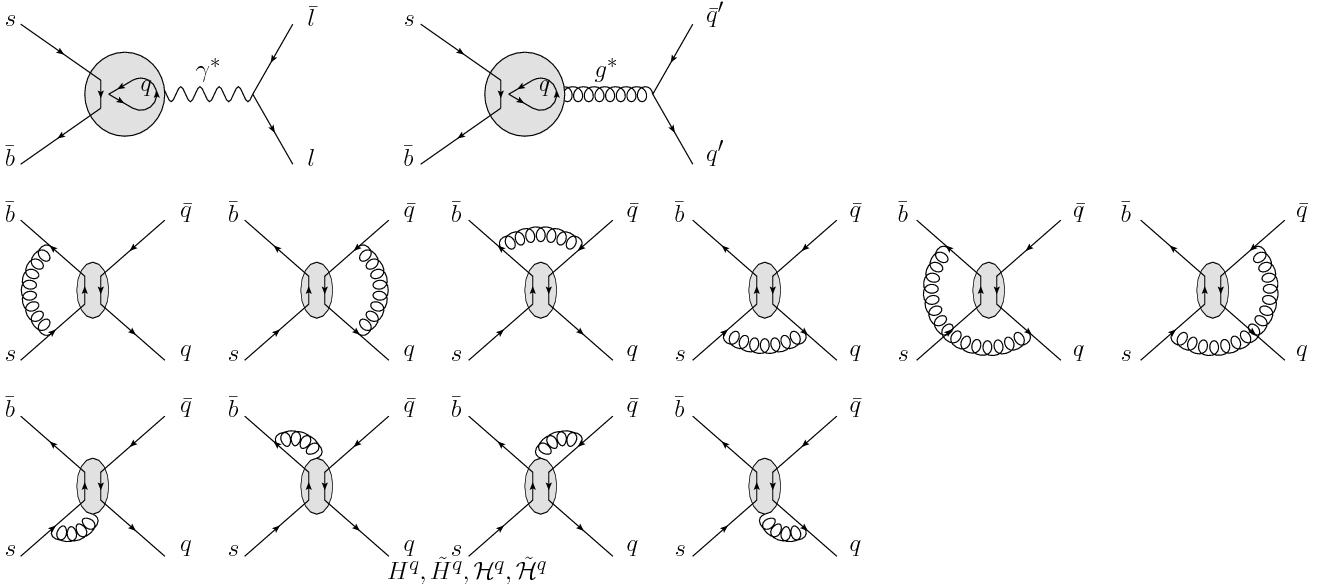}
\end{center}
\caption{Contributions from the $(\bar{b}s)(\bar{q}q)$ operators to the $\bar{b}s\to\bar{l}l$ and $\bar{b}s\to\bar{q}q$ amplitudes. The grey
blob with a quark loop stands for all possible $(\bar{b}s)(\bar{q}q)$ contributions to an off-shell photon/gluon: refer to Fig.\ref{bsqq_bsG}.\label{bsqq_bsqq}}\hrule
\end{figure}
\begin{figure}[d]
\begin{center}
 \includegraphics[width=12.9cm]{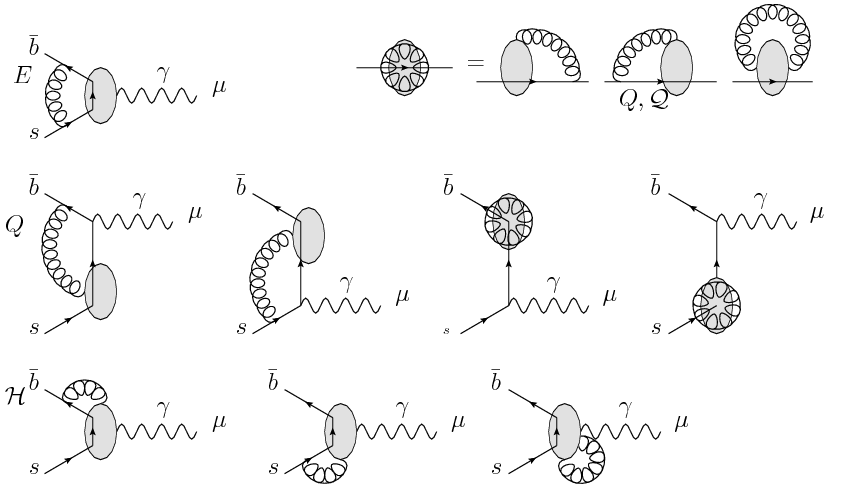}
\end{center}
\caption{Contributions from the $(\bar{b}s)$-Gauge operators ($E$, $Q$ and ${\cal H}\leftrightarrow {\cal H}^S,\tilde{\cal H}^S,{\cal H}^T$) to 
the $\bar{b}s\to\gamma$ amplitude. We also define the $b\to s$ self-energy associated with the operators $Q$ (and ${\cal Q}\leftrightarrow
{\cal Q}^D,\tilde{\cal Q}^D,{\cal Q}^S,\tilde{\cal Q}^S,{\cal Q}^T$) and depicted on this figure, as well as the following ones, by a fermion
line with a grey blob and a gluon bubble.\label{bsG_bsG}}\hrule
\end{figure}\clearpage

 \item $(\bar{b}s)$-Gauge contributions to the $(\bar{b}s)$-Gauge amplitudes.

\noindent Once again, off-shell photons and gluons appear within the calculation of these amplitudes, offering a nice crosscheck of the result, given that 
they must combine into an equation of motion for the corresponding field. Although such terms will not matter for the renormalization, 
$\bar{b}s\to\gamma^*$ and $\bar{b}s\to g^*$ will be relevant for $\bar{b}s\to \bar{l}l/\bar{q}{q}/gg$. Another check, particularly in the
case where two gauge bosons appear in the final state, is simply the projectibility of the result on the basis of operators that we have defined: indeed
the results typically reach this form only after the summation of several diagrams, the coefficients and colour-factors of which must combine
in the appropriate way. Note that only `authentic' dimention $7$ operators need to be included in the amplitudes $\bar{b}s\to\gamma\gamma/\gamma g/gg$:
the presence of $E$ or $Q$ in the divergences of such amplitudes (substracting however the counterterms of $E$ and $Q$) would signal an
instability of the dimension $6$ basis, which would require dimension $7$ counterterms. This evidently does not occur.
\begin{figure}[h]
\begin{center}
 \includegraphics[width=16cm]{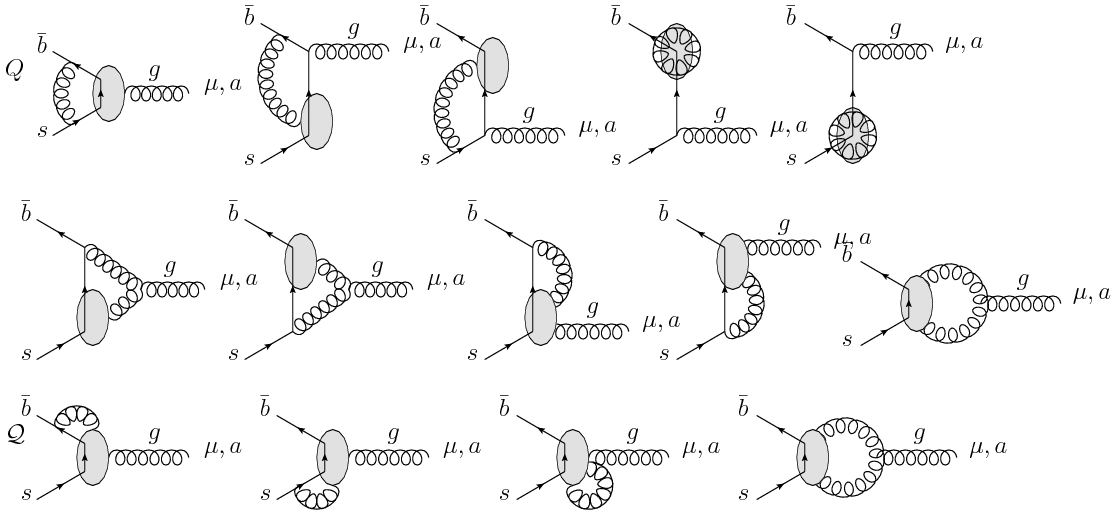}
\end{center}
\caption{Contributions from the $(\bar{b}s)$-Gauge operators ($Q$ and ${\cal Q}$) to the $\bar{b}s\to g$ amplitude.\label{bsG_bsg}}\hrule
\end{figure}
\begin{figure}[ht]
\begin{center}
 \includegraphics[width=16cm]{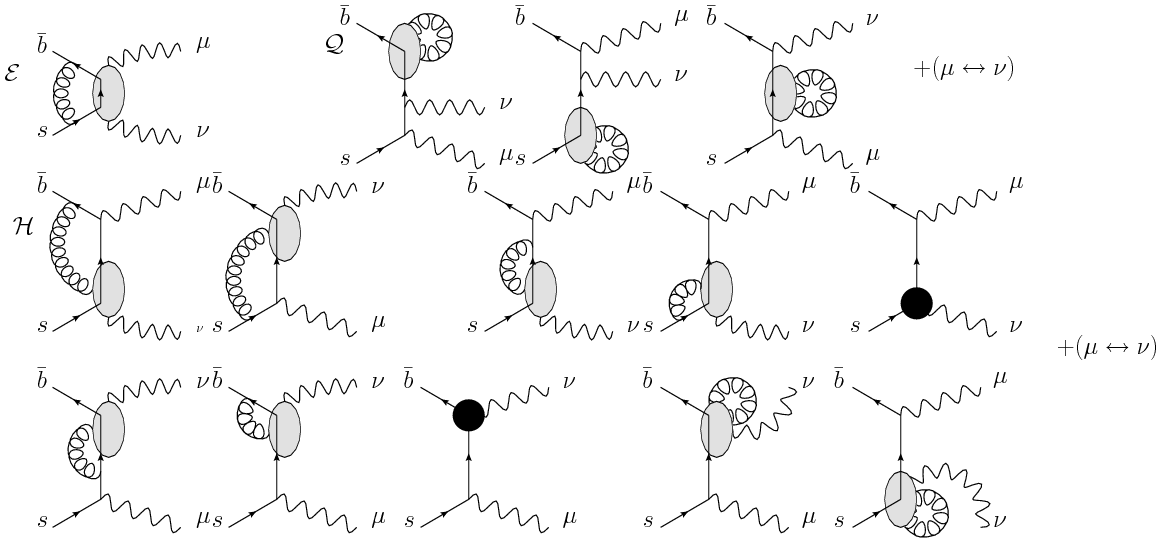}
\end{center}
\caption{Contributions from the $(\bar{b}s)$-Gauge operators (${\cal E}\leftrightarrow{\cal E}^S,\tilde{\cal E}^S,{\cal E}^T$ and ${\cal H}$) to the 
$\bar{b}s\to 2\gamma$ amplitude. The black blobs signal a counterterm from the $\bar{b}s\gamma$ or $\bar{b}sg$ amplitudes.\label{bsG_bsGG}}\hrule
\end{figure}
\begin{figure}[ht]
\begin{center}
 \includegraphics[width=16cm]{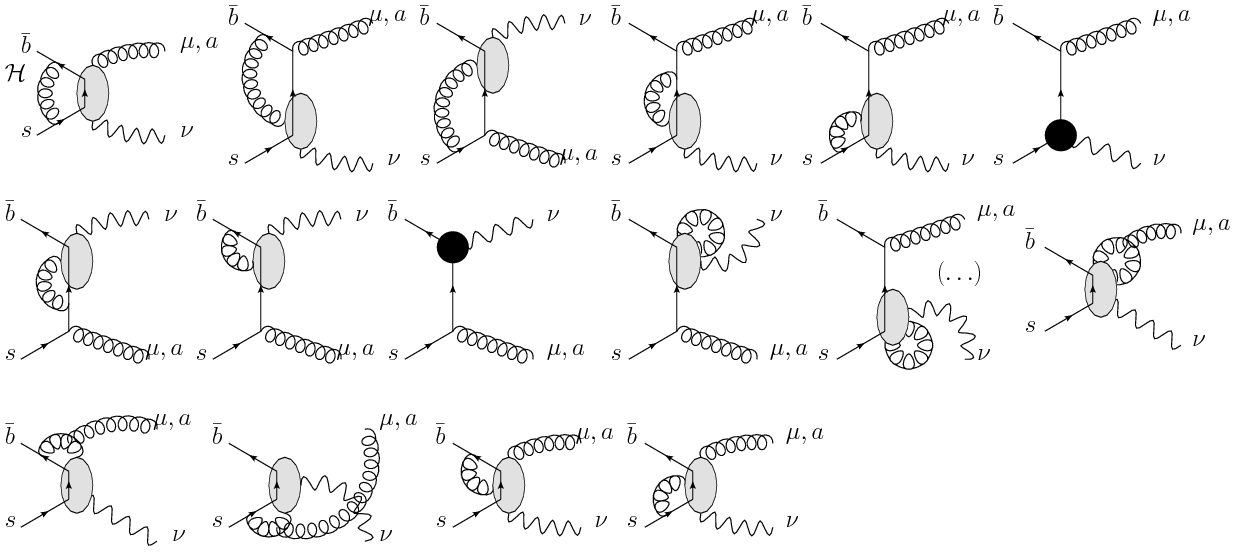}
 \includegraphics[width=16cm]{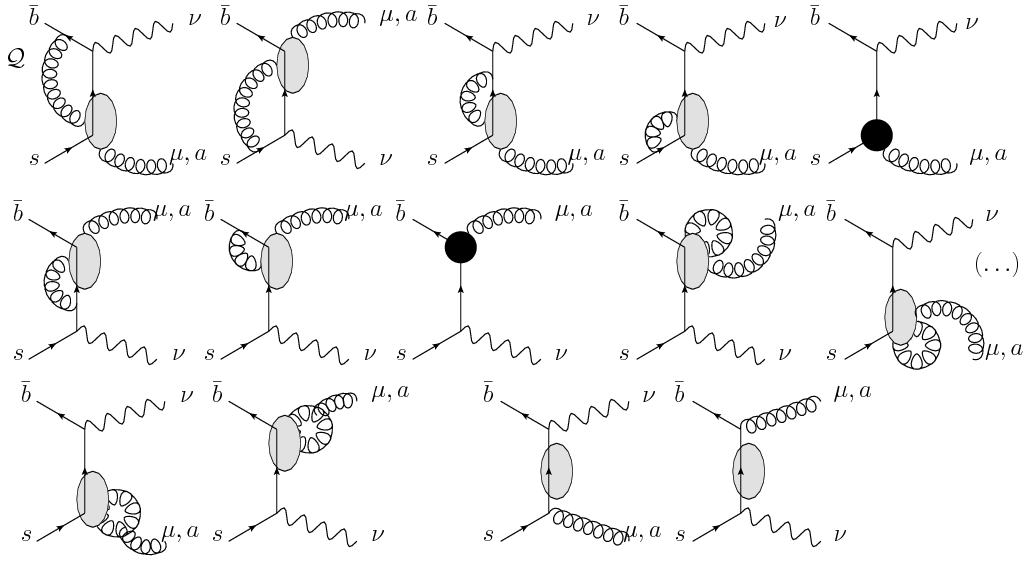}
\end{center}
\caption{Contributions from the $(\bar{b}s)$-Gauge operators (${\cal H}$ and ${\cal Q}$) to the $\bar{b}s\to \gamma g$ amplitude. The $(\ldots)$
stand for additional diagrams involving a gluon tadpole, which give vanishing contributions.\label{bsG_bsgG}}\hrule
\end{figure}
\begin{figure}[ht]
\begin{center}
 \includegraphics[width=16cm]{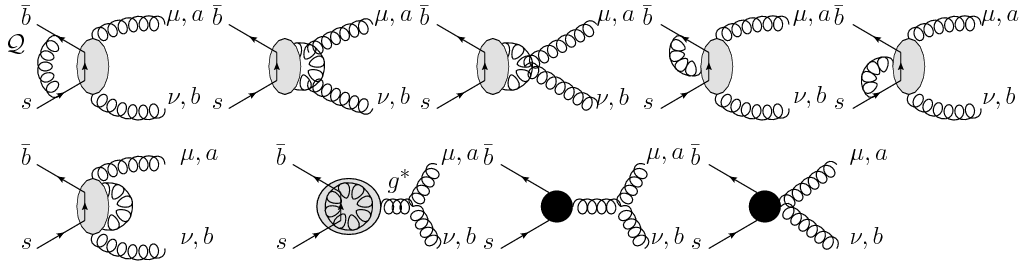}
 \includegraphics[width=16cm]{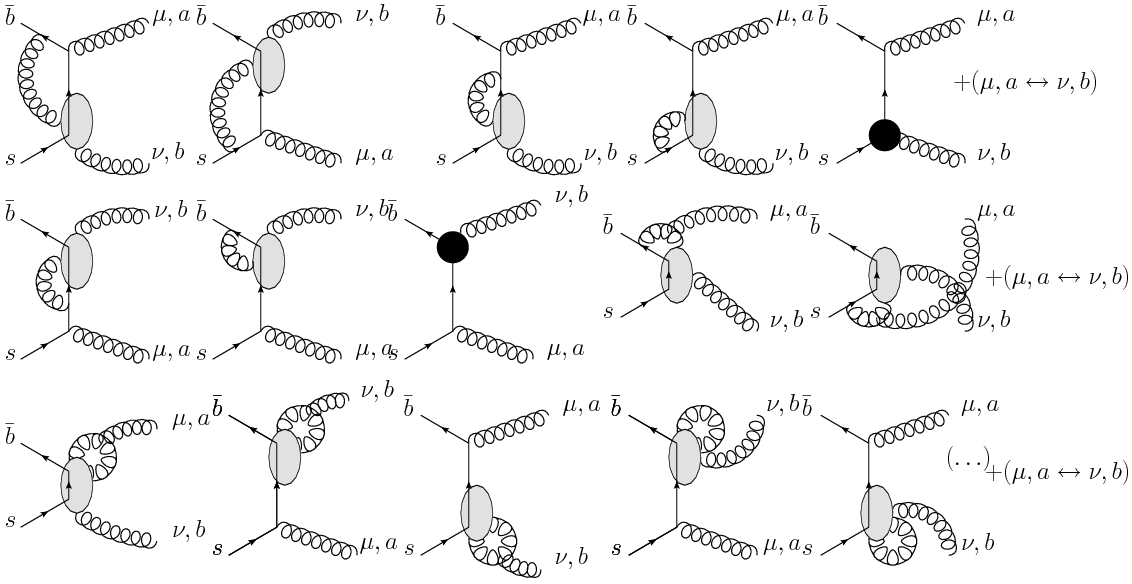}
\end{center}
\caption{Contributions from the $(\bar{b}s)$-Gauge operators (${\cal Q}$) to the $\bar{b}s\to 2g$ amplitude. The grey blob with an inscribed gluon bubble
and attached to a photon/gluon line stands for all the contributions to off-shell photon/gluon production: refer to Fig.\ref{bsG_bsG},\ref{bsG_bsg}. \label{bsG_bsgg}}\hrule
\end{figure}
\clearpage
\item $(\bar{b}s)$-Gauge contributions to the $(\bar{b}s)(\bar{f}f)$ amplitudes.
Such contributions proceed from off-shell photon or gluon exchanges or a gluon `pseudo-box'\footnote{By which we mean the last diagram of Fig.\ref{bsG_bsqq}.} and are 
depicted on Fig.\ref{bsG_bsqq}.
\begin{figure}[h]
\begin{center}
 \includegraphics[width=16cm]{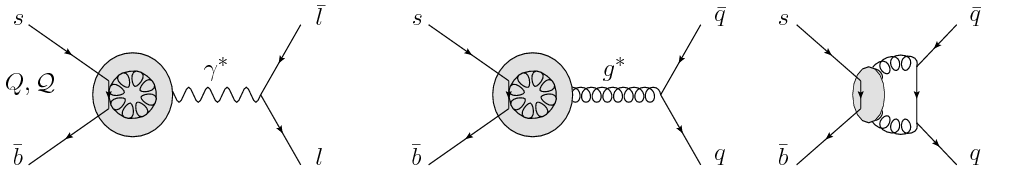}
\end{center}
\caption{Contributions from the $(\bar{b}s)$-Gauge operators ($Q$, ${\cal H}$ and ${\cal Q}$) to the 
$\bar{b}s\to \bar{l}l/\bar{q}q$ amplitude.\label{bsG_bsqq}}\hrule
\end{figure}
\end{itemize}

\noindent Finally, we may cast all these divergent (on-shell) amplitudes into the following matrix form:
\begin{equation}\label{div}
 {\rm div}=\frac{\imath}{4\pi\left(2-\frac{D}{2}\right)}(C){\cal M}(\Gamma)
\end{equation}
where $(C)=(C_{S^l}^{L,R\,L,R},\ldots)$ is a row vector\footnote{We refrain from writing this row vector with a transposition symbol (i.e.\ $(C)^T$), not to make 
already-heavy notations as will appear in the following sections completely unreadable.} collecting all the couplings of the effective Hamiltonian and $(\Gamma)$ is a 
column vector constituted by all the form-factors associated to the operators and intervening in the amplitudes that we have just presented (we only 
need one form-factor per operator since we have not considered redundant amplitudes such as $\bar{b}s\to\bar{f}f\gamma$, etc.). ${\cal M}$ depends only 
on fermion masses, gauge couplings and charges.

\subsection{Renormalization of the operators}
\subsubsection{Basics of the renormalization of the EFT in the MS-scheme}
Let us recall a few basic ingredients\footnote{Those are well-known basics of the QCD renormalization and are available in countless textbooks. We mention
them only for the sake of completeness as well as clarity in our notations.} of the renormalization procedure for our EFT in the MS-scheme.

\noindent Dismissing the effective hamiltonian for the time being, the (dim$\leq4$) EFT is a renormalizable QED$\times$QCD model.
We neglect the QED loop-effects here, since $\alpha\ll\alpha_S$, hence focus on a QCD renormalization only. The fermion
wave-function and mass renormalization constants, $Z_f$ and $Z_{m_f}$ respectively, are determined by requiring that the
counterterm contribution to the $f$ self-energy (with $p$ external momentum), $\Sigma_f^{CT}(p)=\imath(Z_f-1)
\displaystyle{\not}\hspace{0.2mm}p-\imath(Z_fZ_{m_f}-1)$, cancels the divergent loop contributions: $\Sigma_f^{\mbox{\tiny loop\,div}}(p)
=0$ (leptons) or $-\frac{\imath C_2(3)\alpha_S}{4\pi\left(2-\frac{D}{2}\right)}[-\displaystyle{\not}\hspace{0.2mm}p+4m_f]$ (quarks), 
where $C_2(3)=\frac{4}{3}$ denotes the Casimir operator of $SU(3)_c$ in the fundamental representation:
\begin{equation}\label{Zq}
 Z_l=1=Z_{m_l}\hspace{1.5cm};\hspace{1.5cm}\begin{cases}Z_q=1-\frac{C_2(3)\alpha_S}{4\pi\left(2-\frac{D}{2}\right)}+O(\alpha_S^2)\\ Z_{m_q}=1-\frac{3C_2(3)\alpha_S}{4\pi\left(2-\frac{D}{2}\right)}+O(\alpha_S^2)\end{cases}
\end{equation}
The gluon self energy (involving gluons, quarks and ghosts) similarly determines the gluon renormalization constant: in the Feynman gauge,
$Z_3\simeq1-\frac{\alpha_S}{4\pi\left(2-\frac{D}{2}\right)}\left[\frac{2}{3}n_F-\frac{5}{3}N_c\right]$, with $n_F=5$ the number of quark flavours and $N_c=3$ the number of colours. From the quark-gluon vertex, one renormalizes the strong coupling constant $g_S$, resulting in the RGE (from the scale-independence
of the bare coupling) for $\alpha_S\equiv\frac{g_S^2}{4\pi}$, with $\mu$ the renormalization scale:
\begin{multline}\label{ZalpS}
 \frac{d\alpha_S }{d\ln\mu}=-2\left(2-\frac{D}{2}\right)\alpha_S+\beta(\alpha_S)\ \ ;\ \ \beta(\alpha_S)=-\frac{2\beta_0}{4\pi}\alpha_S^2+O(\alpha_S^3)\ \ ;\ \ \beta_0=\frac{11N_c-2n_F}{3}=\frac{23}{3}\\
\Rightarrow\ \ \alpha_S(\mu)\simeq\frac{\alpha_S(\mu_0)}{1+\alpha_S(\mu_0)\frac{2\beta_0}{4\pi}\ln\frac{\mu}{\mu_0}}
\end{multline}
The bare fields $Z_ff$ as well as the bare masses $Z_{m_f}m_f$ must be scale independent (note that the $\mu$-dependence enters $Z_f$
and $Z_{m_f}$ indirectly through $\alpha_S$). One deduces the quark-mass running:
\begin{equation}\label{Zmq}
 \frac{dm_q}{d\ln\mu}=-6C_2(3)\frac{\alpha_S}{4\pi}m_q+O(\alpha_S^2)\ \Leftrightarrow\ \frac{dm_q}{d\ln\alpha_S}=\frac{3C_2(3)}{\beta_0}m_q+O(\alpha_S)\ \Rightarrow\ \frac{m_q(\mu)}{m_q(\mu_0)}=\left(\frac{\alpha_S(\mu)}{\alpha_S(\mu_0)}\right)^{\frac{3C_2(3)}{\beta_0}}
\end{equation}

\noindent Let us come back to the effective hamiltonian. The loop-divergences in the amplitudes, Eq.(\ref{div}), must be compensated by the operator counterterm
contributions: $\imath(C)[Z^C-\mathbb{1}]Z_{\Gamma}(\Gamma)$, with $Z^C$ the operator-renormalization matrix and (the diagonal) $Z_{\Gamma}$
the renormalization constant of the fields (and normalization) entering the form factor $\Gamma$. This determines: $Z^C=[\mathbb{1}-\frac{\cal M}{4\pi\left(2-\frac{D}{2}\right)}]Z_{\Gamma}^{-1}$.
Scale-independence of the bare couplings $(C)Z^C$ (by definition of $Z_{\Gamma}$, $Z_{\Gamma}O_{\Gamma}$ is scale-independent) leads to the RGE
for the couplings $(C)$: 
\begin{equation}\label{RGE0}
 \frac{d(C)}{d\ln\mu}=(C)\gamma_C\ \ \ ;\ \ \ \gamma_C\equiv-\frac{d[Z^C]}{d\ln\mu}(Z^{C})^{-1}\ \ \ ;\ \ \ Z^C=[\mathbb{1}-\frac{\cal M}{4\pi\left(2-\frac{D}{2}\right)}]Z_{\Gamma}^{-1}
\end{equation}
where the $\mu$-dependence of $Z^C$ originates from the dependence of ${\cal M}$ on $g_S$ and quark-masses, as well as from the renormalization constants in $Z_{\Gamma}$.
We have introduced the anomalous dimension matrix $\gamma_C$.

\noindent Considering the normalization of the operators that we introduced at the begining of this section, one may factor out harmoniously $C_2(3)
\alpha_S$ in the matrix ${\cal M}$, leading to:
\begin{equation}\label{divR}
 {\rm div}=\frac{\imath C_2(3)\alpha_S}{4\pi\left(2-\frac{D}{2}\right)}(C)\tilde{\cal M}(\Gamma)
\end{equation}
where $\tilde{\cal M}$ is a constant matrix (depending only on quark-mass ratios $\frac{m_q}{m_b}$).

\noindent At this leading order in QCD, using Eq.\ref{ZalpS}, Eq.\ref{RGE0} may be written explicitly as:
\begin{multline}
\frac{d(C)}{d\ln\mu}=\frac{2\alpha_SC_2(3)}{4\pi}(C)\left[\tilde{\cal D}-\tilde{\cal M}\right]+O(\alpha_S^2)\ \ \Leftrightarrow\ \ \frac{d(C)}{d\ln\alpha_S}=-\frac{C_2(3)}{\beta_0}(C)\left[\tilde{\cal D}-\tilde{\cal M}\right]+O(\alpha_S)\\
\Rightarrow\hspace{1cm}\gamma_C=\frac{2\alpha_SC_2(3)}{4\pi}\left[\tilde{\cal D}-\tilde{\cal M}\right]+O(\alpha_S^2)\hspace{4cm}\null
\end{multline}
where we have introduced the diagonal matrix $\frac{2\alpha_SC_2(3)}{4\pi}\tilde{\cal D}\equiv\frac{dZ_{\Gamma}}{d\ln\mu}$. The corresponding scaling may be extracted
from the field renormalization factors ($Z_q^{1/2}$ for each external quark and $Z_3^{1/2}$ for the external gluons) and the normalization factors
(inserting $g_S$, $\alpha_S$, $m_b$).

\subsubsection{Renormalization Group Equations for the operator basis}\label{RGEsec}
We can now explicitly extract the RGE's for the operators under consideration. The operators are ordered as above in Eq.(\ref{bsllop},\ref{bsqqop},\ref{bsGop}). 
Moreover, as far as the chiralities are concerned, we use the ordering $LL,LR,RR,RL$ for four-fermion operators ($LL,RR$ for the tensors) and $L,R$ for the 
$\bar{b}s$-Gauge operators: we will not write them down explicitly in the following, so as to keep notations as tractable as possible, but remember that the
coefficients come together with one or two chirality indices. The anomalous dimension matrix can be split in several blocks, corresponding to the various types 
(four-quark, semi-leptonic, $\bar{b}s$-gauge) of operators:
\begin{equation}\label{andimmat}
 \gamma_C=\frac{2\alpha_SC_2(3)}{4\pi}\begin{bmatrix}
                                       \tilde{\cal D}^{ll}-\tilde{\cal M}^{ll} & 0 & 0\\
                                        -\tilde{\cal M}^{ql} & \tilde{\cal D}^{qq}-\tilde{\cal M}^{qq} & -\tilde{\cal M}^{qg} \\
					-\tilde{\cal M}^{gl} & -\tilde{\cal M}^{gq} & \tilde{\cal D}^{gg}-\tilde{\cal M}^{gg}
                                      \end{bmatrix}\ \ \ ; \ \ \ (C)=(C^l,C^q,C^g)
\end{equation}
where the meaning of the subindices $l$, $q$ and $g$ should be transparent. For the sake of clarity, we choose to present the various blocks along with the 
specific RGE that they affect, separating the diagonal scaling contributions from the UV-divergent diagrams of section \ref{amplitudes}. However, should one
wish to implement the  whole $251\times251$ $\gamma_C$ matrix directly without considering our detailed study, the corresponding results are gathered in
Appendix \ref{gammaC}.
\begin{itemize}
 \item $(\bar{b}s)(\bar{l}l)$ couplings\footnote{Note that operators with different lepton flavours do not mix, justifying the use of only one index $l$.}:
\end{itemize}
\begin{equation}\label{RGEl}
 \frac{d(C^l)}{d\ln\alpha_S}=-\frac{C_2(3)}{\beta_0}\left\{(C^l)\left[-\frac{1}{4}\mbox{diag}(7\mathbb{1},7\mathbb{1},7\mathbb{1},19\mathbb{1},19\mathbb{1})-\tilde{\cal M}^{ll}\right]-(C^g)\tilde{\cal M}^{gl}-(C^q)\tilde{\cal M}^{ql}\right\}+O(\alpha_S)
\end{equation}
The matrices $\tilde{\cal M}^{ll,gl,ql}$ must be extracted from our calculation of the divergent contributions\footnote{To make our notations more transparent:
each entry in the matrices $\tilde{\cal M}$ corresponds to a $4\times4$, $4\times2$, $2\times4$ or $2\times2$ block in chirality space corresponding to the type
of coefficient it multiplies (row index) and the one it affects (column index). These blocks involve the $4\times4$ or $2\times2$ identities, $\mathbb{1}$, as well 
as several other matrices, $\Sigma$, $\Xi$, $\Phi$, etc., which we define alongside their first site of appearence. When we mention operators alongside the 
matrix $\tilde{\cal M}^{gl}$, in Eq.(\ref{Mgl}), it is simply in order to clarify which type of operator corresponds to the given row (since we cut the matrix to 
skip numerous $0$ entries).}:
\begin{equation}
 \tilde{\cal M}^{ll}=\begin{bmatrix}
                4\mathbb{1} & 0 & 0 & 0 & 0\\
                0 & \mathbb{1} & 0 & 0 & 0\\
                0 & 0 & 0 & 0 & 0\\
                0 & -2\left(\mathbb{1}+\frac{m_s}{m_b}\Sigma\right) & 0 & 0 & 0\\
                0 & 0 & 0 & 0 & \mathbb{1}
               \end{bmatrix}\hspace{2.8cm};\hspace{2.8cm}\Sigma\equiv\begin{matrix}
                                                  0&0&0&1\\
                                                  0&0&1&0\\
                                                  0&1&0&0\\
                                                  1&0&0&0\\
                                                 \end{matrix}
\end{equation}\begin{equation}\label{Mgl}
\tilde{\cal M}^{gl}=-\frac{1}{3}Q_l\begin{bmatrix}
                0 & 0 & 0 & 0 & 0\\
                \null & \null & \ldots & \null & \null\\
                0 & 0 & 0 & 0 & 0\\
                0 & \Xi+\frac{m_s}{m_b}\Phi & 0 & 0 & 0\\
                0 & 2[\Xi'+\frac{m_s}{m_b}\Phi'] & 0 & 0 & 0\\
                0 & 0 & 0 & 0 & 0\\
                \null & \null & \ldots & \null & \null\\
                0 & 0 & 0 & 0 & 0
               \end{bmatrix}\ \ \begin{matrix} \leftarrow E\\ \ldots \\ \leftarrow {\cal E}^T \\ \leftarrow {\cal H}^S \\ \leftarrow \tilde{\cal H}^S \\ \leftarrow {\cal H}^T \\ \ldots \\ \leftarrow {\cal Q}^T\end{matrix}\hspace{1.2cm};\hspace{1.2cm}
\begin{cases}
                                                                                                                                                                    \Xi\equiv\begin{matrix}
                                                                                                                                                                       1 & 1 & 0 & 0\\ 0 & 0 & 1 & 1
                                                                                                                                                                      \end{matrix}\\
                                                                                                                                                                    \Phi\equiv\begin{matrix}
                                                                                                                                                                        0 & 0 & 1 & 1\\1 & 1 & 0 & 0
                                                                                                                                                                      \end{matrix}\\
                                                                                                                                                                    \Xi'\equiv\begin{matrix}
                                                                                                                                                                        1 & 1 & 0 & 0\\0 & 0 & -1 & -1
                                                                                                                                                                      \end{matrix}\\
                                                                                                                                                                    \Phi'\equiv\begin{matrix}
                                                                                                                                                                        0 & 0 & 1 & 1\\-1 & -1 & 0 & 0
                                                                                                                                                                      \end{matrix}
                                                                                                                                                                   \end{cases}
\end{equation}
\begin{equation}
 \tilde{\cal M}^{ql}=\frac{3}{2}Q_lQ_q[\mathbb{1};-\tilde{V}^q_{\mbox{Fierz}}]_{(q=b,s)}\begin{bmatrix}
                0 & 0 & 0 & 0 & 0\\
                0 & -(\mathbb{1}+\tilde{\Sigma}) & 0 & 0 & 0\\
                0 & 0 & 0 & 0 & 0\\
                0 & 0 & 0 & -(\mathbb{1}+\tilde{\Sigma}) & 0\\
                0 & \frac{m_q}{m_b}(\mathbb{1}+\tilde{\Sigma}) & 0 & 0 & 0\\
                0 & 0 & 0 & 0 & 0\\
                0 & -\frac{1}{3}(\mathbb{1}+\tilde{\Sigma}) & 0 & 0 & 0\\
                0 & 0 & 0 & 0 & 0\\
                0 & 0 & 0 & -\frac{1}{3}(\mathbb{1}+\tilde{\Sigma}) & 0\\
                0 & \frac{m_q}{3m_b}(\mathbb{1}+\tilde{\Sigma}) & 0 & 0 & 0\\
               \end{bmatrix}\hspace{0.4cm};\hspace{0.4cm}\tilde{\Sigma}\equiv\begin{matrix}
                                                  0&1&0&0\\
                                                  1&0&0&0\\
                                                  0&0&0&1\\
                                                  0&0&1&0\\
                                                 \end{matrix}
\end{equation}
We define the object $[\mathbb{1};-\tilde{V}^q_{\mbox{Fierz}}]_{(q=b,s)}$ as follows: when $q\neq b,s$ it is simply the ($36\times36$) identity in the space defined 
by the corresponding $(\bar{b}s)(\bar{q}q)$ operators; when $q=b,s$, then it is a ($18\times 36$) matrix built with the ($18\times18$) identity in the left-hand 
subblock and the ($18\times18$) transition-matrix $-\tilde{V}^q_{\mbox{Fierz}}$, defined in Appendix \ref{Fierz}, in the right-hand subblock. Its origin is related to the facts 
that, for $q=b,s$, Fierz identities shorten the list of independent operators (compared to the cases $q\neq b,s$), and that the additional diagrams `in the t-channel', 
which open for $q=b,s$, can be related to contributions that the redundant (omited) operators would have `in the s-channel': refer to our discussion in section \ref{amplitudes}. 

\begin{itemize}
\item $(\bar{b}s)(\bar{q}q)$ couplings\footnote{Note that operators with different quark flavours mix, justifying the use of two quark indices $q$ and $q'$.}:
\end{itemize}
\begin{multline}\label{RGEqq}
 \null\hspace{-0cm}\frac{d(C^{q})}{d\ln\alpha_S}=-\frac{C_2(3)}{\beta_0}\left\{(C^{q'})\left[\delta^{q'q}[\mathbb{1},0]_{(q'=b,s)}\mbox{diag}(5\mathbb{1},5\mathbb{1},5\mathbb{1},2\mathbb{1},2\mathbb{1},5\mathbb{1},5\mathbb{1},5\mathbb{1},2\mathbb{1},2\mathbb{1})\begin{bmatrix}\mathbb{1}\\0\end{bmatrix}_{(q=b,s)}-\tilde{\cal M}^{q'q}\right]\right.\\
\left.\phantom{\left[\delta^{q'q}[\mathbb{1},0]_{(q'=b,s)}\mbox{diag}(5\mathbb{1},5\mathbb{1},5\mathbb{1},2\mathbb{1},2\mathbb{1},5\mathbb{1},5\mathbb{1},5\mathbb{1},2\mathbb{1},2\mathbb{1})\begin{bmatrix}\mathbb{1}\\0\end{bmatrix}_{(q=b,s)}\right]}-(C^g)\tilde{\cal M}^{gq}\right\}+O(\alpha_S)
\end{multline}
The objects $[\mathbb{1},0]_{(q=b,s)}$ and $\begin{bmatrix}\mathbb{1}\\0\end{bmatrix}_{(q=b,s)}$ are again defined as the ($36\times36$) identity when $q\neq b,s$ and
the ($18\times36$ and $36\times18$ respectively) matrices defined by the two ($18\times18$) subblocks corresponding to the identity and the null matrix, when $q=b,s$: 
their meaning should be clear considering the shortened base in the case $q=b,s$, due to Fierz identities.
From our calculation of the divergent contributions, the matrices $\tilde{\cal M}^{q'q,gq}$ read as:
\begin{equation}\label{qqmatrices}
\tilde{\cal M}^{q'q}=\delta^{q'q}[\mathbb{1};0]_{q'=b,s}\tilde{\cal M}^{q'q}_{\mbox{\tiny diag}}\begin{bmatrix}\mathbb{1}\\0\end{bmatrix}_{q=b,s}+[\mathbb{1};-\tilde{V}^{q'}_{\mbox{Fierz}}]_{q'=b,s}\tilde{\cal M}^{q'q}_{\mbox{\tiny univ.}}\begin{bmatrix}\mathbb{1}\\-\tilde{V}^q_{\mbox{Fierz}}\end{bmatrix}_{q=b,s}
\end{equation}\begin{displaymath}
\tilde{\cal M}^{q'q}_{\mbox{\tiny diag}}=\begin{bmatrix}
                8\mathbb{1} & 0 & \frac{1}{32}\Delta & 0 & 0 & 0 & 0 & -\frac{3}{32}\Delta & 0 & 0\\
                0 & 2\mathbb{1}+\frac{3}{4}\Sigma_3 & 0 & 0 & 0 & 0 & -\frac{9}{4}\Sigma_3 & 0 & 0 & 0\\
                24\Delta^T & 0 & 0 & 0 & 0 & -72\Delta^T & 0 & 0 & 0 & 0\\
                A_H & B_H & 0 & \mathbb{1} & 0 & C_H & D_H & 0 & 0 & 0\\
                A_{\tilde{H}} & B_{\tilde{H}} & 0 & 0 & \mathbb{1} & C_{\tilde{H}} & D_{\tilde{H}} & 0 & 0 & 0\\
                \frac{9}{4}\mathbb{1} & 0 & -\frac{3}{64}\Delta & 0 & 0 & \frac{5}{4}\mathbb{1} & 0 & -\frac{7}{64}\Delta & 0 & 0\\
                0 & -\frac{9}{8}(\mathbb{1}+\Sigma_3) & 0 & 0 & 0 & 0 & \frac{1}{8}(43\mathbb{1}-21\Sigma_3) & 0 & 0 & 0\\
                -36\Delta^T & 0 & -\frac{9}{4}\mathbb{1} & 0 & 0 & -84\Delta^T & 0 & \frac{27}{4}\mathbb{1} & 0 & 0\\
                A_{\cal H} & B_{\cal H} & 0 & 0 & 0 & C_{\cal H} & D_{\cal H} & 0 & \mathbb{1} & 0\\
                A_{\cal \tilde{H}} & B_{\cal \tilde{H}} & 0 & 0 & 0 & C_{\cal \tilde{H}} & D_{\cal \tilde{H}} & 0 & 0 & \mathbb{1}
               \end{bmatrix}
\end{displaymath}\begin{displaymath}
\tilde{\cal M}^{q'q}_{\mbox{\tiny univ}}=\frac{1}{4}\begin{bmatrix}
                0 & 0 & 0 & 0 & 0 & 0 & 0 & 0 & 0 & 0\\
                0 & 0 & 0 & 0 & 0 & 0 & 0 & 0 & 0 & 0\\
                0 & 0 & 0 & 0 & 0 & 0 & 0 & 0 & 0 & 0\\
                0 & 0 & 0 & 0 & 0 & 0 & 0 & 0 & 0 & 0\\
                0 & 0 & 0 & 0 & 0 & 0 & 0 & 0 & 0 & 0\\
                0 & 0 & 0 & 0 & 0 & 0 & 0 & 0 & 0 & 0\\
                0 & \frac{1}{3}(\mathbb{1}+\tilde{\Sigma}) & 0 & 0 & 0 & 0 & -(\mathbb{1}+\tilde{\Sigma}) & 0 & 0 & 0\\
                0 & 0 & 0 & 0 & 0 & 0 & 0 & 0 & 0 & 0\\
                0 & 0 & 0 & \frac{1}{3}(\mathbb{1}+\tilde{\Sigma}) & 0 & 0 & 0 & 0 & -(\mathbb{1}+\tilde{\Sigma}) & 0\\
                0 & -\frac{m_q}{3m_b}(\mathbb{1}+\tilde{\Sigma}) & 0 & 0 & 0 & 0 & \frac{m_q}{m_b}(\mathbb{1}+\tilde{\Sigma}) & 0 & 0 & 0
               \end{bmatrix}
\end{displaymath}
\begin{displaymath}
 \begin{cases}
  A_H=\frac{m_q}{4m_b}\left[3(\mathbb{1}+\tilde{\Sigma})+\Sigma_3+E\right]\\
  A_{\tilde{H}}=\frac{1}{4}\left[3\mathbb{1}+\Sigma_3+\frac{m_s}{m_b}(3\Sigma+\tilde{E})\right]\\
  A_{\cal H}=-\frac{3}{8}\frac{m_q}{m_b}\left[3(\mathbb{1}+\tilde{\Sigma})+\Sigma_3+E\right]\\
  A_{\cal \tilde{H}}=-\frac{3}{8}(3\mathbb{1}+\Sigma_3)-\frac{3}{8}\frac{m_s}{m_b}(3\Sigma+\tilde{E})\\
  B_H=\frac{1}{4}\left[-8\mathbb{1}+\Sigma_3+\frac{m_s}{m_b}(-8\Sigma+\tilde{E})\right]\\
  B_{\tilde{H}}=\frac{m_q}{4m_b}\left[-8(\mathbb{1}+\tilde{\Sigma})+\Sigma_3+E\right]\\
  B_{\cal H}=-\frac{3}{8}(3\mathbb{1}+\Sigma_3)-\frac{3}{8}\frac{m_s}{m_b}(3\Sigma+\tilde{E})\\
  B_{\cal \tilde{H}}=-\frac{3}{8}\frac{m_q}{m_b}\left[3(\mathbb{1}+\tilde{\Sigma})+\Sigma_3+E\right]\\
  C_H=-\frac{3}{4}\frac{m_q}{m_b}\left[3(\mathbb{1}+\tilde{\Sigma})+\Sigma_3+E\right]\\
  C_{\tilde{H}}=-\frac{3}{4}\left[(3\mathbb{1}+\Sigma_3)+\frac{m_s}{m_b}(3\Sigma+\tilde{E})\right]\\
  C_{\cal H}=-\frac{7}{8}\frac{m_q}{m_b}\left[3(\mathbb{1}+\tilde{\Sigma})+\Sigma_3+E\right]\\
  C_{\cal \tilde{H}}=-\frac{7}{8}(3\mathbb{1}+\Sigma_3)-\frac{7}{8}\frac{m_s}{m_b}(3\Sigma+\tilde{E})\\
  D_H=-\frac{3}{4}(\Sigma_3+\frac{m_s}{m_b}\tilde{E})\\
  D_{\tilde{H}}=-\frac{3}{4}\frac{m_q}{m_b}(\Sigma_3+E)\\
  D_{\cal H}=\frac{1}{8}(11\mathbb{1}-7\Sigma_3)+\frac{m_s}{8m_b}(11\Sigma-7\tilde{E})\\
  D_{\cal \tilde{H}}=\frac{m_q}{8m_b}\left[11(\mathbb{1}+\tilde{\Sigma})-7(\Sigma_3+E)\right]
 \end{cases}\hspace{0.5cm};\hspace{0.5cm}\begin{cases}
\Sigma_3=\begin{matrix}
          1 & 0 & 0 & 0\\
          0 & -1 & 0 & 0\\
          0 & 0 & 1 & 0\\
          0 & 0 & 0 & -1
         \end{matrix}\\
\null\\
E=\begin{matrix}
          0 & -1 & 0 & 0\\
          1 & 0 & 0 & 0\\
          0 & 0 & 0 & -1\\
          0 & 0 & 1 & 0
         \end{matrix}\\
\null\\
\tilde{E}=\begin{matrix}
          0 & 0 & 0 & -1\\
          0 & 0 & 1 & 0\\
          0 & -1 & 0 & 0\\
          1 & 0 & 0 & 0
         \end{matrix}\\
\null\\
\Delta=\begin{matrix}
          1 & 0 \\
          0 & 0 \\
          0 & 1 \\
          0 & 0
         \end{matrix}
\end{cases}
\end{displaymath}
\begin{multline}
\null\hspace{-0cm} \tilde{\cal M}^{gq}=\begin{bmatrix}
                0 & 0 & 0 & 0 & 0 & 0 & 0 & 0 & 0 & 0\\
                \null & \null & \null & \null & \null & \ldots & \null & \null & \null & \null\\
                0 & 0 & 0 & 0 & 0 & 0 & 0 & 0 & 0 & 0\\
                -12\frac{m_q}{m_b}\Xi & \frac{1}{12}(\Xi+\frac{m_s}{m_b}\Phi) & 0 & 0 & 0 & 0 & -\frac{1}{4}(\Xi+\frac{m_s}{m_b}\Phi) & 0 & 0 & 0\\
                24\frac{m_q}{m_b}\tilde{\Xi} & \frac{1}{6}(\Xi'+\frac{m_s}{m_b}\Phi') & 0 & 0 & 0 & 0 & -\frac{1}{2}(\Xi'+\frac{m_s}{m_b}\Phi') & 0 & 0 & 0\\
                -\frac{11}{8}\frac{m_q}{m_b}\Xi & \frac{7}{144}(\Xi+\frac{m_s}{m_b}\Phi) & 0 & 0 & 0 & -\frac{15}{8}\frac{m_q}{m_b}\Xi & -\frac{7}{48}(\Xi+\frac{m_s}{m_b}\Phi) & 0 & 0 & 0\\
                \frac{11}{4}\frac{m_q}{m_b}\tilde{\Xi} & \frac{7}{72}(\Xi'+\frac{m_s}{m_b}\Phi') & 0 & 0 & 0 & \frac{15}{4}\frac{m_q}{m_b}\tilde{\Xi} & -\frac{7}{24}(\Xi'+\frac{m_s}{m_b}\Phi') & 0 & 0 & 0\\
                0 & 0 & -\frac{3}{64}\frac{m_q}{m_b}\mathbb{1} & 0 & 0 & 0 & 0 & \frac{9}{64}\frac{m_q}{m_b}\mathbb{1} & 0 & 0
               \end{bmatrix}\ \ \begin{matrix} \leftarrow E\\ \ldots \\ \leftarrow {\cal H}^T \\ \leftarrow {\cal Q}^D \\ \leftarrow \tilde{\cal Q}^D \\ \leftarrow {\cal Q}^S \\ \leftarrow \tilde{\cal Q}^S \\ \leftarrow {\cal Q}^T\end{matrix}\\
\times\begin{bmatrix}
       \mathbb{1}\\-\tilde{V}^q_{\mbox{Fierz}}
      \end{bmatrix}_{q=b,s}
\end{multline}
\begin{displaymath}
\tilde{\Xi}\equiv\begin{matrix} 1 & -1 & 0 & 0\\ 0 & 0 & -1 & 1\end{matrix}
\end{displaymath}
The definition and meaning of the notation $\begin{bmatrix}\mathbb{1}\\-\tilde{V}^q_{\mbox{Fierz}}\end{bmatrix}_{q=b,s}$ should be obvious by now.
\begin{itemize}
\item $(\bar{b}s)$-Gauge couplings:
\end{itemize}
\begin{multline}\label{RGEga}
\null\hspace{-0cm}\frac{d(C^{g})}{d\ln\alpha_S}=-\frac{C_2(3)}{\beta_0}\left\{(C^{g})\left[\frac{1}{4}\mbox{diag}(5\mathbb{1},14\mathbb{1},-19\mathbb{1},-19\mathbb{1},-19\mathbb{1},-10\mathbb{1},-10\mathbb{1},-10\mathbb{1},-\mathbb{1},-\mathbb{1},-\mathbb{1},-\mathbb{1},-\mathbb{1})-\tilde{\cal M}^{gg}\right]\right.\\
\left.\phantom{(\tilde{C}^{g})C_2(3)\left[\mbox{diag}(7\mathbb{1},\frac{51}{8}\mathbb{1},\mathbb{1},\mathbb{1},\mathbb{1},\frac{3}{8}\mathbb{1})-\tilde{\cal M}^{gg}\right]}-(C^q)\tilde{\cal M}^{qg}\right\}+O(\alpha_S)
\end{multline}
The matrices $\tilde{\cal M}^{gg,qg}$ proceed from our calculation of the divergent contributions:
\begin{multline}
\tilde{\cal M}^{gg}=\left[\begin{array}{cccccccc}
                0 & 0 & 0 & 0 & 0 & 0 & 0 & 0 \\
                -4Q_d\mathbb{1} & \frac{11}{4}\mathbb{1} & 0 & 0 & 0 & 0 & 0 & 0 \\
                0 & 0 & 4\mathbb{1} & 0 & 0 & 0 & 0 & 0 \\
                0 & 0 & 0 & 4\mathbb{1} & 0 & 0 & 0 & 0 \\
                0 & 0 & 0 & 0 & 0 & 0 & 0 & 0 \\
                -\frac{1}{4}\Omega & 0 & -\frac{2}{3}Q_d\mathbb{1} & -\frac{1}{3}Q_d\sigma_3 & 0 & -\frac{55}{24}\mathbb{1} & \frac{11}{48}\sigma_3 & 0 \\
                \frac{1}{2}\tilde{\Omega} & 0 & -\frac{4}{3}Q_d\sigma_3 & -\frac{2}{3}Q_d\mathbb{1} & 0 & \frac{11}{12}\sigma_3 & -\frac{55}{24}\mathbb{1}& 0 \\
                (1-\frac{m_s^2}{m_b^2})\mathbb{1} & 0 & 0 & 0 & 0 & 0 & 0 & \frac{1}{4}\mathbb{1}\\
                0 & -\frac{3}{8}\Omega & 0 & 0 & 0 & - Q_d\mathbb{1} & - \frac{1}{2}Q_d\sigma_3 & 0 \\
                0 & \frac{3}{4}\tilde{\Omega} & 0 & 0 & 0 & -2 Q_d\sigma_3 & - Q_d\mathbb{1} & 0 \\
                0 & -\frac{7}{32}\Omega & 0 & 0 & 0 & -\frac{7}{12} Q_d\mathbb{1} & -\frac{7}{24} Q_d\sigma_3 & 0 \\
                0 & \frac{7}{16}\tilde{\Omega} & 0 & 0 & 0 & -\frac{7}{6} Q_d\sigma_3 & -\frac{7}{12} Q_d\mathbb{1} & 0 \\
                0 & -\frac{9}{8}(1+\frac{m_s^2}{m_b^2})\mathbb{1} & 0 & 0 & 0 & -\frac{9}{2} Q_d\mathbb{1} & \frac{9}{4} Q_d\sigma_3 & 0 
                \end{array}\right.\\\left.\begin{array}{ccccc}
 0 & 0 & 0 & 0 & 0\\ 0 & 0 & 0 & 0 & 0\\ 0 & 0 & 0 & 0 & 0\\ 0 & 0 & 0 & 0 & 0\\ 0 & 0 & 0 & 0 & 0\\ 0 & 0 & 0 & 0 & 0\\ 0 & 0 & 0 & 0 & 0\\ 0 & 0 & 0 & 0 & 0\\ \frac{17}{2}\mathbb{1} & 0 & -\mathbb{1} & -\frac{1}{2}\sigma_3 & \frac{9}{2}\mathbb{1} \\ 0 & \frac{17}{2}\mathbb{1} & -2\sigma_3 & -\mathbb{1} & -9\sigma_3 \\ \frac{21}{16}\mathbb{1} & -\frac{3}{32}\sigma_3 & \frac{379}{24}\mathbb{1} & -\frac{41}{48}\sigma_3 & \frac{11}{8}\mathbb{1} \\ -\frac{3}{8}\sigma_3 & \frac{21}{16}\mathbb{1} & -\frac{41}{12}\sigma_3 & \frac{379}{24}\mathbb{1} & -\frac{11}{4}\sigma_3 \\ \frac{3}{2}\mathbb{1} & -\frac{3}{4}\sigma_3 & \frac{45}{2}\mathbb{1} & -\frac{45}{4}\sigma_3 & -\frac{9}{4}\mathbb{1}
\end{array}\right]
\end{multline}
\begin{displaymath}
 \begin{cases}\Omega\equiv(1+\frac{m_s^2}{m_b^2})\mathbb{1}-\frac{2}{3}\frac{m_s}{m_b}\sigma_1\\ \tilde{\Omega}\equiv(1+\frac{m_s^2}{m_b^2})\sigma_3-\frac{2}{3}\frac{m_s}{m_b}\varepsilon\end{cases}\ \ ;\ \ \sigma_1\equiv\begin{matrix}0&1\\1&0\end{matrix}\ \ ;\ \ \sigma_3\equiv\begin{matrix}1&0\\0&-1\end{matrix}\ \ ;\ \ \varepsilon\equiv\begin{matrix}0&-1\\1&0\end{matrix}
\end{displaymath}
\begin{equation}\label{qgmatrix}
 \tilde{\cal M}^{qg}=[\mathbb{1};-\tilde{V}^q_{\mbox{Fierz}}]_{q=b,s}\begin{bmatrix}
                      0 & 0 & 0 & \null\\
                      0 & 0 & 0 & \null\\
                      18Q_q\frac{m_q}{m_b}\mathbb{1}&0&0&\null\\
                      0 & 0 & 0 & \null\\
                      0 & 0 & 0 & \ldots\\
                      0 & 0 & 0 & \null\\
                      0 & 0 & 0 & \null\\
                      6Q_q\frac{m_q}{m_b}\mathbb{1} & 6\frac{m_q}{m_b}\mathbb{1}&0&\null\\
                      0 & 0 & 0 & \null\\
                      0 & 0 & 0 & \null
                     \end{bmatrix}
\end{equation}

\section{Solving the RGE's}
Having presented, in the previous section, what is meant to be the main result of this paper, i.e.\ the RGE's for all operators of mass-dimension $\leq7$
intervening in $b\to s$ transitions, we shall now propose a partial solution for these RGE's, partial in the sense that we will not provide a general solution
for operators of dimension $6$ but only recover the anomalous-dimension matrix for those operators that are usually considered. On the contrary, for
dimension $7$ operators, we will present the full result. This section shall also provide us with the opportunity to discuss the limits of our approach
and how our extended analysis may be combined with the far-more advanced one of dimension-$6$ vector operators.

\subsection{Dimension $6$ operators}
First of all, we wish to check whether we can recover the usual leading-order anomalous-dimension matrix for dimension $6$ vector operators. We thus
consider the basis proposed in Eq.(2.15) of \cite{Grinstein:1987vj}. These can be viewed as specific linear combinations of ours (including 
also a rescaling), so that we should be able to read the corresponding anomalous-dimension matrix $\gamma_C^{[1-6]}$ from our results. For the time being, we
forget about the dimension $7$ entries.

\noindent For the four-quark operators $O_{1-6}$ of \cite{Grinstein:1987vj}, we obtain:
\begin{equation}
 \gamma_C^{[1-6]}=\frac{g_S^2}{8\pi^2}\begin{bmatrix}
 -1 & 3 & 0 & 0 & 0 & 0\\
 3 & -1 & -1/9 & 1/3 & -1/9 & 1/3\\
 0 & 0 & -11/9 & 11/3 & -2/9 & 2/3\\
 0 & 0 & 22/9 & 2/3 & -5/9 & 5/3\\
 0 & 0 & 0 & 0 & 1 & -3\\
 0 & 0 & -5/9 & 5/3 & -5/9 & -19/3
\end{bmatrix}
\end{equation}
which coincides with the first $6\times6$ subblock of Eq.(3.4) of this same reference.

\noindent Similarly, we consider the so-called electroweak four-quark operators $Q_{7-10}$ of Eq.(VII.2) of \cite{Buchalla:1995vs} and recover
the matrix elements presented in Table XIV of this same reference.

\noindent Let us turn to the magnetic and chromo-magnetic operators. In our approach, they satisfy the RGE, with $(C_G)\equiv(C_E,C_Q)^{L,R}$:
\begin{align}
 \frac{d(C_G)}{d\ln\alpha_S} &=-\frac{C_2(3)}{\beta_0}\left\{(C_G)\begin{bmatrix} \frac{5}{4} & 0 \\ 4Q_d & \frac{3}{4} \end{bmatrix}
-(C^q)\tilde{\cal M}^{qg}\right\}\nonumber\\
&\simeq-\frac{C_2(3)}{\beta_0}\left\{({C}_G)\begin{bmatrix} \frac{5}{4} & 0 \\ -\frac{4}{3} & \frac{3}{4} \end{bmatrix}
+({C}_S^b,{C}_T^b)\begin{bmatrix} \frac{1}{16}\Delta & -\frac{3}{16}\Delta \\ 7\mathbb{1} & -3\mathbb{1} \end{bmatrix}\right\}
\end{align}
where we have neglected $\frac{m_q}{m_b}\ll1$ for $q\neq b$ and replaced $\tilde{V}^b_{\mbox{\tiny Fierz}}$ explicitly. The contributions from the $4$-quark operators 
that we obtain here have no equivalent in the usual approach: the corresponding operators are simply
not considered in the classical case, so that it is pointless for us to keep track of those here\footnote{They are symbolically replaced by $(\ldots)$ in 
Eq.(\ref{bsGscale}). Note that $\Delta$ was defined in the previous 
subsection and appears through $\tilde{V}^q_{\mbox{\tiny Fierz}}$: see Eqs.(\ref{qqmatrices},\ref{qgmatrix}).}. Therefore:
\begin{equation}\label{bsGscale}\begin{cases}
 {C}_E(\mu)=\left(\frac{\alpha_S(\mu)}{\alpha_S(\mu_0)}\right)^{-\frac{5C_2(3)}{4\beta_0}}{C}_E(\mu_0)+\frac{8}{3}\left[\left(\frac{\alpha_S(\mu)}{\alpha_S(\mu_0)}\right)^{-\frac{3C_2(3)}{4\beta_0}}-\left(\frac{\alpha_S(\mu)}{\alpha_S(\mu_0)}\right)^{-\frac{5C_2(3)}{4\beta_0}}\right]{C}_Q(\mu_0)+(\ldots)\\
 {C}_Q(\mu)=\left(\frac{\alpha_S(\mu)}{\alpha_S(\mu_0)}\right)^{-\frac{3C_2(3)}{4\beta_0}}{C}_Q(\mu_0)+(\ldots)
\end{cases}\end{equation}
To recover the usual scaling of the operators $O_{7,8}$ \cite{Grinstein:1987vj}, one has to multiply those coefficients by $\frac{m_b}{\alpha_S}$, which 
leads to the additional factor $\left(\frac{\alpha_S(\mu)}{\alpha_S(\mu_0)}\right)^{-\frac{11C_2(3)}{4\beta_0}}$. This is, again, consistent
with Eq.(3.4) of \cite{Grinstein:1987vj}, for the subblock $7-8$.

\noindent Yet, one notices at once an apparent discrepancy with Eq.(3.4) of \cite{Grinstein:1987vj}: operators $O_{1-6}$ do not lead to divergences
in the $O_{7,8}$ directions. This is not surprising though, in the sense that the corresponding off-diagonal elements appear at the two-loop QCD level
while we considered only the one-loop divergences. One may wonder why such two-loop effects in \cite{Grinstein:1987vj} are competitive with the diagonal
one-loop elements of the $O_{7-8}$ subblock and, if so, whether it endangers the validity of our one-loop analysis. The explanation is to be found in 
the relative order of the SM matching for the four-quark and magnetic operators (considering our normalization in Eq.(\ref{bsqqop},\ref{bsGop})): if both 
were of the same order, then the two-loop mixing would only generate a subleading effect, of relative importance $O(\alpha_S)$, which, for consistency, could
be neglected. However, if one refers to the magnitude of the matching elements that we sketch at the end of section \ref{list}, which is relevant for e.g.\ 
the SM, one observes that the matching coefficients of the magnetic operators are already of order $\alpha_S$ (translating the fact that it arises at the loop 
level even though the $\alpha_S$ factor itself is only an artefact of the relative normalization of the operators), so that the two-loop mixing effect,
related to the $O(\alpha_S^0)$ four-quark matching, becomes competitive: in other words, the naive hierarchy has been destabilized by the SM matching. 
More pragmatically, this situation in the SM is related to the fact that the closure of the charm-quark loop does not entail any $\alpha_S$ suppression. 
In our naive approach, we would naturally miss such an effect. Nevertheless, one also observes from the estimates at the end of section \ref{list} that 
this unbalance among matching conditions does only concern the dimension 6 four-quark operators $O_{1-6}$\footnote{Actually $O_2$ only, at the level of the 
matching conditions, but, $O_2$ closing on the whole $O_{1-6}$ basis under renormalization, the whole subset needs to be included.}, at least in a SM-like 
model: the corresponding matrix elements are therefore already known and may be included straightforwardly within our analysis. This potentially misleading 
aspect of our naive approach should, however, be kept in mind in order to combine our findings consistently with the usual dimension $6$ analysis, directed 
at realistic high-energy models.

\noindent Let us finally consider the semi-leptonic operators $(C^l)=({C}_S^l,{C}_V^l,{C}_T^l)$. They satisfy the RGE:
\begin{equation}
 \frac{d(C^l)}{d\ln\alpha_S}=\frac{C_2(3)}{\beta_0}\left\{(C^l)\frac{1}{4}\begin{bmatrix} 23\mathbb{1} & 0 & 0\\ 0 & 11\mathbb{1} & 0\\ 0 & 0 & 7\mathbb{1} \end{bmatrix}+(C^q)\tilde{\cal M}_{ql}\right\}
\end{equation}
leading to:
\begin{equation}\begin{cases}
{C}_S^l(\mu)=\left(\frac{\alpha_S(\mu)}{\alpha_S(\mu_0)}\right)^{\frac{23C_2(3)}{4\beta_0}}{C}_S^l(\mu_0)\\
{C}_V^l(\mu)=\left(\frac{\alpha_S(\mu)}{\alpha_S(\mu_0)}\right)^{\frac{11C_2(3)}{4\beta_0}}{C}_V^l(\mu_0)+(\ldots)\\
{C}_T^l(\mu)=\left(\frac{\alpha_S(\mu)}{\alpha_S(\mu_0)}\right)^{\frac{7C_2(3)}{4\beta_0}}{C}_T^l(\mu_0)
\end{cases}\end{equation}
$(\ldots)$ stands for the inhomogeneous terms due to four-quark operators which we discuss later. If one factors out $\alpha_S^{-1}$ (for $S^l$) or 
$\alpha_S^{-1}m_b$ (for $V^l,T^l$) in the definition of the operators, the scaling of the coefficients is modified by factors 
$\left(\frac{\alpha_S(\mu)}{\alpha_S(\mu_0)}\right)^{-\frac{23C_2(3)}{4\beta_0}}$ and $\left(\frac{\alpha_S(\mu)}{\alpha_S(\mu_0)}\right)^{-
\frac{11C_2(3)}{4\beta_0}}$ (respectively), leading to the commonly known non-running of scalar and vector coefficients: see e.g.\ \cite{Bobeth:2001sq}. 
Let us now consider the inhomogeneous terms in the classical case of $O_9\equiv(\bar{b}\gamma^{\mu}P_Ls)(\bar{l}\gamma_{\mu}l)$ and
$O_{10}\equiv(\bar{b}\gamma^{\mu}P_Ls)(\bar{l}\gamma_{\mu}\gamma_5l)$. Then we find that the off-diagonal elements of the anomalous-dimension matrix,
translating the mixing with operators $O_{1-6}$,
are given by $\gamma_C^{i7}=\frac{g_S^2}{8\pi^2}\left[-\frac{4}{3},-\frac{4}{9},-\frac{2}{9},\frac{10}{9},-\frac{2}{3},-\frac{2}{9}\right]$, 
$\gamma_C^{i8}=0$, $i=1,\ldots,6$. Up to a factor $2$ accounting for the relative normalization of the operators, these matrix elements coincide with
Eq.(VIII.11) of \cite{Buchalla:1995vs}.

\noindent We have thus verified that, except for the mixing of four-quark vector operators with the magnetic and chromo-magnetic operators, which is
a two-loop effect, our calculation covered consistently all the leading-order elements of the anomalous-dimension matrix for those operators that had
been considered in the literature.

\noindent Before turning to the solution for dimension $7$ running, it is worthwhile discussing how dimension $7$ corrections to these dimension $6$ 
RGE's should be implemented. First note that we are talking here about corrections of order $\frac{m_b}{M_Z}$ at most, so that it would make little 
sense to include them if the dimension $6$ coefficients were considered at leading order only: dimension $7$ effects will make sense numerically 
only if dimension $6$ contributions are known up to $O(\alpha_S)$ or even $O(\alpha_S^2)$ (depending on the matching conditions). A second remark 
comments on the absence of dimension $6$ effects in the RGE's of dimension $7$ operators (in other words, the anomalous dimension matrix is block-triangular): 
this naturally proceeds from the fact that the dimension $6$ basis is stable and does not require dimension $7$ counterterms to cancel its divergences 
(this can be derived from simple power-counting). The dimension $7$ RGE's can therefore be considered separately while their mixing-effect is injected 
directly as a small inhomogeneous term in the dimension $6$ RGE's.

\subsection{Dimension 7 RGE's}
We present the solution for the RGE's of dimension $7$ operators. We stress again that those are independent from the dimension $6$ running, allowing
us to solve them separately. The block-triangular shape appearing in $\gamma^{\mbox{\tiny dim. 7}}_C$, the restriction of the anomalous dimension matrix 
to dimension 7 operators, invites for a splitting into subblock RGE's with inhomogeneous terms: this observation makes an explicit solution tractable
analytically, which we perform for the sake of completeness. Note that the methodology, consisting in splitting the anomalous-dimension matrix according
to its block triangular shape and resulting in inhomogeneous linear equations, is a recurring feature in EFT's and has been used, e.g.\ in \cite{Herrlich:1996vf}
(although in a different context; see section 3.4.2 of this reference).

\subsubsection{$(\bar{b}s)$-Gauge operators}
\null\hspace{2mm}{\em Gluonic operators}; $(C_{\cal Q})\equiv(C_{{\cal Q}^D},{C}_{\tilde{\cal Q}^D},{C}_{{\cal Q}^S},{C}_{\tilde{\cal Q}^S},{C}_{{\cal Q}^T})^{L,R}$:\newline
The RGE is a simple homogeneous 1st-order differential equation:
\begin{multline}\label{RGEQ}
 \frac{d({C}_{\cal Q})}{d\ln\alpha_S}=\frac{C_2(3)}{\beta_0}({C}_{\cal Q})\begin{bmatrix}
\frac{35}{4}\mathbb{1} & 0 & -\mathbb{1} & -\frac{1}{2}\sigma_3 & \frac{9}{2}\mathbb{1}\\
0 & \frac{35}{4}\mathbb{1} & -2\sigma_3 & -\mathbb{1} & -9\sigma_3\\
\frac{21}{16}\mathbb{1} & -\frac{3}{32}\sigma_3 & \frac{385}{24}\mathbb{1} & -\frac{41}{48}\sigma_3 & \frac{11}{8}\mathbb{1}\\
-\frac{3}{8}\sigma_3 & \frac{21}{16}\mathbb{1} & -\frac{41}{12}\sigma_3 & \frac{385}{24}\mathbb{1} & -\frac{11}{4}\sigma_3 \\
\frac{3}{2}\mathbb{1} & -\frac{3}{4}\sigma_3 & \frac{45}{2}\mathbb{1} & -\frac{45}{4}\sigma_3 & -2\mathbb{1}
                                                                         \end{bmatrix}\equiv\frac{C_2(3)}{\beta_0}({C}_{\cal Q}){\cal A}_{\cal QQ}\\
\Rightarrow\ \ ({C}_{\cal Q})(\mu)=({C}_{Q})(\mu_0)\exp\left[\frac{C_2(3)}{\beta_0}{\cal A}_{\cal QQ}\ln\frac{\alpha_S(\mu)}{\alpha_S(\mu_0)}\right]=({C}_{\cal Q})(\mu_0){\cal O}_{\cal Q}\left(\frac{\alpha_S(\mu)}{\alpha_S(\mu_0)}\right)^{\frac{C_2(3)}{\beta_0}{\cal D}_{\cal Q}}{\cal O}_{\cal Q}^{-1}
\end{multline}
with ${\cal A}_{\cal QQ}\equiv {\cal O}_{\cal Q}{\cal D}_{\cal Q}{\cal O}_{\cal Q}^{-1} $, where ${\cal D}_{\cal Q}$ is a diagonal matrix:
\newline ${\cal D}_{\cal Q}=\mbox{diag}\left[-\frac{19}{4}\mathbb{1},\frac{1}{8}(117-\sqrt{3073})\mathbb{1},\frac{1}{24}(277-\sqrt{3193})\mathbb{1},\frac{1}{24}(277+\sqrt{3193})\mathbb{1},\frac{1}{8}(117+\sqrt{3073})\mathbb{1}\right]$
\newline The details of this diagonalization can be found in Appendix \ref{diago}.

\null\hspace{2mm}{\em Hybrid operators}; $({C}_{\cal H})\equiv({C}_{{\cal H}^S},{C}_{\tilde{\cal H}^S},{C}_{{\cal H}^T})^{L,R}$:
\begin{multline}\label{RGEH}
\frac{d({C}_{\cal H})}{d\ln\alpha_S}=\frac{C_2(3)}{\beta_0}\left\{({C}_{\cal H})\begin{bmatrix}
\frac{5}{24}\mathbb{1} & \frac{11}{48}\sigma_3 & 0 \\
\frac{11}{12}\sigma_3 & \frac{5}{24}\mathbb{1}& 0 \\
 0 & 0 & \frac{11}{4}\mathbb{1}
\end{bmatrix}-Q_d({C}_{\cal Q})\begin{bmatrix}
\mathbb{1} & \frac{1}{2} \sigma_3 & 0 \\
2\sigma_3 & \mathbb{1} & 0 \\
\frac{7}{12} \mathbb{1} & \frac{7}{24} \sigma_3 & 0 \\
\frac{7}{2} \sigma_3 & \frac{7}{12} \mathbb{1} & 0 \\
\frac{9}{2} \mathbb{1} & -\frac{9}{4} \sigma_3 & 0
\end{bmatrix}\right\}\\
\equiv\frac{C_2(3)}{\beta_0}\left\{({C}_{\cal H}){\cal A_{HH}}-({C}_{\cal Q}){\cal A_{QH}}\right\}
\end{multline}
\begin{multline}
 \Rightarrow ({C}_{\cal H})(\mu)=\left\{({C}_{\cal H})(\mu_0)-\frac{C_2(3)}{\beta_0}\int_0^{\ln\left(\frac{\alpha_S(\mu)}{\alpha_S(\mu_0)}\right)}{({C}_{\cal Q})(\mu'){\cal A_{QH}}\exp\left[-\frac{C_2(3)}{\beta_0}{\cal A}_{\cal HH}\ln\frac{\alpha_S(\mu')}{\alpha_S(\mu_0)}\right]\,d\ln\left(\frac{\alpha_S(\mu')}{\alpha_S(\mu_0)}\right)}\right\}\\
\times\exp\left[\frac{C_2(3)}{\beta_0}{\cal A}_{\cal HH}\ln\frac{\alpha_S(\mu)}{\alpha_S(\mu_0)}\right]
\end{multline}
The exponentiation of ${\cal A}_{\cal HH}$ can be achieved without difficulty through its diagonalization: ${\cal A}_{\cal HH}={\cal O_H D_H O}^{-1}_{\cal H}$, 
with ${\cal D_H}\equiv\mbox{diag}\left[-\frac{1}{4}\mathbb{1},\frac{2}{3}\mathbb{1},\frac{11}{4}\mathbb{1}\right]$ (refer to Appendix \ref{diago}).

\null\hspace{2mm}{\em Photonic operators}; $({C}_{\cal E})\equiv({C}_{{\cal E}^S},{C}_{\tilde{\cal E}^S},{C}_{{\cal E}^T})^{L,R}$:
\begin{equation}
\frac{d({C}_{\cal E})}{d\ln\alpha_S}=\frac{C_2(3)}{\beta_0}\left\{({C}_{\cal E})\frac{1}{4}\begin{bmatrix}
35\mathbb{1} & 0 & 0 \\
0 & 35\mathbb{1} & 0  \\
0 & 0 & 19\mathbb{1}
\end{bmatrix}-({C}_{\cal H})Q_d\begin{bmatrix}
\frac{2}{3}\mathbb{1} & \frac{1}{3}\sigma_3 & 0 \\
\frac{4}{3}\sigma_3 & \frac{2}{3}\mathbb{1} & 0 \\
 0 & 0 & 0
\end{bmatrix}\right\}
\end{equation}
\begin{multline}
 \Rightarrow ({C}_{\cal E})(\mu)=({C}_{\cal E})(\mu_0)\begin{bmatrix}\left(\frac{\alpha_S(\mu)}{\alpha_S(\mu_0)}\right)^{\frac{35C_2(3)}{4\beta_0}}\mathbb{1}&0&0\\0&\left(\frac{\alpha_S(\mu)}{\alpha_S(\mu_0)}\right)^{\frac{35C_2(3)}{4\beta_0}}\mathbb{1}&0\\0&0&\left(\frac{\alpha_S(\mu)}{\alpha_S(\mu_0)}\right)^{\frac{19C_2(3)}{4\beta_0}}\mathbb{1}\end{bmatrix}\\
-\frac{C_2(3)Q_d}{\beta_0}\left(\frac{\alpha_S(\mu)}{\alpha_S(\mu_0)}\right)^{\frac{35C_2(3)}{4\beta_0}}\int_0^{\ln\left(\frac{\alpha_S(\mu)}{\alpha_S(\mu_0)}\right)}{({C}_{\cal H})(\mu')\left(\frac{\alpha_S(\mu')}{\alpha_S(\mu_0)}\right)^{-\frac{35C_2(3)}{4\beta_0}}\,d\ln\left(\frac{\alpha_S(\mu')}{\alpha_S(\mu_0)}\right)}\begin{bmatrix}
\frac{2}{3}\mathbb{1} & \frac{1}{3}\sigma_3 & 0 \\
\frac{4}{3}\sigma_3 & \frac{2}{3}\mathbb{1} & 0 \\
 0 & 0 & 0
\end{bmatrix}
\end{multline}

\noindent Interestingly, certain of those operators have a negative anomalous dimension, meaning that they will grow at low
energy, and thus become important. Note however that this negative dimension may simply be an artefact of artificial factors $\alpha_S^{-1}$
in the normalization of the operators: the matching conditions in a definite model are therefore likely to correct this feature by introducing 
suppression factors. Otherwise, such growing directions in the dimension $7$ anomalous-dimension matrix could lead to significant 
perturbation of the dimension $6$ results. While the computation of the anomalous dimension is particularly simple, e.g.\ in the case 
of the photonic operators (leaving little room for false moves), the prospect of dimension 7 contributions enhanced at low energy should be 
considered however with caution.

\subsubsection{Dimension 7 $(\bar{b}s)(\bar{q}q)$ operators}
$({C}^q_H)\equiv({C}^q_{H^q},{C}^q_{\tilde{H}^q},{C}^q_{{\cal H}^q},{C}^q_{\tilde{\cal H}^q})^{L,R\,L,R}$:
\begin{multline}\label{RGEHq}
 \frac{d({C}^q_H)}{d\ln\alpha_S}=-\frac{C_2(3)}{\beta_0}\left\{({C}^q_H)-({C}^{q'}_H)\frac{1}{4}[\mathbb{1};-\tilde{V}^{q'}_{\mbox{Fierz}}]_{q'=b,s}\begin{bmatrix}
                0 & 0 & 0 & 0 \\
                0 & 0 & 0 & 0 \\
                \frac{1}{3}(\mathbb{1}+\tilde{\Sigma}) & 0 & -(\mathbb{1}+\tilde{\Sigma}) & 0\\
                0 & 0 & 0 & 0
               \end{bmatrix}\begin{bmatrix}\mathbb{1}\\-\tilde{V}^q_{\mbox{Fierz}}\end{bmatrix}_{q=b,s}\right\}\\\equiv-\frac{C_2(3)}{\beta_0}({C}^q_H)\left[\mathbb{1}+{\cal A}_{qq}\right]
\end{multline}
\begin{displaymath}
 \Rightarrow ({C}^q_H)(\mu)=({C}^q_H)(\mu_0)\exp\left[-\frac{C_2(3)}{\beta_0}\left(\mathbb{1}+{\cal A}_{qq}\right)\ln\frac{\alpha_S(\mu')}{\alpha_S(\mu_0)}\right]
\end{displaymath}
${\cal A}_{qq}$ can be explicitly diagonalized: ${\cal A}_{qq}={\cal O}_q\mbox{diag}(0,0,0,0,0,0,0,0,0,0,0,0,0,0,0,\frac{8}{3}{\cal D}_{24}){\cal O}_q^{-1}$ (refer to Appendix \ref{diago}).

\subsubsection{Dimension 7 $(\bar{b}s)(\bar{l}l)$ operators}
$({C}^l_H)\equiv({C}^l_{H^l},{C}^l_{\tilde{H}^l})$:
\begin{equation}
 \frac{d({C}^l_H)}{d\ln\alpha_S}=\frac{C_2(3)}{4\beta_0}({C}^{l}_H)\begin{bmatrix} 19\mathbb{1} & 0 \\ 0 & 23\mathbb{1} \end{bmatrix}\\
-\frac{2Q_lQ_q}{\beta_0}({C}_H^q)[\mathbb{1};-\tilde{V}^{q}_{\mbox{Fierz}}]_{q=b,s}\begin{bmatrix}\mathbb{1}+\tilde{\Sigma} & 0\\ 0 & 0\\
\frac{1}{3}(\mathbb{1}+\tilde{\Sigma}) & 0\\ 0 & 0\end{bmatrix}
\end{equation}
\begin{multline}
 \Rightarrow ({C}^l_H)(\mu)=({C}^l_H)(\mu_0)\begin{bmatrix}\left(\frac{\alpha_S(\mu)}{\alpha_S(\mu_0)}\right)^{\frac{19C_2(3)}{4\beta_0}}\mathbb{1}&0\\0&\left(\frac{\alpha_S(\mu)}{\alpha_S(\mu_0)}\right)^{\frac{23C_2(3)}{4\beta_0}}\mathbb{1}\end{bmatrix}\\
\null\hspace{-1cm}-\frac{2Q_lQ_q}{\beta_0}\left(\frac{\alpha_S(\mu)}{\alpha_S(\mu_0)}\right)^{\frac{19C_2(3)}{4\beta_0}}\int_0^{\ln\left(\frac{\alpha_S(\mu)}{\alpha_S(\mu_0)}\right)}{({C}_H^q)(\mu')\left(\frac{\alpha_S(\mu')}{\alpha_S(\mu_0)}\right)^{-\frac{19C_2(3)}{4\beta_0}}\,d\ln\left(\frac{\alpha_S(\mu')}{\alpha_S(\mu_0)}\right)}\\
\times[\mathbb{1};-\tilde{V}^{q}_{\mbox{Fierz}}]_{q=b,s}\begin{bmatrix}\mathbb{1}+\tilde{\Sigma} & 0\\ 0 & 0\\
\frac{1}{3}(\mathbb{1}+\tilde{\Sigma}) & 0\\ 0 & 0\end{bmatrix}
\end{multline}
Note that the positive eigenvalues (negative anomalous dimensions) change sign once the artificial factor $\alpha_S^{-1}$ is factored away from the operators.

\noindent This concludes the solution of all dimension $7$ RGE's. We will now consider a simple example engaging these dimension $7$ effects.

\section{A simple application}
\subsection{Setup}
Let us assume that leading new-physics effects would arise in the possible (though unlikely) form of the following dimension 6 SM operator \cite{Grzadkowski:2010es}:
\begin{equation}
 {\cal L}_{\mbox{\tiny NP}}=-\frac{\imath K_{sb}}{M_Z^2}H^{\dag}\frac{g'B_{\mu\nu}+2gW_{\mu\nu}^a\tau^a}{\sqrt{g^2+{g'}^2}}\bar{s}\sigma^{\mu\nu}P_L\begin{pmatrix}t\\b\end{pmatrix}+h.c.
\end{equation}
where $g'$ and $g$ are the $U(1)_Y$ and $SU(2)_L$ coupling constants, $B_{\mu\nu}$ and $W_{\mu\nu}^a\tau^a$ the corresponding field-strength tensors ($\tau^a$ correspond 
to the $SU(2)_L$ generators), while $H$ is the Higgs doublet (with negative hypercharge). $K_{sb}$ is a complex coefficient encoding new-physics effects. At low-energy,
it results in a modified $Z-b-s$ coupling:
\begin{equation}
 \frac{\imath K_{sb}^{*}\mbox{\em v}}{M_Z^2}\left[\bar{b}\sigma^{\mu\nu}P_Rs\right](\partial_{\mu}Z_{\nu}-\partial_{\nu}Z_{\mu})+h.c.
\end{equation}
where $v=(2\sqrt{2}G_F)^{-1/2}$ is the electroweak vacuum expectation value. Such a term could generate flavour effects at the $Z$-pole, resulting in a perturbation of
the precision measurements ($Z\to\bar{b}s,\bar{s}b$ being mistaken experimentally for a $Z\to\bar{b}b$ decay).

\noindent Considering the modified $Z$-coupling, we compute the associated matching conditions, at tree-level, for the operators of the $b\to s$ transition:
\begin{equation}\label{matching}
 \begin{cases}
  \delta^{\mbox{\tiny NP}}C_{V^f}^{L\,L}(M_Z)=-4\sqrt{2}(I_3^f-Q^fs_W^2)\frac{m_sK_{sb}^*}{M_Z^3}\simeq0\\
  \delta^{\mbox{\tiny NP}}C_{V^f}^{L\,R}(M_Z)=4\sqrt{2}Q^fs_W^2\frac{m_sK_{sb}^*}{M_Z^3}\simeq0\\
  \delta^{\mbox{\tiny NP}}C_{V^f}^{R\,R}(M_Z)=-4\sqrt{2}Q^fs_W^2\frac{m_b(M_Z)K_{sb}^*}{M_Z^3}\\
  \delta^{\mbox{\tiny NP}}C_{V^f}^{R\,L}(M_Z)=4\sqrt{2}(I_3^f-Q^fs_W^2)\frac{m_b(M_Z)K_{sb}^*}{M_Z^3}\\
 \end{cases}\hspace{1cm};\hspace{1cm}
 \begin{cases}
  \delta^{\mbox{\tiny NP}}C_{H^f}^{L\,L}(M_Z)=0\\
  \delta^{\mbox{\tiny NP}}C_{H^f}^{L\,R}(M_Z)=0\\
  \delta^{\mbox{\tiny NP}}C_{H^f}^{R\,R}(M_Z)=4\sqrt{2}Q^fs_W^2\frac{K_{sb}^*}{M_Z^3}\\
  \delta^{\mbox{\tiny NP}}C_{H^f}^{R\,L}(M_Z)=-4\sqrt{2}(I_3^f-Q^fs_W^2)\frac{K_{sb}^*}{M_Z^3}\\
 \end{cases}
\end{equation}
where we here use the following normalization of the operators ($f$ stands for any of the low-energy quarks and leptons):
\begin{equation}
 (V^f)^{L,R\ L,R}=(\bar{b}\gamma^{\mu}P_{L,R}s)(\bar{f}\gamma_{\mu}P_{L,R}f)\ \ \ ;\ \ \ 
(H^f)^{L,R\ L,R}=[\bar{b}\imath(\overrightarrow{D}-\overleftarrow{D})^{\mu}P_{L,R}s](\bar{f}\gamma_{\mu}P_{L,R}f)
\end{equation}

\noindent Note, at this point, that the classical framework of dimension $6$ vector-type operators would not allow for a consistent description of these new-physics
effects: albeit the $(V^f)^{L,R\ L,R}$ operators receive a contribution, the associated effect is of the same order as that of the dimension $7$ $(H^f)^{L,R\ L,R}$
operators. One should thus rely on our extended analysis.

\subsection{\boldmath $BR(B_s\to l^+l^-)$}
\noindent Now let us consider the decay $B_s\to l^+l^-$. This rate is one of the traditional search channels for new-physics. Evidence for its observation (in the case 
$l=\mu$) has been reported a few months ago at LHCb \cite{Aaij:2012nna}, evidencing a good agreement with the SM, hence constraining new-physics effects more tightly. 
For a recent review, refer to \cite{Buras:2013uqa}. The well-known SM-matching provides \cite{Buchalla:1995vs} (the function $Y$ is defined in this reference):
\begin{equation}
 C_{V^l}^{L\,L}(M_Z)\simeq-\frac{\sqrt{2}G_F\alpha}{\pi s_W^2}V_{ts}^*V_{tb}Y(\frac{m_t^2}{M_W^2})\simeq-6.2\cdot10^{-9}\equiv C^{\mbox{\tiny SM}}_{V^l} 
\end{equation}
We (safely) neglect dimension $7$ effects associated with the SM and assume that only new physics associated with $K_{sb}$ may lead to a significant deviation.

\noindent Then, we consider the following matrix elements (in terms of the $B_s$-meson decay constant $f_B$, and its four-momentum $P_{\mu}$):
\begin{align}
 &\left<0\right|\bar{b}\gamma_5s\left|B_s\right>=-\imath f_B\frac{M_B^2}{m_b+m_s} & \left<0\right|\bar{b}s\left|B_s\right>=0\hspace{1.65cm}\nonumber\\
 &\left<0\right|\bar{b}\gamma_{\mu}\gamma_5s\left|B_s\right>=\imath f_B P_{\mu} & \left<0\right|\bar{b}\gamma_{\mu}s\left|B_s\right>=0\hspace{1.3cm}\\
 &\left<0\right|\bar{b}\imath(\overrightarrow{D}-\overleftarrow{D})_{\mu}\gamma_5s\left|B_s\right>=\imath f_B(m_b-m_s) P_{\mu} & \left<0\right|\bar{b}\imath(\overrightarrow{D}-\overleftarrow{D})_{\mu}s\left|B_s\right>=0\nonumber
\end{align}
Here we have followed the conventions of \cite{Ali:1999mm} and derived the result for the matrix elements $\left<0\right|\bar{b}\imath(\overrightarrow{D}-\overleftarrow{D})_{\mu}\mathbb{1}/\gamma_5s\left|B_s\right>$.
One then derives (where we consider only the operators $V^l$ and $H^l$, relevant in this particular example):
\begin{equation}\label{BRBll}
 BR(B_s\to l^+l^-)\simeq\frac{f_B^2M_Bm_l^2}{32\pi\Gamma_{B}}\sqrt{1-4\frac{m_l^2}{M_B^2}}\left|C_{V^l}^{L\,R}-C_{V^l}^{L\,L}+C_{V^l}^{R\,L}-C_{V^l}^{R\,R}+m_b\left(C_{H^l}^{L\,R}-C_{H^l}^{L\,L}+C_{H^l}^{R\,L}-C_{H^l}^{R\,R}\right)\right|^2(\mu_b)
\end{equation}
Note that for the combinations $C_{V^l}^{L,R\,L}-C_{V^l}^{L,R\,R}$ and $C_{H^l}^{L,R\,L}-C_{H^l}^{L,R\,R}$, the mixing to the four-quark operators cancels in the 
RGE (Eq.(\ref{RGEl}), due to the $\mathbb{1}+\tilde{\Sigma}$ in $\tilde{\cal M}^{ql}$) so that we obtain simple scalings for both these quantities: 
$\left(\frac{\alpha_S(\mu_b)}{\alpha_S(M_Z)}\right)^{0}=1$ for $C_{V^l}^{L,R\,L}-C_{V^l}^{L,R\,R}$ and $\left(\frac{\alpha_S(\mu_b)}{\alpha_S(M_Z)}\right)^{-\frac{C_2(3)}{\beta_0}}=\left(\frac{\alpha_S(\mu_b)}{\alpha_S(M_Z)}\right)^{-\frac{4}{23}}$ 
for $C_{H^l}^{L,R\,L}-C_{H^l}^{L,R\,R}$. Note however the off-diagonal term in $\tilde{\cal M}^{ll}$ mixing $V^l$ and $H^l$ operators. The RGE for the vector coefficients
hence reads:
\begin{multline}\label{runleptons}
 \frac{d\ }{d\ln\alpha_S}[C_{V^l}^{L,R\,L}-C_{V^l}^{L,R\,R}]=-\frac{2C_2(3)}{\beta_0}m_b[C_{H^l}^{L,R\,L}-C_{H^l}^{L,R\,R}]\ \ \ \Rightarrow\\
 [C_{V^l}^{L,R\,L}-C_{V^l}^{L,R\,R}](\mu_b)=[C_{V^l}^{L,R\,L}-C_{V^l}^{L,R\,R}](M_Z)-\frac{2C_2(3)}{\beta_0}m_b(M_Z)[C_{H^l}^{L,R\,L}-C_{H^l}^{L,R\,R}](M_Z)\int_0^{\ln\left(\frac{\alpha_S(\mu_b)}{\alpha_S(M_Z)}\right)}{e^{\frac{2C_2(3)}{\beta_0}x}\,dx}\\
=[C_{V^l}^{L,R\,L}-C_{V^l}^{L,R\,R}](M_Z)-m_b(M_Z)[C_{H^l}^{L,R\,L}-C_{H^l}^{L,R\,R}](M_Z)\left[\left(\frac{\alpha_S(\mu_b)}{\alpha_S(M_Z)}\right)^{\frac{2C_2(3)}{\beta_0}}-1\right]
\end{multline}

\noindent It is also more convenient to work with a low-energy $b$ mass: $m_b=m_b(M_Z)\left(\frac{\alpha_S(\mu_b)}{\alpha_S(M_Z)}\right)^{\frac{3C_2(3)}{\beta_0}}$. 
Note that this replacement does not concern the explicit factor $m_b$ multiplying the $H^l$ coefficients in Eq.(\ref{BRBll}). Replacing the coefficients at the $M_Z$
scale by the matching conditions, we then observe a complete cancellation of the new-physics contributions:
\begin{equation}
BR(B_s\to l^+l^-)\simeq\frac{f_B^2M_Bm_l^2}{32\pi\Gamma_{B}}\sqrt{1-4\frac{m_l^2}{M_B^2}}\left|C^{\mbox{\tiny SM}}_{V^l}\right|^2
\end{equation}
The scenario under investigation hence receives no constraint from $BR(B_s\to l^+l^-)$. Although this is what the naive tree-level calculation would have predicted, it
is a non-trivial result to observe that this feature is preserved by the resummation of the leading logarithms through the RGE's: inconsistently neglecting the $H$ 
contributions or the $V-H$ mixing, for instance, would have generated limits of the order $|K_{sb}|\lsim10^{-3}$. Note however that the cancellation of the new-physics 
effects is tightly related to the form of the matching conditions of Eq.(\ref{matching}) and that a perturbation of Eq.(\ref{matching}), e.g.\ through the implementation 
of additional dimension $6$ operators, would lead to relevant limits again.

\subsection{\boldmath $BR(B\to K\nu\bar{\nu})$}
Many other $b\to s$ transitions are of course relevant to constrain $K_{sb}$. Let us consider another simple example: the decay $B\to K\nu\bar{\nu}$ is bounded by the 
limit $BR(B^-\to K^-+{\rm Inv.})<13\cdot10^{-6}$ (90\% C.L.) from BABAR \cite{delAmoSanchez:2010bk}. The SM provides $BR(B^-\to K^-\nu\bar{\nu})
\simeq(4.4^{+1.3+0.8+0.0}_{-1.1-0.7-0.7})\cdot10^{-6}$: refer to \cite{Buchalla:2010jv} for a summary. The relevant matching coefficient for the SM can be read in 
\cite{Buchalla:1995vs} (the function $X$ is defined in this reference):
\begin{equation}
 C_{V^{\nu}}^{L\,L}(M_Z)\simeq\frac{\sqrt{2}G_F\alpha}{\pi s_W^2}V_{ts}^*V_{tb}X(\frac{m_t^2}{M_W^2})\simeq1\cdot10^{-8}\equiv C^{\mbox{\tiny SM}}_{V^{\nu}} 
\end{equation}

\noindent Concerning the decay constants, we again follow the conventions of \cite{Ali:1999mm} and derive the matrix element corresponding to the $H$-type operators:
\begin{align}
 &<K(P_K)|\bar{s}\gamma_{\mu}b|B(P_B)>=[P_B+P_K]_{\mu}f_+(s)+\frac{M_B^2-M_K^2}{s}[P_B-P_K]_{\mu}(f_0(s)-f_+(s))\ \ \ s\equiv(P_B-P_K)^2\nonumber\\
 &<K(P_K)|\bar{s}\frac{\imath}{2}\sigma_{\mu\nu}(P_B-P_K)^{\nu}b|B(P_B)>=\imath\left\{[P_B+P_K]_{\mu}\frac{s}{M_B+M_K}-[P_B-P_K]_{\mu}(M_B-M_K)\right\}f_T(s)\nonumber\\
 &<K(P_K)|\bar{s}\imath(\overrightarrow{D}-\overleftarrow{D})_{\mu}b|B(P_B)>\simeq[P_B+P_K]_{\mu}\left\{m_bf_+(s)-\frac{s f_T(s)}{M_B+M_K}\right\}\\
 &\hspace{5cm}+[P_B-P_K]_{\mu}\left\{m_b\frac{M_B^2-M_K^2}{s}\left(f_0(s)-f_+(s)\right)+(M_B-M_K)f_T(s)\right\}\nonumber
\end{align}
For the form factors, we follow the discussion in \cite{Buchalla:2010jv} closely, employing the parametrization of \cite{Becirevic:1999kt} and the approximate
relation $f_T(s)/f_+(s)\simeq\frac{M_B+M_K}{M_B}$.

\noindent The estimation of the branching ratio is now straightforward ($N_{\nu}=3$ is the number of neutrino flavours):
\begin{multline}
 BR(B\to K\nu\bar{\nu})=\frac{N_{\nu}/3}{512\pi^3M_B^3\Gamma_B}\int_0^{(M_B-M_K)^2}{ds\left[s^2-2s(M_B^2+M_K^2)+(M_B^2-M_K^2)^2\right]^{3/2}}\\
\times\left|\left[C_{V^{\nu}}^{L\,L}+C_{V^{\nu}}^{R\,L}+m_b(C_{H^{\nu}}^{L\,L}+C_{H^{\nu}}^{R\,L})\right]-\frac{s f_T(s)}{M_B+M_K}(C_{H^{\nu}}^{L\,L}+C_{H^{\nu}}^{R\,L})\right|^2(\mu_b)
\end{multline}
where we will use $M_B=5.27925(17)$~GeV, $\tau_B=(1.641\pm0.008)\cdot10^{-12}$~s, $M_K=0.493677(16)$~GeV \cite{Beringer:1900zz}.

\noindent The running of the coefficients is again very simple as the neutrality of the neutrinos ensures no mixing with the four-quark operators. Similarly to Eq.(\ref{runleptons}),
we have:
\begin{align}
 &[C_{V^{\nu}}^{L\,L}+C_{V^{\nu}}^{R\,L}](\mu_b)=[C_{V^{\nu}}^{L\,L}+C_{V^{\nu}}^{R\,L}](M_Z)-m_b(M_Z)[C_{H^{\nu}}^{L\,L}+C_{H^{\nu}}^{R\,L}](M_Z)\left[\left(\frac{\alpha_S(\mu_b)}{\alpha_S(M_Z)}\right)^{\frac{2C_2(3)}{\beta_0}}-1\right]\nonumber\\
 &[C_{H^{\nu}}^{L\,L}+C_{H^{\nu}}^{R\,L}](\mu_b)=\left(\frac{\alpha_S(\mu_b)}{\alpha_S(M_Z)}\right)^{-\frac{C_2(3)}{\beta_0}}[C_{H^{\nu}}^{L\,L}+C_{H^{\nu}}^{R\,L}](M_Z)
\end{align}
Going to our explicit matching conditions of Eq.(\ref{matching}), we observe that new-physics effects here persist in the part multiplying $f_T(s)$. Numerically, we find 
$K_{sb}$ (which we assume to be real) in the range $[-4,1.5]\cdot10^{-3}$. The naive estimate $K_{sb}\sim10^{-3}\sim \frac{M_Z^2}{\Lambda_{NP}^2}$ returns 
$\Lambda_{NP}\sim$~TeV; this scale can be lowered further if one assumes that the new-physics operators should be loop-suppressed, follow Minimal Flavour Violation, etc.
The bound on new physics appearing in this fashion is therefore relatively loose. Note however that several other observables in the $b\to s$ sector should be considered 
before drawing any conclusion concerning the viability of $K_{sb}\sim10^{-3}$. Since our focus in this section was merely to consider a concrete, yet simple, case where the 
inclusion of the dimension 7 RGE's was relevant, we shall not pursue this analysis further.

\vspace{2cm}
\noindent Let us briefly summarize our achievements. we have established the most general basis of (on-shell) operators, for the EFT describing $b\to s$ transitions, 
up to mass-dimension $7$. After computing the associated ultraviolet QCD divergences at leading order, we have derived the corresponding RGE's. Comparison with the 
existing studies concerning dimension $6$ operators proved satisfactory. We have also solved all the RGE's describing the evolution of pure dimension $7$ operators:
interestingly, we observed that some directions exhibited negative anomalous dimensions, which could lead to an enhancement effect at low energy. Note however that 
this property depends on the normalization of the operators so that the matching conditions in a definite high-energy model are likely to regulate it. We finally used 
this analysis to constrain, using the measurement of the decay $B_s^0\to\mu^+\mu^-$ at LHCb and the limit set by BABAR on $B\to K\nu\bar{\nu}$, a very naive extension of 
the SM resulting from the addition of a dimension $6$ SM-operator. More generally, let us recall that the inclusion of dimension $7$ effects has little relevance in 
e.g.\ the SM, where such operators receive extra-suppression due to the very-constrained pattern of flavour-violation. Beyond the SM, requiring New-Physics to project 
only on operators of higher-mass dimension would be an elegant way to circumvent the strong limits on non-standard flavour violation: however, such models would have 
to be designed on purpose.

\section*{Acknowledgements}
This work has been supported by the Collaborative Research Center SFB676 of the DFG, ``Particles, Strings, and the Early Universe'', as well as the the BMBF grant 
05H12VKF. The authors wish to thank U.~Nierste for constructive discussions and comments. F.D.\ also acknowledges useful discussions with T.~Ewerth and M.~Wiebusch.

\appendix
\section{Fierz identities}\label{Fierz}
We derive/recover the following Fierz identities\footnote{Note that, here, we consider the identities only from the point of view of the Clifford 
algebra, without including the $-$ sign resulting from the anticommutation of fermion fields.}:
\begin{itemize}
 \item Scalar identities:
\begin{equation}
 \begin{cases}
  (\bar{\psi}'P_L\psi)(\bar{\chi}'P_L\chi)=\frac{1}{2}(\bar{\psi}'P_L\chi)(\bar{\chi}'P_L\psi)-\frac{1}{32}(\bar{\psi}'\sigma^{\mu\nu}P_L\chi)(\bar{\chi}'\sigma_{\mu\nu}P_L\psi)\\
  (\bar{\psi}'P_R\psi)(\bar{\chi}'P_R\chi)=\frac{1}{2}(\bar{\psi}'P_R\chi)(\bar{\chi}'P_R\psi)-\frac{1}{32}(\bar{\psi}'\sigma^{\mu\nu}P_R\chi)(\bar{\chi}'\sigma_{\mu\nu}P_R\psi)\\
  (\bar{\psi}'P_L\psi)(\bar{\chi}'P_R\chi)=\frac{1}{2}(\bar{\psi}'\gamma^{\mu}P_R\chi)(\bar{\chi}'\gamma_{\mu}P_L\psi)\\
  (\bar{\psi}'P_R\psi)(\bar{\chi}'P_L\chi)=\frac{1}{2}(\bar{\psi}'\gamma^{\mu}P_L\chi)(\bar{\chi}'\gamma_{\mu}P_R\psi)
 \end{cases}
\end{equation}
 \item Vector identities:
\begin{equation}
 \begin{cases}
  (\bar{\psi}'\gamma^{\mu}P_L\psi)(\bar{\chi}'\gamma_{\mu}P_L\chi)=-(\bar{\psi}'\gamma^{\mu}P_L\chi)(\bar{\chi}'\gamma_{\mu}P_L\psi)\\
  (\bar{\psi}'\gamma^{\mu}P_R\psi)(\bar{\chi}'\gamma_{\mu}P_R\chi)=-(\bar{\psi}'\gamma^{\mu}P_R\chi)(\bar{\chi}'\gamma_{\mu}P_R\psi)\\
  (\bar{\psi}'\gamma^{\mu}P_L\psi)(\bar{\chi}'\gamma_{\mu}P_R\chi)=2(\bar{\psi}'P_R\chi)(\bar{\chi}'P_L\psi)\\
  (\bar{\psi}'\gamma^{\mu}P_R\psi)(\bar{\chi}'\gamma_{\mu}P_L\chi)=2(\bar{\psi}'P_L\chi)(\bar{\chi}'P_R\psi)
 \end{cases}
\end{equation}
 \item Tensor identities:
\begin{equation}
 \begin{cases}
  (\bar{\psi}'\sigma^{\mu\nu}P_L\psi)(\bar{\chi}'\sigma_{\mu\nu}P_L\chi)=-24(\bar{\psi}'P_L\chi)(\bar{\chi}'P_L\psi)-\frac{1}{2}(\bar{\psi}'\sigma^{\mu\nu}P_L\chi)(\bar{\chi}'\sigma_{\mu\nu}P_L\psi)\\
  (\bar{\psi}'\sigma^{\mu\nu}P_R\psi)(\bar{\chi}'\sigma_{\mu\nu}P_R\chi)=-24(\bar{\psi}'P_R\chi)(\bar{\chi}'P_R\psi)-\frac{1}{2}(\bar{\psi}'\sigma^{\mu\nu}P_R\chi)(\bar{\chi}'\sigma_{\mu\nu}P_R\psi)
 \end{cases}
\end{equation}
 \item Hybrid identities:
\begin{equation}
 \begin{cases}
  (\bar{\psi}'P_L\imath D^{\mu}\psi)(\bar{\chi}'\gamma_{\mu}P_L\chi)=\frac{1}{2}(\bar{\psi}'P_L\chi)(\bar{\chi}'P_R\imath\displaystyle{\not}\hspace{0.2mm}D\psi)-\frac{1}{4}(\bar{\psi}'\sigma^{\mu\nu}P_L\chi)(\bar{\chi}'\gamma_{\nu}P_L\imath D_{\mu}\psi)\\
  (\bar{\psi}'P_R\imath D^{\mu}\psi)(\bar{\chi}'\gamma_{\mu}P_R\chi)=\frac{1}{2}(\bar{\psi}'P_R\chi)(\bar{\chi}'P_L\imath\displaystyle{\not}\hspace{0.2mm}D\psi)-\frac{1}{4}(\bar{\psi}'\sigma^{\mu\nu}P_R\chi)(\bar{\chi}'\gamma_{\nu}P_R\imath D_{\mu}\psi)\\
  (\bar{\psi}'P_L\imath D^{\mu}\psi)(\bar{\chi}'\gamma_{\mu}P_R\chi)=(\bar{\psi}'\gamma^{\mu}P_R\chi)(\bar{\chi}'P_L\imath D_{\mu}\psi)-\frac{1}{2}(\bar{\psi}'\gamma^{\mu}P_R\chi)(\bar{\chi}'\gamma_{\mu}P_R\imath\displaystyle{\not}\hspace{0.2mm}D\psi)\\
  (\bar{\psi}'P_R\imath D^{\mu}\psi)(\bar{\chi}'\gamma_{\mu}P_L\chi)=(\bar{\psi}'\gamma^{\mu}P_L\chi)(\bar{\chi}'P_R\imath D_{\mu}\psi)-\frac{1}{2}(\bar{\psi}'\gamma^{\mu}P_L\chi)(\bar{\chi}'\gamma_{\mu}P_L\imath\displaystyle{\not}\hspace{0.2mm}D\psi)
 \end{cases}
\end{equation}
One can straightforwardly generalize these results to the case where the covariant derivative acts on the first spinor, instead of the second. Moreover, with a little bit of algebra:
\begin{multline}
  \null\hspace{-1cm}(\bar{\psi}'\sigma^{\mu\nu}P_{L,R}\chi)(\bar{\chi}'\gamma_{\nu}P_{L,R}\imath D_{\mu}\psi)=-[\bar{\psi}'\imath(\overrightarrow{D}-\imath\overleftarrow{D})_{\mu}P_{L,R}\chi](\bar{\chi}'\gamma^{\mu}P_{L,R}\psi)-[\bar{\psi}'\imath(\overrightarrow{D}-\imath\overleftarrow{D})_{\mu}\gamma_5P_{L,R}\chi](\bar{\chi}'\gamma^{\mu}\gamma_5P_{L,R}\psi)\\
-[\bar{\psi}'(\imath\overleftarrow{\displaystyle{\not}\hspace{0.2mm}D}\gamma^{\mu}P_{L,R}-\gamma^{\mu}P_{R,L}\imath\overrightarrow{\displaystyle{\not}\hspace{0.2mm}D})\chi](\bar{\chi}'\gamma_{\mu}P_{L,R}\psi)-[\bar{\psi}'(\imath\overleftarrow{\displaystyle{\not}\hspace{0.2mm}D}\gamma^{\mu}\gamma_5P_{L,R}+\gamma^{\mu}\gamma_5P_{R,L}\imath\overrightarrow{\displaystyle{\not}\hspace{0.2mm}D})\chi](\bar{\chi}'\gamma_{\mu}\gamma_5P_{L,R}\psi)\\
-\frac{1}{4}(\bar{\psi}'\sigma^{\mu\nu}P_{L,R}\chi)[\bar{\chi}'(\overleftarrow{\displaystyle{\not}\hspace{0.2mm}D}\sigma_{\mu\nu}P_{L,R}+\sigma_{\mu\nu}P_{R,L}\overrightarrow{\displaystyle{\not}\hspace{0.2mm}D})\psi]-\frac{1}{4}\imath\partial_{\rho}[(\bar{\psi}'\sigma_{\mu\nu}P_{L,R}\chi)(\bar{\chi}'\gamma^{\rho}\gamma^{\nu})\gamma^{\mu}P_{L,R}\psi]
\end{multline}
\end{itemize}

\noindent Consequently, for $q=b,s$, ${\cal O}^q_{L,R\ L,R}={\cal S,V,T,H,\tilde{H}}^q_{L,R\ L,R}$ can be expressed in terms of $O^q_{L,R\ L,R}=S,V,T,H,\\\tilde{H}^q_{L,R\ L,R}$ as 
${\cal O}^q_{L,R\ L,R}=-\tilde{V}^q_{\mbox{Fierz}}O^q_{L,R\ L,R}$ (the sign originates from the necessity of anticommuting fermion fields), and 
reciprocally. Chiralities are ordered as $LL,LR,RR,RL$.
{\small\begin{equation}
\null\hspace{-1.5cm}\tilde{V}^b_{\mbox{Fierz}}=\begin{bmatrix}
  \begin{matrix}
\frac{1}{2} & 0 & 0 & 0 &\null\\
0 & 0 & 0 & 0 &\null\\
0 & 0 & \frac{1}{2} & 0 &\null\\
0 & 0 & 0 & 0 &\null
  \end{matrix} &   \begin{matrix}
0 & 0 & 0 & 0 &\null\\
0 & \frac{1}{2} & 0 & 0 &\null\\
0 & 0 & 0 & 0 &\null\\
0 & 0 & 0 & \frac{1}{2} &\null
  \end{matrix} &   \begin{matrix}
-\frac{1}{32} & 0 &\null\\
0 & 0 &\null\\
0 & -\frac{1}{32} &\null\\
0 & 0 &\null
  \end{matrix} & 0 & 0\\
\begin{matrix}
 \null&\null&\null&\null&\null
\end{matrix}\\
  \begin{matrix}
0 & 0 & 0 & 0 &\null\\
0 & 2 & 0 & 0 &\null\\
0 & 0 & 0 & 0 &\null\\
0 & 0 & 0 & 2 &\null
  \end{matrix} &   \begin{matrix}
-1 & 0 & 0 & 0 &\null\\
0 & 0 & 0 & 0 &\null\\
0 & 0 & -1 & 0 &\null\\
0 & 0 & 0 & 0 &\null
  \end{matrix} &  0 & 0 & 0\\
\begin{matrix}
 \null&\null&\null&\null&\null
\end{matrix}\\
\begin{matrix}
-24 & 0 & 0 & 0 &\null\\
0 & 0 & -24 & 0 &\null
  \end{matrix} & 0 &  \begin{matrix}
-\frac{1}{2} & 0 &\null\\
0 & -\frac{1}{2} &\null
  \end{matrix} &   0 & 0\\
\begin{matrix}
 \null&\null&\null&\null&\null
\end{matrix}\\
\begin{matrix}
-\frac{1}{2} & 0 & 0 & \frac{m_s}{m_b} &\null\\
-\frac{1}{2} & 1 & 0 & 0 &\null\\
0 & \frac{m_s}{m_b} & -\frac{1}{2} & 0 &\null\\
0 & 0 & -\frac{1}{2} & 1 &\null
  \end{matrix} &   \begin{matrix}
-1 & -\frac{1}{2} & 0 & 0 &\null\\
0 & -\frac{1}{2} & -\frac{m_s}{m_b} & 0 &\null\\
0 & 0 & -1 & -\frac{1}{2} &\null\\
-\frac{m_s}{m_b} & 0 & 0 & -\frac{1}{2} &\null
  \end{matrix} &   \begin{matrix}
-\frac{1}{32} & 0 &\null\\
-\frac{1}{32} & 0 &\null\\
0 & -\frac{1}{32} &\null\\
0 & -\frac{1}{32} &\null
  \end{matrix} & \begin{matrix}
0 & 0 & 0 & 0 &\null\\
0 & 1 & 0 & 0 &\null\\
0 & 0 & 0 & 0 &\null\\
0 & 0 & 0 & 1 &\null
  \end{matrix} & \begin{matrix}
1 & 0 & 0 & 0 &\null\\
0 & 0 & 0 & 0 &\null\\
0 & 0 & 1 & 0 &\null\\
0 & 0 & 0 & 0 &\null
  \end{matrix}\\
\begin{matrix}
 \null&\null&\null&\null&\null
\end{matrix}\\
\begin{matrix}
-\frac{1}{2} & 1 & 0 & 0 &\null\\
0 & 1 & -\frac{m_s}{2m_b} & 0 &\null\\
0 & 0 & -\frac{1}{2} & 1 &\null\\
-\frac{m_s}{2m_b} & 0 & 0 & 1 &\null
  \end{matrix} &   \begin{matrix}
-1 & 0 & 0 & -\frac{m_s}{2m_b} &\null\\
-1 & -\frac{1}{2} & 0 & 0 &\null\\
0 & -\frac{m_s}{2m_b} & -1 & 0 &\null\\
0 & 0 & -1 & -\frac{1}{2} &\null
  \end{matrix} &   \begin{matrix}
-\frac{1}{32} & 0 &\null\\
0 & -\frac{m_s}{32m_b} &\null\\
0 & -\frac{1}{32} &\null\\
-\frac{m_s}{32m_b} & 0 &\null
  \end{matrix} & \begin{matrix}
1 & 0 & 0 & 0 &\null\\
0 & 0 & 0 & 0 &\null\\
0 & 0 & 1 & 0 &\null\\
0 & 0 & 0 & 0 &\null
  \end{matrix} & \begin{matrix}
0 & 0 & 0 & 0 &\null\\
0 & 1 & 0 & 0 &\null\\
0 & 0 & 0 & 0 &\null\\
0 & 0 & 0 & 1 &\null
  \end{matrix}
\end{bmatrix}
\end{equation}
\begin{equation}
\null\hspace{-1cm}\tilde{V}^s_{\mbox{Fierz}}=\begin{bmatrix}
  \begin{matrix}
\frac{1}{2} & 0 & 0 & 0 &\null\\
0 & 0 & 0 & 0 &\null\\
0 & 0 & \frac{1}{2} & 0 &\null\\
0 & 0 & 0 & 0 &\null
  \end{matrix} &   \begin{matrix}
0 & 0 & 0 & 0 &\null\\
0 & 0 & 0 & \frac{1}{2} &\null\\
0 & 0 & 0 & 0 &\null\\
0 & \frac{1}{2} & 0 & 0 &\null
  \end{matrix} &   \begin{matrix}
-\frac{1}{32} & 0 &\null\\
0 & 0 &\null\\
0 & -\frac{1}{32} &\null\\
0 & 0 &\null
  \end{matrix} & 0 & 0\\
\begin{matrix}
 \null&\null&\null&\null&\null
\end{matrix}\\
  \begin{matrix}
0 & 0 & 0 & 0 &\null\\
0 & 0 & 0 & 2 &\null\\
0 & 0 & 0 & 0 &\null\\
0 & 2 & 0 & 0 &\null
  \end{matrix} &   \begin{matrix}
-1 & 0 & 0 & 0 &\null\\
0 & 0 & 0 & 0 &\null\\
0 & 0 & -1 & 0 &\null\\
0 & 0 & 0 & 0 &\null
  \end{matrix} &  0 & 0 & 0\\
\begin{matrix}
 \null&\null&\null&\null&\null
\end{matrix}\\
\begin{matrix}
-24 & 0 & 0 & 0 &\null\\
0 & 0 & -24 & 0 &\null
  \end{matrix} & 0 &  \begin{matrix}
-\frac{1}{2} & 0 &\null\\
0 & -\frac{1}{2} &\null
  \end{matrix} &   0 & 0\\
\begin{matrix}
 \null&\null&\null&\null&\null
\end{matrix}\\
\begin{matrix}
-\frac{m_s}{2m_b} & \frac{m_s}{m_b} & 0 & 0 &\null\\
-\frac{m_s}{2m_b} & 0 & 0 & 1 &\null\\
0 & 0 & -\frac{m_s}{2m_b} & \frac{m_s}{m_b} &\null\\
0 & 1 & -\frac{m_s}{2m_b} & 0 &\null
  \end{matrix} &   \begin{matrix}
-1 & 0 & 0 & -\frac{m_s}{2m_b} &\null\\
0 & 0 & -\frac{m_s}{m_b} & -\frac{m_s}{2m_b} &\null\\
0 & -\frac{m_s}{2m_b} & -1 & 0 &\null\\
-\frac{m_s}{m_b} & -\frac{m_s}{2m_b} & 0 & 0 &\null
  \end{matrix} &   \begin{matrix}
-\frac{m_s}{32m_b} & 0 &\null\\
-\frac{m_s}{32m_b} & 0 &\null\\
0 & -\frac{m_s}{32m_b} &\null\\
0 & -\frac{m_s}{32m_b} &\null
  \end{matrix} & \begin{matrix}
1 & 0 & 0 & 0 &\null\\
0 & 0 & 0 & 0 &\null\\
0 & 0 & 1 & 0 &\null\\
0 & 0 & 0 & 0 &\null
  \end{matrix} & \begin{matrix}
0 & 0 & 0 & 0 &\null\\
0 & 0 & 0 & 1 &\null\\
0 & 0 & 0 & 0 &\null\\
0 & 1 & 0 & 0 &\null
  \end{matrix}\\
\begin{matrix}
 \null&\null&\null&\null&\null
\end{matrix}\\
\begin{matrix}
-\frac{1}{2} & 0 & 0 & \frac{m_s}{m_b} &\null\\
0 & 0 & -\frac{m_s}{2m_b} & \frac{m_s}{m_b} &\null\\
0 & \frac{m_s}{m_b} & -\frac{1}{2} & 0 &\null\\
-\frac{m_s}{2m_b} & \frac{m_s}{m_b} & 0 & 0 &\null
  \end{matrix} &   \begin{matrix}
-\frac{m_s}{m_b} & -\frac{m_s}{2m_b} & 0 & 0 &\null\\
-\frac{m_s}{m_b} & 0 & 0 & -\frac{1}{2} &\null\\
0 & 0 & -\frac{m_s}{m_b} & -\frac{m_s}{2m_b} &\null\\
0 & -\frac{1}{2} & -\frac{m_s}{m_b} & 0 &\null
  \end{matrix} &   \begin{matrix}
-\frac{1}{32} & 0 &\null\\
0 & -\frac{m_s}{32m_b} &\null\\
0 & -\frac{1}{32} &\null\\
-\frac{m_s}{32m_b} & 0 &\null
  \end{matrix} & \begin{matrix}
0 & 0 & 0 & 0 &\null\\
0 & 0 & 0 & 1 &\null\\
0 & 0 & 0 & 0 &\null\\
0 & 1 & 0 & 0 &\null
  \end{matrix} & \begin{matrix}
1 & 0 & 0 & 0 &\null\\
0 & 0 & 0 & 0 &\null\\
0 & 0 & 1 & 0 &\null\\
0 & 0 & 0 & 0 &\null
  \end{matrix}
\end{bmatrix}
\end{equation}}

\section{Anomalous Dimension Matrix}\label{gammaC}
Considering that RGE's in EFT's already represent a significantly equiped industry, we have decided to collect here our results for the anomalous dimension 
matrix $\gamma_C$, so that one may easily implement this matrix without having to refer to our discussion in section \ref{RGEsec} and sequels, oriented at a
more progressive, if exhaustive, solution. Note however that, due to the size of the matrix, we are bound to present it block after block. Remember that our 
operators are ordered as in Eq.(\ref{bsllop},\ref{bsqqop},\ref{bsGop}) and that the chiralities follow the ordering $LL$, $LR$, $RR$, $RL$, or $L$, $R$ 
(depending on the number of chirality indices the operator carries). The definition of the blocks in chirality space is reminded at the very end of this section.
Recalling Eq.(\ref{andimmat}):
\begin{equation}
\gamma_C=\begin{bmatrix}
\gamma_C^{ll} & 0 & 0\\
\gamma_C^{ql} & \gamma_C^{qq} & \gamma_C^{qg} \\
\gamma_C^{gl} & \gamma_C^{gq} & \gamma_C^{gg}
\end{bmatrix}
\end{equation}

\begin{itemize}
\item{\em Subblock $\gamma_C^{ll}$}
\end{itemize}
We remind the reader that operators with different lepton flavours do not mix (at least as long as massless neutrinos are considered) so that the subblock 
$\gamma_C^{ll}$ is diagonal in lepton-flavour space:
\begin{equation}
\gamma_C^{ll}=\begin{bmatrix}
\tilde{\gamma}_C^{ee} & 0 & 0 & 0 & 0 & 0 \\
0 & \tilde{\gamma}_C^{\mu\mu} & 0 & 0 & 0 & 0\\
0 & 0 & \tilde{\gamma}_C^{\tau\tau} & 0 & 0 & 0\\
0 & 0 & 0 & \tilde{\gamma}_C^{\nu_e\nu_e} & 0 & 0\\
0 & 0 & 0 & 0 & \tilde{\gamma}_C^{\nu_{\mu}\nu_{\mu}} & 0\\
0 & 0 & 0 & 0 & 0 & \tilde{\gamma}_C^{\nu_{\tau}\nu_{\tau}}
\end{bmatrix}
\end{equation}
From Eq.(\ref{RGEl}) and sequels, we read, for any lepton flavour $l$ (but note that the basis of operators is shortened in the neutrino case, due to the absence of 
low-energy right-handed neutrinos):
\begin{equation}
 \tilde{\gamma}_C^{ll}=\frac{2\alpha_SC_2(3)}{4\pi}\begin{bmatrix}
                -\frac{23}{4}\mathbb{1} & 0 & 0 & 0 & 0\\
                0 & -\frac{11}{4}\mathbb{1} & 0 & 0 & 0\\
                0 & 0 & -\frac{7}{4} & 0 & 0\\
                0 & 2\left(\mathbb{1}+\frac{m_s}{m_b}\Sigma\right) & 0 & -\frac{19}{4} & 0\\
                0 & 0 & 0 & 0 & -\frac{23}{4}\mathbb{1}
               \end{bmatrix}
\end{equation}

\begin{itemize}
\item{\em Subblock $\gamma_C^{ql}$}
\end{itemize}
\begin{equation}
\gamma_C^{ql}=\begin{bmatrix}
\tilde{\gamma}_C^{ue} & \tilde{\gamma}_C^{u\mu} & \tilde{\gamma}_C^{u\tau} & \tilde{\gamma}_C^{u\nu_e} & \tilde{\gamma}_C^{u\nu_{\mu}} & \tilde{\gamma}_C^{u\nu_{\tau}} \\
\tilde{\gamma}_C^{de} & \tilde{\gamma}_C^{d\mu} & \tilde{\gamma}_C^{d\tau} & \tilde{\gamma}_C^{d\nu_e} & \tilde{\gamma}_C^{d\nu_{\mu}} & \tilde{\gamma}_C^{d\nu_{\tau}} \\
\tilde{\gamma}_C^{se} & \tilde{\gamma}_C^{s\mu} & \tilde{\gamma}_C^{s\tau} & \tilde{\gamma}_C^{s\nu_e} & \tilde{\gamma}_C^{s\nu_{\mu}} & \tilde{\gamma}_C^{s\nu_{\tau}} \\
\tilde{\gamma}_C^{ce} & \tilde{\gamma}_C^{c\mu} & \tilde{\gamma}_C^{c\tau} & \tilde{\gamma}_C^{c\nu_e} & \tilde{\gamma}_C^{c\nu_{\mu}} & \tilde{\gamma}_C^{c\nu_{\tau}} \\
\tilde{\gamma}_C^{be} & \tilde{\gamma}_C^{b\mu} & \tilde{\gamma}_C^{b\tau} & \tilde{\gamma}_C^{b\nu_e} & \tilde{\gamma}_C^{b\nu_{\mu}} & \tilde{\gamma}_C^{b\nu_{\tau}} 
\end{bmatrix}
\end{equation}
From Eq.(\ref{RGEl}) and sequels, we read, for  any lepton flavour $l$ and $q=u,d,c$:
\begin{equation}
\tilde{\gamma}_C^{ql}=-\frac{2\alpha_SC_2(3)}{4\pi}\frac{3}{2}Q_lQ_q\begin{bmatrix}
                0 & 0 & 0 & 0 & 0\\
                0 & -(\mathbb{1}+\tilde{\Sigma}) & 0 & 0 & 0\\
                0 & 0 & 0 & 0 & 0\\
                0 & 0 & 0 & -(\mathbb{1}+\tilde{\Sigma}) & 0\\
                0 & \frac{m_q}{m_b}(\mathbb{1}+\tilde{\Sigma}) & 0 & 0 & 0\\
                0 & 0 & 0 & 0 & 0\\
                0 & -\frac{1}{3}(\mathbb{1}+\tilde{\Sigma}) & 0 & 0 & 0\\
                0 & 0 & 0 & 0 & 0\\
                0 & 0 & 0 & -\frac{1}{3}(\mathbb{1}+\tilde{\Sigma}) & 0\\
                0 & \frac{m_q}{3m_b}(\mathbb{1}+\tilde{\Sigma}) & 0 & 0 & 0\\
               \end{bmatrix}
\end{equation}
For $q=b,s$, we have:
\begin{equation}
\tilde{\gamma}_C^{ql}=\frac{2\alpha_SC_2(3)}{4\pi}\frac{3}{2}Q_lQ_q\left\{-\begin{bmatrix}
                0 & 0 & 0 & 0 & 0\\
                0 & -(\mathbb{1}+\tilde{\Sigma}) & 0 & 0 & 0\\
                0 & 0 & 0 & 0 & 0\\
                0 & 0 & 0 & -(\mathbb{1}+\tilde{\Sigma}) & 0\\
                0 & \frac{m_q}{m_b}(\mathbb{1}+\tilde{\Sigma}) & 0 & 0 & 0\\
\end{bmatrix}+\tilde{V}^q_{\mbox{Fierz}}\begin{bmatrix}
                0 & 0 & 0 & 0 & 0\\
                0 & -\frac{1}{3}(\mathbb{1}+\tilde{\Sigma}) & 0 & 0 & 0\\
                0 & 0 & 0 & 0 & 0\\
                0 & 0 & 0 & -\frac{1}{3}(\mathbb{1}+\tilde{\Sigma}) & 0\\
                0 & \frac{m_q}{3m_b}(\mathbb{1}+\tilde{\Sigma}) & 0 & 0 & 0\\
               \end{bmatrix}
\right\}
\end{equation}

\begin{itemize}
\item{\em Subblock $\gamma_C^{gl}$}
\end{itemize}
\begin{equation}
\gamma_C^{gl}=\begin{bmatrix}
\tilde{\gamma}_C^{ge} & \tilde{\gamma}_C^{g\mu} & \tilde{\gamma}_C^{g\tau} & \tilde{\gamma}_C^{g\nu_e} & \tilde{\gamma}_C^{g\nu_{\mu}} & \tilde{\gamma}_C^{g\nu_{\tau}} 
\end{bmatrix}
\end{equation}
From Eq.(\ref{RGEl}) and sequels, we read, for  any lepton flavour $l$:
\begin{equation}
\tilde{\gamma}_C^{gl}=\frac{2\alpha_SC_2(3)}{4\pi}\frac{1}{3}Q_l\begin{bmatrix}
                0 & 0 & 0 & 0 & 0\\
                0 & 0 & 0 & 0 & 0\\
                0 & 0 & 0 & 0 & 0\\
                0 & 0 & 0 & 0 & 0\\
                0 & 0 & 0 & 0 & 0\\
                0 & \Xi+\frac{m_s}{m_b}\Phi & 0 & 0 & 0\\
                0 & 2[\Xi'+\frac{m_s}{m_b}\Phi'] & 0 & 0 & 0\\
                0 & 0 & 0 & 0 & 0\\
                0 & 0 & 0 & 0 & 0\\
                0 & 0 & 0 & 0 & 0\\
                0 & 0 & 0 & 0 & 0\\
                0 & 0 & 0 & 0 & 0\\
                0 & 0 & 0 & 0 & 0
               \end{bmatrix}
\end{equation}

\begin{itemize}
\item{\em Subblock $\gamma_C^{qq}$}
\end{itemize}
\begin{equation}
\gamma_C^{ql}=\begin{bmatrix}
\tilde{\gamma}_C^{uu} & \tilde{\gamma}_C^{ud} & \tilde{\gamma}_C^{us} & \tilde{\gamma}_C^{uc} & \tilde{\gamma}_C^{ub} \\
\tilde{\gamma}_C^{du} & \tilde{\gamma}_C^{dd} & \tilde{\gamma}_C^{ds} & \tilde{\gamma}_C^{dc} & \tilde{\gamma}_C^{db} \\
\tilde{\gamma}_C^{su} & \tilde{\gamma}_C^{sd} & \tilde{\gamma}_C^{ss} & \tilde{\gamma}_C^{sc} & \tilde{\gamma}_C^{sb} \\
\tilde{\gamma}_C^{cu} & \tilde{\gamma}_C^{cd} & \tilde{\gamma}_C^{cs} & \tilde{\gamma}_C^{cc} & \tilde{\gamma}_C^{cb} \\
\tilde{\gamma}_C^{bu} & \tilde{\gamma}_C^{bd} & \tilde{\gamma}_C^{bs} & \tilde{\gamma}_C^{bc} & \tilde{\gamma}_C^{bb} 
\end{bmatrix}
\end{equation}
For two quark flavours $q$ and $q'$, we read from Eq.(\ref{RGEqq}) and sequels,

* if $q$ and $q'\neq b,s$:
\begin{equation}
 \tilde{\gamma}_C^{q'q}=\frac{2\alpha_SC_2(3)}{4\pi}\left\{\delta^{q'q}\mbox{diag}(5\mathbb{1},5\mathbb{1},5\mathbb{1},2\mathbb{1},2\mathbb{1},5\mathbb{1},5\mathbb{1},5\mathbb{1},2\mathbb{1},2\mathbb{1})-
\delta^{q'q}\tilde{\cal M}^{q'q}_{\mbox{\tiny diag}}-\tilde{\cal M}^{q'q}_{\mbox{\tiny univ.}}\right\}
\end{equation}

* if $q$ and $q'=b$ or $s$:
\begin{equation}
 \tilde{\gamma}_C^{q'q}=\frac{2\alpha_SC_2(3)}{4\pi}\left\{\delta^{q'q}\mbox{diag}(5\mathbb{1},5\mathbb{1},5\mathbb{1},2\mathbb{1},2\mathbb{1})-
\delta^{q'q}[\mathbb{1},0]\tilde{\cal M}^{q'q}_{\mbox{\tiny diag}}\begin{bmatrix}\mathbb{1}\\0\end{bmatrix}-[\mathbb{1},-\tilde{V}^{q'}_{\mbox{Fierz}}]\tilde{\cal M}^{q'q}_{\mbox{\tiny univ.}}\begin{bmatrix}\mathbb{1}\\-\tilde{V}^q_{\mbox{Fierz}}\end{bmatrix}\right\}
\end{equation}

* if $q \neq b,s$ and $q'=b$ or $s$:
\begin{equation}
 \tilde{\gamma}_C^{q'q}=-\frac{2\alpha_SC_2(3)}{4\pi}[\mathbb{1},-\tilde{V}^{q'}_{\mbox{Fierz}}]\tilde{\cal M}^{q'q}_{\mbox{\tiny univ.}}
\end{equation}

* if $q=b$ or $s$ and $q'\neq b,s$:
\begin{equation}
 \tilde{\gamma}_C^{q'q}=-\frac{2\alpha_SC_2(3)}{4\pi}\tilde{\cal M}^{q'q}_{\mbox{\tiny univ.}}\begin{bmatrix}\mathbb{1}\\-\tilde{V}^q_{\mbox{Fierz}}\end{bmatrix}
\end{equation}
with:
\begin{displaymath}
\tilde{\cal M}^{q'q}_{\mbox{\tiny diag}}=\begin{bmatrix}
                8\mathbb{1} & 0 & \frac{1}{32}\Delta & 0 & 0 & 0 & 0 & -\frac{3}{32}\Delta & 0 & 0\\
                0 & 2\mathbb{1}+\frac{3}{4}\Sigma_3 & 0 & 0 & 0 & 0 & -\frac{9}{4}\Sigma_3 & 0 & 0 & 0\\
                24\Delta^T & 0 & 0 & 0 & 0 & -72\Delta^T & 0 & 0 & 0 & 0\\
                A_H & B_H & 0 & \mathbb{1} & 0 & C_H & D_H & 0 & 0 & 0\\
                A_{\tilde{H}} & B_{\tilde{H}} & 0 & 0 & \mathbb{1} & C_{\tilde{H}} & D_{\tilde{H}} & 0 & 0 & 0\\
                \frac{9}{4}\mathbb{1} & 0 & -\frac{3}{64}\Delta & 0 & 0 & \frac{5}{4}\mathbb{1} & 0 & -\frac{7}{64}\Delta & 0 & 0\\
                0 & -\frac{9}{8}(\mathbb{1}+\Sigma_3) & 0 & 0 & 0 & 0 & \frac{1}{8}(43\mathbb{1}-21\Sigma_3) & 0 & 0 & 0\\
                -36\Delta^T & 0 & -\frac{9}{4}\mathbb{1} & 0 & 0 & -84\Delta^T & 0 & \frac{27}{4}\mathbb{1} & 0 & 0\\
                A_{\cal H} & B_{\cal H} & 0 & 0 & 0 & C_{\cal H} & D_{\cal H} & 0 & \mathbb{1} & 0\\
                A_{\cal \tilde{H}} & B_{\cal \tilde{H}} & 0 & 0 & 0 & C_{\cal \tilde{H}} & D_{\cal \tilde{H}} & 0 & 0 & \mathbb{1}
               \end{bmatrix}
\end{displaymath}\begin{displaymath}
\tilde{\cal M}^{q'q}_{\mbox{\tiny univ}}=\frac{1}{4}\begin{bmatrix}
                0 & 0 & 0 & 0 & 0 & 0 & 0 & 0 & 0 & 0\\
                0 & 0 & 0 & 0 & 0 & 0 & 0 & 0 & 0 & 0\\
                0 & 0 & 0 & 0 & 0 & 0 & 0 & 0 & 0 & 0\\
                0 & 0 & 0 & 0 & 0 & 0 & 0 & 0 & 0 & 0\\
                0 & 0 & 0 & 0 & 0 & 0 & 0 & 0 & 0 & 0\\
                0 & 0 & 0 & 0 & 0 & 0 & 0 & 0 & 0 & 0\\
                0 & \frac{1}{3}(\mathbb{1}+\tilde{\Sigma}) & 0 & 0 & 0 & 0 & -(\mathbb{1}+\tilde{\Sigma}) & 0 & 0 & 0\\
                0 & 0 & 0 & 0 & 0 & 0 & 0 & 0 & 0 & 0\\
                0 & 0 & 0 & \frac{1}{3}(\mathbb{1}+\tilde{\Sigma}) & 0 & 0 & 0 & 0 & -(\mathbb{1}+\tilde{\Sigma}) & 0\\
                0 & -\frac{m_q}{3m_b}(\mathbb{1}+\tilde{\Sigma}) & 0 & 0 & 0 & 0 & \frac{m_q}{m_b}(\mathbb{1}+\tilde{\Sigma}) & 0 & 0 & 0
               \end{bmatrix}
\end{displaymath}

\begin{itemize}
\item{\em Subblock $\gamma_C^{gq}$}
\end{itemize}
\begin{equation}
\gamma_C^{gq}=\begin{bmatrix}
\tilde{\gamma}_C^{gu} & \tilde{\gamma}_C^{gd} & \tilde{\gamma}_C^{gs} & \tilde{\gamma}_C^{gc} & \tilde{\gamma}_C^{gb}
\end{bmatrix}
\end{equation}
For a quark flavour $q=u,d,s$, we read from Eq.(\ref{RGEqq}) and sequels:
\begin{equation}
\tilde{\gamma}_C^{gq}=-\frac{2\alpha_SC_2(3)}{4\pi}\begin{bmatrix}
                0 & 0 & 0 & 0 & 0 & 0 & 0 & 0 & 0 & 0\\
                0 & 0 & 0 & 0 & 0 & 0 & 0 & 0 & 0 & 0\\
                0 & 0 & 0 & 0 & 0 & 0 & 0 & 0 & 0 & 0\\
                0 & 0 & 0 & 0 & 0 & 0 & 0 & 0 & 0 & 0\\
                0 & 0 & 0 & 0 & 0 & 0 & 0 & 0 & 0 & 0\\
                0 & 0 & 0 & 0 & 0 & 0 & 0 & 0 & 0 & 0\\
                0 & 0 & 0 & 0 & 0 & 0 & 0 & 0 & 0 & 0\\
                0 & 0 & 0 & 0 & 0 & 0 & 0 & 0 & 0 & 0\\
                -12\frac{m_q}{m_b}\Xi & \frac{1}{12}(\Xi+\frac{m_s}{m_b}\Phi) & 0 & 0 & 0 & 0 & -\frac{1}{4}(\Xi+\frac{m_s}{m_b}\Phi) & 0 & 0 & 0\\
                24\frac{m_q}{m_b}\tilde{\Xi} & \frac{1}{6}(\Xi'+\frac{m_s}{m_b}\Phi') & 0 & 0 & 0 & 0 & -\frac{1}{2}(\Xi'+\frac{m_s}{m_b}\Phi') & 0 & 0 & 0\\
                -\frac{11}{8}\frac{m_q}{m_b}\Xi & \frac{7}{144}(\Xi+\frac{m_s}{m_b}\Phi) & 0 & 0 & 0 & -\frac{15}{8}\frac{m_q}{m_b}\Xi & -\frac{7}{48}(\Xi+\frac{m_s}{m_b}\Phi) & 0 & 0 & 0\\
                \frac{11}{4}\frac{m_q}{m_b}\tilde{\Xi} & \frac{7}{72}(\Xi'+\frac{m_s}{m_b}\Phi') & 0 & 0 & 0 & \frac{15}{4}\frac{m_q}{m_b}\tilde{\Xi} & -\frac{7}{24}(\Xi'+\frac{m_s}{m_b}\Phi') & 0 & 0 & 0\\
                0 & 0 & -\frac{3}{64}\frac{m_q}{m_b}\mathbb{1} & 0 & 0 & 0 & 0 & \frac{9}{64}\frac{m_q}{m_b}\mathbb{1} & 0 & 0
               \end{bmatrix}
\end{equation}
For $q=b,s$, we have:
\begin{multline}
\tilde{\gamma}_C^{gq}=\frac{2\alpha_SC_2(3)}{4\pi}\times\\\left\{-\begin{bmatrix}
                0 & 0 & 0 & 0 & 0 \\
                0 & 0 & 0 & 0 & 0 \\
                0 & 0 & 0 & 0 & 0 \\
                0 & 0 & 0 & 0 & 0 \\
                0 & 0 & 0 & 0 & 0 \\
                0 & 0 & 0 & 0 & 0 \\
                0 & 0 & 0 & 0 & 0 \\
                0 & 0 & 0 & 0 & 0 \\
                -12\frac{m_q}{m_b}\Xi & \frac{1}{12}(\Xi+\frac{m_s}{m_b}\Phi) & 0 & 0 & 0 \\
                24\frac{m_q}{m_b}\tilde{\Xi} & \frac{1}{6}(\Xi'+\frac{m_s}{m_b}\Phi') & 0 & 0 & 0 \\
                -\frac{11}{8}\frac{m_q}{m_b}\Xi & \frac{7}{144}(\Xi+\frac{m_s}{m_b}\Phi) & 0 & 0 \\
                \frac{11}{4}\frac{m_q}{m_b}\tilde{\Xi} & \frac{7}{72}(\Xi'+\frac{m_s}{m_b}\Phi') & 0 & 0 & 0 \\
                0 & 0 & -\frac{3}{64}\frac{m_q}{m_b}\mathbb{1} & 0 & 0 
\end{bmatrix}+\begin{bmatrix}
                 0 & 0 & 0 & 0 & 0\\
                 0 & 0 & 0 & 0 & 0\\
                 0 & 0 & 0 & 0 & 0\\
                 0 & 0 & 0 & 0 & 0\\
                 0 & 0 & 0 & 0 & 0\\
                 0 & 0 & 0 & 0 & 0\\
                 0 & 0 & 0 & 0 & 0\\
                 0 & 0 & 0 & 0 & 0\\
                 0 & -\frac{1}{4}(\Xi+\frac{m_s}{m_b}\Phi) & 0 & 0 & 0\\
                 0 & -\frac{1}{2}(\Xi'+\frac{m_s}{m_b}\Phi') & 0 & 0 & 0\\
                -\frac{15}{8}\frac{m_q}{m_b}\Xi & -\frac{7}{48}(\Xi+\frac{m_s}{m_b}\Phi) & 0 & 0 & 0\\
                 \frac{15}{4}\frac{m_q}{m_b}\tilde{\Xi} & -\frac{7}{24}(\Xi'+\frac{m_s}{m_b}\Phi') & 0 & 0 & 0\\
                 0 & 0 & \frac{9}{64}\frac{m_q}{m_b}\mathbb{1} & 0 & 0
               \end{bmatrix}\tilde{V}^q_{\mbox{Fierz}}
\right\}
\end{multline}

\begin{itemize}
\item{\em Subblock $\gamma_C^{qg}$}
\end{itemize}
\begin{equation}
\gamma_C^{qg}=\begin{bmatrix}
\tilde{\gamma}_C^{ug} \\
\tilde{\gamma}_C^{dg} \\
\tilde{\gamma}_C^{sg} \\
\tilde{\gamma}_C^{cg} \\
\tilde{\gamma}_C^{bg}
\end{bmatrix}
\end{equation}
For a quark flavour $q=u,d,s$, we read from Eq.(\ref{RGEga}) and sequels:
\begin{equation}
\tilde{\gamma}_C^{qg}=-\frac{2\alpha_SC_2(3)}{4\pi}\begin{bmatrix}
                      0 & 0 & 0 & \null\\
                      0 & 0 & 0 & \null\\
                      18Q_q\frac{m_q}{m_b}\mathbb{1}&0&0&\null\\
                      0 & 0 & 0 & \null\\
                      0 & 0 & 0 & \ldots\\
                      0 & 0 & 0 & \null\\
                      0 & 0 & 0 & \null\\
                      6Q_q\frac{m_q}{m_b}\mathbb{1} & 6\frac{m_q}{m_b}\mathbb{1}&0&\null\\
                      0 & 0 & 0 & \null\\
                      0 & 0 & 0 & \null
                     \end{bmatrix}
\end{equation}
For $q=b,s$, we have:
\begin{equation}
\tilde{\gamma}_C^{gq}=\frac{2\alpha_SC_2(3)}{4\pi}\left\{-\begin{bmatrix}
                      0 & 0 & 0 & \null\\
                      0 & 0 & 0 & \null\\
                      18Q_q\frac{m_q}{m_b}\mathbb{1}&0&0&\ldots\\
                      0 & 0 & 0 & \null\\
                      0 & 0 & 0 & \null\\
\end{bmatrix}+\tilde{V}^q_{\mbox{Fierz}}\begin{bmatrix}
                 0 & 0 & 0 & \null\\
                      0 & 0 & 0 & \null\\
                      6Q_q\frac{m_q}{m_b}\mathbb{1} & 6\frac{m_q}{m_b}\mathbb{1}&0&\ldots\\
                      0 & 0 & 0 & \null\\
                      0 & 0 & 0 & \null
               \end{bmatrix}
\right\}
\end{equation}

\begin{itemize}
\item{\em Subblock $\gamma_C^{gg}$}
\end{itemize}
\begin{multline}
\gamma_C^{gg}=\frac{2\alpha_SC_2(3)}{4\pi}\left[\begin{array}{cccccccc}
                \frac{5}{4}\mathbb{1} & 0 & 0 & 0 & 0 & 0 & 0 & 0 \\
                4Q_d\mathbb{1} & \frac{3}{4}\mathbb{1} & 0 & 0 & 0 & 0 & 0 & 0 \\
                0 & 0 & -\frac{35}{4}\mathbb{1} & 0 & 0 & 0 & 0 & 0 \\
                0 & 0 & 0 & -\frac{35}{4}\mathbb{1} & 0 & 0 & 0 & 0 \\
                0 & 0 & 0 & 0 & -\frac{19}{4}\mathbb{1} & 0 & 0 & 0 \\
                \frac{1}{4}\Omega & 0 & \frac{2}{3}Q_d\mathbb{1} & \frac{1}{3}Q_d\sigma_3 & 0 & -\frac{5}{24}\mathbb{1} & -\frac{11}{48}\sigma_3 & 0 \\
                -\frac{1}{2}\tilde{\Omega} & 0 & \frac{4}{3}Q_d\sigma_3 & \frac{2}{3}Q_d\mathbb{1} & 0 & -\frac{11}{12}\sigma_3 & -\frac{5}{24}\mathbb{1}& 0 \\
                -(1-\frac{m_s^2}{m_b^2})\mathbb{1} & 0 & 0 & 0 & 0 & 0 & 0 & -\frac{11}{4}\mathbb{1}\\
                0 & \frac{3}{8}\Omega & 0 & 0 & 0 &  Q_d\mathbb{1} &  \frac{1}{2}Q_d\sigma_3 & 0 \\
                0 & -\frac{3}{4}\tilde{\Omega} & 0 & 0 & 0 & 2 Q_d\sigma_3 &  Q_d\mathbb{1} & 0 \\
                0 & \frac{7}{32}\Omega & 0 & 0 & 0 & \frac{7}{12} Q_d\mathbb{1} & \frac{7}{24} Q_d\sigma_3 & 0 \\
                0 & -\frac{7}{16}\tilde{\Omega} & 0 & 0 & 0 & \frac{7}{6} Q_d\sigma_3 & \frac{7}{12} Q_d\mathbb{1} & 0 \\
                0 & \frac{9}{8}(1+\frac{m_s^2}{m_b^2})\mathbb{1} & 0 & 0 & 0 & \frac{9}{2} Q_d\mathbb{1} & -\frac{9}{4} Q_d\sigma_3 & 0 
                \end{array}\right.\\\left.\begin{array}{ccccc}
 0 & 0 & 0 & 0 & 0\\ 0 & 0 & 0 & 0 & 0\\ 0 & 0 & 0 & 0 & 0\\ 0 & 0 & 0 & 0 & 0\\ 0 & 0 & 0 & 0 & 0\\ 0 & 0 & 0 & 0 & 0\\ 0 & 0 & 0 & 0 & 0\\ 0 & 0 & 0 & 0 & 0\\ -\frac{35}{4}\mathbb{1} & 0 & \mathbb{1} & \frac{1}{2}\sigma_3 & -\frac{9}{2}\mathbb{1} \\ 0 & -\frac{35}{4}\mathbb{1} & 2\sigma_3 & \mathbb{1} & 9\sigma_3 \\ -\frac{21}{16}\mathbb{1} & \frac{3}{32}\sigma_3 & -\frac{385}{24}\mathbb{1} & \frac{41}{48}\sigma_3 & -\frac{11}{8}\mathbb{1} \\ \frac{3}{8}\sigma_3 & -\frac{21}{16}\mathbb{1} & \frac{41}{12}\sigma_3 & -\frac{385}{24}\mathbb{1} & \frac{11}{4}\sigma_3 \\ -\frac{3}{2}\mathbb{1} & \frac{3}{4}\sigma_3 & -\frac{45}{2}\mathbb{1} & \frac{45}{4}\sigma_3 & 2\mathbb{1}
\end{array}\right]
\end{multline}

\noindent We finally collect the definition of the various relevant subblocks in chirality space (with exception of the identity $\mathbb{1}$ and the null matrix $0$):
\begin{equation}
\Sigma\equiv\begin{matrix}
0&0&0&1\\
0&0&1&0\\
0&1&0&0\\
1&0&0&0\\
\end{matrix}\ \ ;\ \ \tilde{\Sigma}\equiv\begin{matrix}
0&1&0&0\\
1&0&0&0\\
0&0&0&1\\
0&0&1&0\\
\end{matrix}\ \ ;\ \ \begin{cases}
\Xi\equiv\begin{matrix}
1 & 1 & 0 & 0\\ 0 & 0 & 1 & 1
\end{matrix}\\ \null\\
\Phi\equiv\begin{matrix}
0 & 0 & 1 & 1\\1 & 1 & 0 & 0
\end{matrix}
\end{cases}\ \ ;\ \ \begin{cases}
\Xi'\equiv\begin{matrix}
1 & 1 & 0 & 0\\0 & 0 & -1 & -1
\end{matrix}\\ \null\\
\Phi'\equiv\begin{matrix}
0 & 0 & 1 & 1\\-1 & -1 & 0 & 0
\end{matrix}
\end{cases}
\end{equation}
\begin{displaymath}
 \tilde{\Xi}\equiv\begin{matrix} 1 & -1 & 0 & 0\\ 0 & 0 & -1 & 1\end{matrix}\ \ ;\ \ \sigma_1\equiv\begin{matrix}0&1\\1&0\end{matrix}\ \ ;\ \ \sigma_3\equiv\begin{matrix}1&0\\0&-1\end{matrix}\ \ ;\ \ \varepsilon\equiv\begin{matrix}0&-1\\1&0\end{matrix}
\end{displaymath}
\begin{displaymath}
\Sigma_3=\begin{matrix}\ \ ;\ \ 
          1 & 0 & 0 & 0\\
          0 & -1 & 0 & 0\\
          0 & 0 & 1 & 0\\
          0 & 0 & 0 & -1
         \end{matrix}\ \ ;\ \ 
E=\begin{matrix}
          0 & -1 & 0 & 0\\
          1 & 0 & 0 & 0\\
          0 & 0 & 0 & -1\\
          0 & 0 & 1 & 0
         \end{matrix}\ \ ;\ \ 
\tilde{E}=\begin{matrix}
          0 & 0 & 0 & -1\\
          0 & 0 & 1 & 0\\
          0 & -1 & 0 & 0\\
          1 & 0 & 0 & 0
         \end{matrix}\ \ ;\ \ 
\Delta=\begin{matrix}
          1 & 0 \\
          0 & 0 \\
          0 & 1 \\
          0 & 0
         \end{matrix}
\end{displaymath}
\begin{displaymath}
 \begin{cases}
  A_H=\frac{m_q}{4m_b}\left[3(\mathbb{1}+\tilde{\Sigma})+\Sigma_3+E\right]\\
  A_{\tilde{H}}=\frac{1}{4}\left[3\mathbb{1}+\Sigma_3+\frac{m_s}{m_b}(3\Sigma+\tilde{E})\right]\\
  A_{\cal H}=-\frac{3}{8}\frac{m_q}{m_b}\left[3(\mathbb{1}+\tilde{\Sigma})+\Sigma_3+E\right]\\
  A_{\cal \tilde{H}}=-\frac{3}{8}(3\mathbb{1}+\Sigma_3)-\frac{3}{8}\frac{m_s}{m_b}(3\Sigma+\tilde{E})
 \end{cases}\ \ \ ;\ \ \ \begin{cases}
  B_H=\frac{1}{4}\left[-8\mathbb{1}+\Sigma_3+\frac{m_s}{m_b}(-8\Sigma+\tilde{E})\right]\\
  B_{\tilde{H}}=\frac{m_q}{4m_b}\left[-8(\mathbb{1}+\tilde{\Sigma})+\Sigma_3+E\right]\\
  B_{\cal H}=-\frac{3}{8}(3\mathbb{1}+\Sigma_3)-\frac{3}{8}\frac{m_s}{m_b}(3\Sigma+\tilde{E})\\
  B_{\cal \tilde{H}}=-\frac{3}{8}\frac{m_q}{m_b}\left[3(\mathbb{1}+\tilde{\Sigma})+\Sigma_3+E\right]
 \end{cases}
\end{displaymath}
\begin{displaymath}\null\hspace{0.4cm}\begin{cases}
  C_H=-\frac{3}{4}\frac{m_q}{m_b}\left[3(\mathbb{1}+\tilde{\Sigma})+\Sigma_3+E\right]\\
  C_{\tilde{H}}=-\frac{3}{4}\left[(3\mathbb{1}+\Sigma_3)+\frac{m_s}{m_b}(3\Sigma+\tilde{E})\right]\\
  C_{\cal H}=-\frac{7}{8}\frac{m_q}{m_b}\left[3(\mathbb{1}+\tilde{\Sigma})+\Sigma_3+E\right]\\
  C_{\cal \tilde{H}}=-\frac{7}{8}(3\mathbb{1}+\Sigma_3)-\frac{7}{8}\frac{m_s}{m_b}(3\Sigma+\tilde{E})
 \end{cases}\ \ \ ;\ \ \ \begin{cases}
  D_H=-\frac{3}{4}(\Sigma_3+\frac{m_s}{m_b}\tilde{E})\\
  D_{\tilde{H}}=-\frac{3}{4}\frac{m_q}{m_b}(\Sigma_3+E)\\
  D_{\cal H}=\frac{1}{8}(11\mathbb{1}-7\Sigma_3)+\frac{m_s}{8m_b}(11\Sigma-7\tilde{E})\\
  D_{\cal \tilde{H}}=\frac{m_q}{8m_b}\left[11(\mathbb{1}+\tilde{\Sigma})-7(\Sigma_3+E)\right]
 \end{cases}
\end{displaymath}
\begin{displaymath}
\Omega\equiv(1+\frac{m_s^2}{m_b^2})\mathbb{1}-\frac{2}{3}\frac{m_s}{m_b}\sigma_1\ \ ;\ \ \tilde{\Omega}\equiv(1+\frac{m_s^2}{m_b^2})\sigma_3-\frac{2}{3}\frac{m_s}{m_b}\varepsilon
\end{displaymath}

\section{Diagonalization of the anomalous-dimension matrix for dimension $7$ operators}\label{diago}
\subsection{Gluonic operators}
We consider the matrix ${\cal A}_{\cal QQ}$ defined in Eq.(\ref{RGEQ}):
\begin{equation}
 {\cal A}_{\cal QQ}=\begin{bmatrix}
\frac{35}{4}\mathbb{1} & 0 & -\mathbb{1} & -\frac{1}{2}\sigma_3 & \frac{9}{2}\mathbb{1}\\
0 & \frac{35}{4}\mathbb{1} & -2\sigma_3 & -\mathbb{1} & -9\sigma_3\\
\frac{21}{16}\mathbb{1} & -\frac{3}{32}\sigma_3 & \frac{385}{24}\mathbb{1} & -\frac{41}{48}\sigma_3 & \frac{11}{8}\mathbb{1}\\
-\frac{3}{8}\sigma_3 & \frac{21}{16}\mathbb{1} & -\frac{41}{12}\sigma_3 & \frac{385}{24}\mathbb{1} & -\frac{11}{4}\sigma_3 \\
\frac{3}{2}\mathbb{1} & -\frac{3}{4}\sigma_3 & \frac{45}{2}\mathbb{1} & -\frac{45}{4}\sigma_3 & -2\mathbb{1}
                                                                         \end{bmatrix}={\cal O_Q D_Q O}^{-1}_{\cal Q}
\end{equation}
Its diagonalization happens to be accidentally tractable:
\begin{multline}
 {\cal D}_{\cal Q}=\mbox{diag}\left[-\frac{19}{4}\mathbb{1},\frac{1}{8}(117-\sqrt{3073})\mathbb{1},\frac{1}{24}(277-\sqrt{3193})\mathbb{1},\frac{1}{24}(277+\sqrt{3193})\mathbb{1},\frac{1}{8}(117+\sqrt{3073})\mathbb{1}\right]\\
{\cal O_Q}=\begin{bmatrix}
1 & -\frac{1}{12}(47+\sqrt{3073}) & -\frac{1}{54}(67+\sqrt{3193} & \frac{1}{2} & \frac{1}{24}(-47+\sqrt{3073})\\
-2 & \frac{1}{6}(47+\sqrt{3073}) & -\frac{1}{27}(67+\sqrt{3193}) & 1 & \frac{1}{12}(47-\sqrt{3073})\\
\frac{7}{60} & 1 & \frac{1}{2} & -\frac{1}{96}(67+\sqrt{3193}) & \frac{1}{2}\\
-\frac{7}{30} & -2 & 1 & -\frac{1}{48}(67+\sqrt{3193}) & -1\\
-3 & 2 & 0 & 0 & 1            
           \end{bmatrix}
\end{multline}
so that the exponentiation is fully determined.

\subsection{Hybrid $(\bar{b}s)$-Gauge operators}
We consider the matrix ${\cal A}_{\cal HH}$ defined in Eq.(\ref{RGEH}):
\begin{equation}
 {\cal A}_{\cal HH}=
\begin{bmatrix}
\frac{5}{24}\mathbb{1} & \frac{11}{48}\sigma_3 & 0 \\
\frac{11}{12}\sigma_3 & \frac{5}{24}\mathbb{1}& 0 \\
 0 & 0 & \frac{11}{4}\mathbb{1}
\end{bmatrix}={\cal O_H D_H O}^{-1}_{\cal H}
\end{equation}
Its diagonalization is straigtforward:
\begin{equation}
 {\cal O_H}\equiv\frac{1}{\sqrt{2}}\begin{bmatrix}
                  \mathbb{1} & \frac{1}{2}\sigma_3 & 0 \\
                  -{2}\sigma_3 & \mathbb{1} & 0 \\
                  0 & 0 & \sqrt{2}\mathbb{1}
                 \end{bmatrix}\ \ ;\ \ {\cal D_H}\equiv\mbox{diag}\left[-\frac{1}{4}\mathbb{1},\frac{2}{3}\mathbb{1},\frac{11}{4}\mathbb{1}\right]
\end{equation}
allowing for a simple exponentiation:
\begin{multline}
 \exp\left[\frac{C_2(3)}{\beta_0}{\cal A}_{\cal HH}\ln\frac{\alpha_S(\mu)}{\alpha_S(\mu_0)}\right]=\left(\frac{\alpha_S(\mu)}{\alpha_S(\mu_0)}\right)^{\frac{5C_2(3)}{24\beta_0}}\begin{bmatrix}
\cosh\vartheta(\mu)\mathbb{1} & \frac{1}{2}\sinh\vartheta(\mu)\sigma_3 & 0\\
-2\sinh\vartheta(\mu)\sigma_3 & \cosh\vartheta(\mu)\mathbb{1} & 0\\
 0 & 0 & \left(\frac{\alpha_S(\mu)}{\alpha_S(\mu_0)}\right)^{\frac{61C_2(3)}{24\beta_0}}\mathbb{1}
\end{bmatrix}\\
\ \ ; \ \ \vartheta(\mu)\equiv\frac{11C_2(3)}{24\beta_0}\ln\frac{\alpha_S(\mu)}{\alpha_S(\mu_0)}
\end{multline}

\subsection{$(\bar{b}s)(\bar{q}q)$ sector}
We consider the matrix ${\cal A}_{qq}$ defined in Eq.(\ref{RGEHq}):
\begin{multline}
 {\cal A}_{qq}=\frac{1}{4}[\mathbb{1};-\tilde{V}^{q'}_{\mbox{Fierz}}]_{q'=b,s}\begin{bmatrix}
                0 & 0 & 0 & 0 \\
                0 & 0 & 0 & 0 \\
                -\frac{1}{3}(\mathbb{1}+\tilde{\Sigma}) & 0 & \mathbb{1}+\tilde{\Sigma} & 0\\
                0 & 0 & 0 & 0
               \end{bmatrix}\begin{bmatrix}\mathbb{1}\\-\tilde{V}^q_{\mbox{Fierz}}\end{bmatrix}_{q=b,s}\\
={\cal O}_q\mbox{diag}(0,0,0,0,0,0,0,0,0,0,0,0,0,0,0,\frac{8}{3}{\cal D}_{24}){\cal O}_q^{-1}
\end{multline}

\noindent Let us introduce the following blocks:
\begin{equation}
{\cal D}_{24}\equiv\begin{matrix}0&0&0&0\\0&1&0&0\\0&0&0&0\\0&0&0&1\end{matrix} \ \ \ \ \ \ ;\ \ \ \ \ \ 
{\cal R}\equiv\frac{1}{\sqrt{2}}\begin{matrix}-1&0&0&1\\1&0&0&1\\0&1&-1&0\\0&1&1&0\end{matrix}\ \ ;\ \ 
\end{equation}
\begin{displaymath}
{\cal U}^s\equiv\frac{1}{\sqrt{2}}\begin{matrix}1&0&0&-1&0&1&0&0\\-1&0&0&0&0&-4&0&0\\0&-1&1&0&0&0&0&1\\0&0&-1&0&0&0&0&-4\\0&0&0&0&1&0&0&0\\0&-1&-1&0&0&0&0&0\\0&0&0&0&0&0&1&0\\-1&0&0&-1&0&0&0&0\end{matrix}\ \ ;\ \ 
{\cal U}^b\equiv\frac{1}{\sqrt{2}}\begin{matrix}-1&0&0&0&0&-4&0&0\\1&0&0&-1&0&1&0&0\\0&0&-1&0&0&0&0&-4\\0&-1&1&0&0&0&0&1\\-1&0&0&-1&0&0&0&0\\0&0&0&0&0&0&-1&0\\0&-1&-1&0&0&0&0&0\\0&0&0&0&-1&0&0&0\end{matrix}
\end{displaymath}
\begin{multline}
 {\cal O}_q=\begin{bmatrix}
\begin{matrix}
 \mathbb{1} & 0 & 0 & 0 & \null\\
 0 & \mathbb{1} & 0 & 0 & \null\\
 \frac{\mathbb{1}}{3} & 0 & {\cal R} & 0 & \null\\
 0 & 0 & 0 & \mathbb{1} & \null\\
 \null & \null & \null & \null & \null
\end{matrix}&
\begin{matrix}
 0 & 0 & 0 & 0 & \null\\
 0 & 0 & 0 & 0 & \null\\
 0 & 0 & 0 & 0 & \null\\
 0 & 0 & 0 & 0 & \null\\
 \null & \null & \null & \null & \null
\end{matrix}&
\begin{matrix}
 0 & 0 & 0 & 0 & \null\\
 0 & 0 & 0 & 0 & \null\\
 0 & 0 & 0 & 0 & \null\\
 0 & 0 & 0 & 0 & \null\\
 \null & \null & \null & \null & \null
\end{matrix}&
\begin{matrix}
 0 & 0 & \null\\
 0 & 0 & \null\\
 0 & 0 & \null\\
 0 & 0 & \null\\
 \null & \null & \null 
\end{matrix}&
\begin{matrix}
 0 & 0 & \null\\
 0 & 0 & \null\\
 0 & 0 & \null\\
 0 & 0 & \null\\
 \null & \null & \null 
\end{matrix}\\
\begin{matrix}
 0 & 0 & 0 & 0 & \null\\
 0 & 0 & 0 & 0 & \null\\
 0 & 0 & 0 & 0 & \null\\
 0 & 0 & 0 & 0 & \null\\
 \null & \null & \null & \null & \null
\end{matrix}&
\begin{matrix}
 \mathbb{1} & 0 & 0 & 0 & \null\\
 0 & \mathbb{1} & 0 & 0 & \null\\
 \frac{\mathbb{1}}{3} & 0 & {\cal R} & 0 & \null\\
 0 & 0 & 0 & \mathbb{1} & \null\\
 \null & \null & \null & \null & \null
\end{matrix}&
\begin{matrix}
 0 & 0 & 0 & 0 & \null\\
 0 & 0 & 0 & 0 & \null\\
 0 & 0 & 0 & 0 & \null\\
 0 & 0 & 0 & 0 & \null\\
 \null & \null & \null & \null & \null
\end{matrix}&
\begin{matrix}
 0 & 0 & \null\\
 0 & 0 & \null\\
 0 & 0 & \null\\
 0 & 0 & \null\\
 \null & \null & \null 
\end{matrix}&
\begin{matrix}
 0 & 0 & \null\\
 0 & 0 & \null\\
 0 & 0 & \null\\
 0 & 0 & \null\\
 \null & \null & \null 
\end{matrix}\\
\begin{matrix}
 0 & 0 & 0 & 0 & \null\\
 0 & 0 & 0 & 0 & \null\\
 0 & 0 & 0 & 0 & \null\\
 0 & 0 & 0 & 0 & \null\\
 \null & \null & \null & \null & \null
\end{matrix}&
\begin{matrix}
 0 & 0 & 0 & 0 & \null\\
 0 & 0 & 0 & 0 & \null\\
 0 & 0 & 0 & 0 & \null\\
 0 & 0 & 0 & 0 & \null\\
 \null & \null & \null & \null & \null
\end{matrix}&
\begin{matrix}
 \mathbb{1} & 0 & 0 & 0 & \null\\
 0 & \mathbb{1} & 0 & 0 & \null\\
 \frac{\mathbb{1}}{3} & 0 & {\cal R} & 0 & \null\\
 0 & 0 & 0 & \mathbb{1} & \null\\
 \null & \null & \null & \null & \null
\end{matrix}&
\begin{matrix}
 0 & 0 & \null\\
 0 & 0 & \null\\
 0 & 0 & \null\\
 0 & 0 & \null\\
 \null & \null & \null 
\end{matrix}&
\begin{matrix}
 0 & 0 & \null\\
 0 & 0 & \null\\
 0 & 0 & \null\\
 0 & 0 & \null\\
 \null & \null & \null 
\end{matrix}\\
\begin{matrix}
 0 & 0 & 0 & 0 & \null\\
 0 & 0 & 0 & 0 & \null\\
 \null & \null & \null & \null & \null
\end{matrix}&
\begin{matrix}
 0 & 0 & 0 & 0 & \null\\
 0 & 0 & 0 & 0 & \null\\
 \null & \null & \null & \null & \null
\end{matrix}&
\begin{matrix}
 0 & 0 & 0 & 0 & \null\\
 0 & 0 & 0 & 0 & \null\\
 \null & \null & \null & \null & \null
\end{matrix}&
{\cal U}^s&
\begin{matrix}
 0 & 0 & \null\\
 0 & 0 & \null\\
 \null & \null & \null 
\end{matrix}\\
\begin{matrix}
 0 & 0 & 0 & 0 & \null\\
 0 & 0 & 0 & 0 & \null
\end{matrix}&
\begin{matrix}
 0 & 0 & 0 & 0 & \null\\
 0 & 0 & 0 & 0 & \null
\end{matrix}&
\begin{matrix}
 0 & 0 & 0 & 0 & \null\\
 0 & 0 & 0 & 0 & \null
\end{matrix}&
\begin{matrix}
 0 & 0 & \null\\
 0 & 0 & \null
\end{matrix}&
{\cal U}^b
\end{bmatrix}\\
\times\begin{bmatrix}
\begin{matrix}
 \mathbb{1} & 0 & 0 & 0 & \null\\
 0 & \mathbb{1} & 0 & 0 & \null\\
 0 & 0 & \mathbb{1} & 0 & \null\\
 0 & 0 & 0 & \mathbb{1} & \null\\
 \null & \null & \null & \null & \null
\end{matrix}&
\begin{matrix}
 0 & 0 & 0 & 0 & \null\\
 0 & 0 & 0 & 0 & \null\\
 0 & 0 & 0 & 0 & \null\\
 0 & 0 & 0 & 0 & \null\\
 \null & \null & \null & \null & \null
\end{matrix}&
\begin{matrix}
 0 & 0 & 0 & 0 & \null\\
 0 & 0 & 0 & 0 & \null\\
 0 & 0 & 0 & 0 & \null\\
 0 & 0 & 0 & 0 & \null\\
 \null & \null & \null & \null & \null
\end{matrix}&
\begin{matrix}
 0 & 0 & \null\\
 0 & 0 & \null\\
 {\cal D}_{24} & 0 &  \null\\
 0 & 0 & \null\\
 \null & \null & \null 
\end{matrix}&
\begin{matrix}
 0 & 0 & \null\\
 0 & 0 & \null\\
 {\cal D}_{24} &  0 & \null\\
 0 & 0 & \null\\
 \null & \null & \null 
\end{matrix}\\
\begin{matrix}
 0 & 0 & 0 & 0 & \null\\
 0 & 0 & 0 & 0 & \null\\
 0 & 0 & -{\cal D}_{24} & 0 & \null\\
 0 & 0 & 0 & 0 & \null\\
 \null & \null & \null & \null & \null
\end{matrix}&
\begin{matrix}
 \mathbb{1} & 0 & 0 & 0 & \null\\
 0 & \mathbb{1} & 0 & 0 & \null\\
 0 & 0 & \mathbb{1} & 0 & \null\\
 0 & 0 & 0 & \mathbb{1} & \null\\
 \null & \null & \null & \null & \null
\end{matrix}&
\begin{matrix}
 0 & 0 & 0 & 0 & \null\\
 0 & 0 & 0 & 0 & \null\\
 0 & 0 & 0 & 0 & \null\\
 0 & 0 & 0 & 0 & \null\\
 \null & \null & \null & \null & \null
\end{matrix}&
\begin{matrix}
 0 & 0 & \null\\
 0 & 0 & \null\\
 {\cal D}_{24} & 0 &  \null\\
 0 & 0 & \null\\
 \null & \null & \null 
\end{matrix}&
\begin{matrix}
 0 & 0 & \null\\
 0 & 0 & \null\\
 {\cal D}_{24} & 0 &  \null\\
 0 & 0 & \null\\
 \null & \null & \null 
\end{matrix}\\
\begin{matrix}
 0 & 0 & 0 & 0 & \null\\
 0 & 0 & 0 & 0 & \null\\
 0 & 0 & 0 & 0 & \null\\
 0 & 0 & 0 & 0 & \null\\
 \null & \null & \null & \null & \null
\end{matrix}&
\begin{matrix}
 0 & 0 & 0 & 0 & \null\\
 0 & 0 & 0 & 0 & \null\\
 0 & 0 & -{\cal D}_{24} & 0 & \null\\
 0 & 0 & 0 & 0 & \null\\
 \null & \null & \null & \null & \null
\end{matrix}&
\begin{matrix}
 \mathbb{1} & 0 & 0 & 0 & \null\\
 0 & \mathbb{1} & 0 & 0 & \null\\
 0 & 0 & \mathbb{1}-{\cal D}_{24} & 0 & \null\\
 0 & 0 & 0 & \mathbb{1} & \null\\
 \null & \null & \null & \null & \null
\end{matrix}&
\begin{matrix}
 0 & 0 & \null\\
 0 & 0 & \null\\
 {\cal D}_{24} & 0 &  \null\\
 0 & 0 & \null\\
 \null & \null & \null 
\end{matrix}&
\begin{matrix}
 0 & 0 & \null\\
 0 & 0 & \null\\
 {\cal D}_{24} & 0 &  \null\\
 0 & 0 & \null\\
 \null & \null & \null 
\end{matrix}\\
\begin{matrix}
 0 & 0 & 0 & 0 & \null\\
 0 & 0 & 0 & 0 & \null\\
 \null & \null & \null & \null & \null
\end{matrix}&
\begin{matrix}
 0 & 0 & 0 & 0 & \null\\
 0 & 0 & 0 & 0 & \null\\
 \null & \null & \null & \null & \null
\end{matrix}&
\begin{matrix}
 0 & 0 & {\cal D}_{24} & 0 & \null\\
 0 & 0 & 0 & 0 & \null\\
 \null & \null & \null & \null & \null
\end{matrix}&
\begin{matrix}
 \mathbb{1}-\frac{16}{7}{\cal D}_{24} & 0 & \null\\
 0 & \mathbb{1} & \null\\
 \null & \null & \null 
\end{matrix}&
\begin{matrix}
 0 & 0 & \null\\
  0 & {\cal D}_{24} & \null\\
 \null & \null & \null 
\end{matrix}\\
\begin{matrix}
 0 & 0 & 0 & 0 & \null\\
 0 & 0 & 0 & 0 & \null
\end{matrix}&
\begin{matrix}
 0 & 0 & 0 & 0 & \null\\
 0 & 0 & 0 & 0 & \null
\end{matrix}&
\begin{matrix}
 0 & 0 & -{\cal D}_{24} & 0 & \null\\
 0 & 0 & 0 & 0 & \null
\end{matrix}&
\begin{matrix}
 -\frac{9}{7}{\cal D}_{24} & 0 & \null\\
  0 & 0 & \null
\end{matrix}&
\begin{matrix}
 \mathbb{1} & 0 & \null\\
 0 & \mathbb{1} & \null
\end{matrix}
\end{bmatrix}
\end{multline}
${\cal O}_q$ turns out to diagonalize ${\cal A}_{qq}$ satisfactorily.

\end{document}